\documentclass[fleqn,usenatbib, onecolumn, untrackchanges]{aastex6}
\usepackage{newtxtext,newtxmath}
\usepackage[T1]{fontenc}
\usepackage{ae,aecompl}
\usepackage{graphicx}	
\usepackage{amsmath}	
\usepackage{amssymb}	
\usepackage{hyperref}
\usepackage{makecell}

\definecolor{darkblue}{rgb}{0.0,0.0,0.5}
\hypersetup{ colorlinks, citecolor=darkblue, filecolor=darkblue, linkcolor=darkblue, urlcolor=blue } 

\newcommand{\be}{\begin{equation}}
\newcommand{\ee}{\end{equation}}

\usepackage{ulem}

\usepackage{color}

\title{Inconsistency with ultra-high energy cosmic ray acceleration within synchrotron model gamma-ray bursts}

\hypersetup{final}
\begin{document}
\label{firstpage}

\title[The limited contribution of GRBs to UHECRs]{The limited contribution of low- and high-luminosity gamma-ray bursts to ultra-high energy cosmic rays}

\author{Filip Samuelsson\altaffilmark{1,2}, Damien B\'egu\'e\altaffilmark{3}, Felix Ryde\altaffilmark{1,2}, Asaf Pe'er\altaffilmark{4,5}}
\shortauthors{Samuelsson et al.}
\shorttitle{The limited contribution of GRBs to UHECRs}

\altaffiltext{1}{ Department of Physics, KTH Royal Institute of Technology, AlbaNova, SE-106 91 Stockholm, Sweden} 
\altaffiltext{2}{ The Oskar Klein Centre for Cosmoparticle Physics}
\altaffiltext{3}{ Max-Planck-Institut f{\"u}r extraterrestrische Physik, Giessenbachstrasse, D-85748 Garching, Germany} 
\altaffiltext{4}{ Department of Physics, University College Cork, Cork, Ireland} 
\altaffiltext{5}{ Department of Physics, Bar-Ilan University, Ramat-Gan 52900, Israel}


\begin{abstract}
The acceleration site for ultra-high energy cosmic rays \added{(UHECR)} is still an open question despite extended research. 
In this paper, 
we reconsider the prompt phase of gamma-ray bursts \added{(GRBs)} as a possible candidate for this acceleration and constrain the maximum proton energy in optically thin synchrotron and photospheric models, using properties of the prompt photon spectra. 
We find that neither of the models favour acceleration of protons to $10^{20}$ eV in high-luminosity bursts. 
We repeat the calculations for low-luminosity \replaced{gamma-ray bursts}{GRBs} \added{(llGRBs)} considering both protons and completely stripped iron and find that the highest obtainable energies are $< 10^{19}$ eV and $< 10^{20}$ eV for protons and iron respectively, regardless of the model. We conclude therefore that for \replaced{standard}{our fiducial} parameters, GRBs, including low-luminosity bursts, contribute little to none to the UHECR observed.
\replaced{We further constrain any association between ultra-high energy cosmic rays and low-luminosity gamma-ray bursts. We find that it would require very efficient acceleration, favor photospheric origin of the prompt gamma-rays, result in high prompt optical spectral fluxes, and demand small fractional energy given to a small fraction of accelerated electrons, challenging low-luminosity gamma-ray bursts as sources of $10^{20}$ eV iron.}{We further constrain the conditions necessary for an association between UHECR and llGRBs and find that iron can be accelerated to $10^{20}$ eV in photospheric models, given very efficiency acceleration and/or a small fractional energy given to a small fraction of accelerated electrons. This will necessarily result in high prompt optical fluxes, and the detection of such a signal could therefore be an indication of successful UHECR acceleration at the source.}

\end{abstract}
\begin{keywords}
{gamma-ray bursts --- ultra-high energy cosmic rays}
\end{keywords}

\maketitle

\section{Introduction}
\label{IntroductionSection}

The acceleration sites of ultra-high energy cosmic rays (UHECR) 
with energies $E > \ $few$\ \times 10^{18}$ eV is still unknown despite several decades of research. While supernova remnants seem to be confirmed cosmic ray (CR) accelerators (see e.g. \citet{Koyama1995,Neronov2017}), they can only generate particles with energies up to the knee at $E_{\rm knee} \sim 10^{15.5}$ eV \citep{Lagage1983}. The most energetic CRs with energies above $\sim 10^{18}$ eV have such large gyroradii that they can diffuse out of the galactic disk, leading to the assumption that the ones we observe have an extragalactic origin \citep{Budnik2008, Thoudam2016}. The recent detection of a high-energy neutrino from a flaring blazar has shed new light on the high-energy cosmic ray problem \citep{IceCubeCollaboration2018}. Yet, modeling of the blazar spectral energy distribution (SED) in a single-zone model seem to disfavor UHECR acceleration \citep{Keivani2018}. Studying UHECR production in connection with the electromagnetic spectrum is especially timely now in the multi-messenger era, and understanding UHECR is the next step towards completing the picture of the CR spectrum.

\citet{Waxman1995} (hereafter W95) was early to suggest that the prompt phase of cosmic gamma-ray bursts (GRBs) could accelerate particles to the highest observed energies ($E \sim 3\times 10^{20}$ eV). Additionally, the UHECR flux observed at Earth could also be explained as long as GRBs emit an energy in UHERC comparable to that emitted in $\gamma$-ray photons, however this relies on the distribution of GRBs being roughly constant towards low redshift \citep{Stecker2000}. Today, GRBs are still viewed among the prime candidates for UHECR acceleration \citep{Baerwald2015, Globus2015, Biehl2018, Zhang2018}, together with active galactic nuclei (AGN) \citep{Murase2012}, millisecond pulsars \citep{Blasi2000}, tidal disruption events \citep{Biehl2017}, and starburst galaxies (\citet{PierreAugerCollaboration2018}; see however \citet{Sudoh2018}). 

Accelerating UHECR to the highest energies observed puts severe constraints on the physical conditions at the acceleration site, for instance on the comoving magnetic field $B'$, see e.g. W95. Under the plausible assumption that electrons are also accelerated at the site, the emitted synchrotron spectrum from these electrons will be imprinted with these constraints. By using known properties of the observed photon spectra, we evaluate the compatibility of UHECR acceleration with GRB observations. 
We do not address the kinetic details of particle acceleration or their escape. 

GRBs are most commonly described by the fireball model \citep{Paczynski1986,Rees&Meszaros1992,Rees&Meszaros1994,Piran1993}, where initial gravitational energy is converted into thermal energy and the high radiation pressure accelerates the outflow. Alternatively, the outflow can be accelerated magnetically, by magnetic reconnection \citep{Drenkhahn2002} or launched by the Blandford-Znajek mechanism \citep{BlandfordZnajek1977}. In each case, the outflow reaches its maximum velocity at the saturation radius, where most of the energy is in kinetic form. This kinetic energy must then be dissipated to create the prompt $\gamma$-ray radiation that we observe. There is still an ongoing debate regarding the mechanism behind the prompt emission. Leading models include photospheric emission of the outflow when it becomes transparent to Compton scattering at the photosphere \citep{Goodman1986, Paczynski1986, Ryde2004, Rees&Meszaros2005, Peer2005, Beloborodov2010, Ruffini2013, Vereshchagin2014, Peer2015}, internal shocks \citep{Rees&Meszaros1994, Daigne1998, Kobayashi1999}, and energy dissipation by magnetic reconnection \citep{Usov1992,Spruit&Daigne&Drenkhahn2001, Lyutikov2003, Giannos2006, Zhang&Yan2011, Begue2017}. Provided that dissipation takes place far enough from the central engine, shocks and magnetic reconnection result in the emission of optically thin synchrotron radiation.

In synchrotron models, electrons are either in a slow cooling (SC) regime, where they lose most of their energy due to the adiabatic expansion of the ejecta, or in a fast cooling (FC) regime, where the faster synchrotron losses dominate. This results in a cooling break at $\gamma_{\rm c}$; the electron Lorentz factor for which these two loss processes are equal. The emission region is characterized to be SC or FC depending on if the minimum electron Lorentz factor $\gamma_{\rm m}$ is less than or greater than $\gamma_{\rm c}$ respectively. 
In the context of synchrotron models of GRBs in the prompt emission phase, one expects the emission to be in the FC regime. Indeed, an inherent problem with SC emission is its inefficiency in converting the kinetic energy of the outflow into radiation. However, spectral fits to GRB prompt phases show that about 2/3 of GRBs have a harder low energy slope than what FC synchrotron emission can produce, leading  to the "synchrotron line of death" problem \citep{Crider1997, Preece1998}. Physical synchrotron models have been fitted directly to the data of single pulse bursts, see e.g. \citet{Burgess2018}, showing that synchrotron could, in these cases, be the emission mechanism. Yet the fits require that $\gamma_{\rm c}$ be close to $\gamma_{\rm m}$, the so-called marginally fast cooling regime.

Photospheric models of GRBs can account for the hard low-energy slope of the observed spectrum, as well as the clustering of the peak energy \citep{Peer2005}. 
While many bursts show clear signs of a photospheric component modeled by a black body, e.g. \citet{Ryde2004}, the emission is never purely thermal, maybe with rare exceptions such as GRB 090902B and GRB 100507 \citep{Ryde2010, Ryde2017}. However, photospheric models have problems explaining the softest observed spectra, unless the outflow moves slowly with bulk Lorentz factor $\Gamma$ in the order of a few tens, \citep{Vurm2013,Beloborodov2013}, as well as bursts with long variability times 
\citep{Fishman&Meegan1995, Margutti2011}. The latter can be reconciled if the variability time scale is set by the behavior of the central engine.




All models have different attributes that one can use in order to characterize the magnetic field in the emitting region and as such constrain the maximum proton energy. In addition, to give predictions for the observed electromagnetic signal in these models given UHECR acceleration, is a powerful tool that has so far been largely overlooked. 
In this paper, we use known properties of the GRB spectrum to put constraints on the highest obtainable cosmic ray energies at the source ($< 10^{20}$ eV), but we also give predictions of what electromagnetic signatures could be indicative of successful UHECR acceleration (high spectral fluxes in the optical band).

The paper is structured as follows. In Section \ref{MagneticFieldSection}, we derive what constraints proton acceleration to the highest energies observed puts on the comoving magnetic field. In Section \ref{Sec:Results}, we evaluate if these constraints can be reconciled with observation for synchrotron radiation models (Subsection \ref{Sec:SynchrotronPrompt}) and photospheric emission models (Subsection \ref{Sec:PhotosphericEmission}). We then investigate low-luminosity GRBs as UHECR sources,  considering both protons and iron in Section \ref{Sec:LowLum}. Finally, we discuss the dependency of the  results on the parameters in Section \ref{Sec:Discussion} and finish with a conclusion in Section \ref{Sec:Conslusion}. 


\section{Requirements on magnetic field for UHECR acceleration}\label{MagneticFieldSection}

To estimate the maximum energy a particle can reach, we compare its acceleration time $t'_{\rm acc}$ to the different time scales of energy losses. Following W95, we consider energy losses from synchrotron cooling, adiabatic expansion, and photo-hadronic interaction, with corresponding typical time scales $t'_{\rm sync}$, $t'_{\rm ad}$, and $t'_{p\gamma}$ respectively. Here and in the remainder of the paper, primed quantities are expressed in the frame comoving with the outflow. We do not discuss cooling by the Bethe-Heitler process, as it is shown to be weakly constraining, see \citet{Denton2018}. In addition, energy losses from hadron-hadron collisions are less relevant than photo-hadronic collisions despite the larger cross section of the former, due to the much lower baryon density compared to photon density in the outflow \citep{Hummer2010}. The discussion of this section follows closely that of W95, and is presented for completeness.




The acceleration time scale for diffusive shock acceleration for a strong shock is 
\begin{equation}\label{t_acc}
	t'_{\text{acc},i} = \frac{E}{\eta c Z_i e B' \Gamma}.
\end{equation}
Here, $E$ is the measured UHECR energy which has to be divided by the bulk Lorentz factor of the flow $\Gamma$ to get the corresponding energy in the rest frame of the outflow, $\eta$ is the acceleration efficiency, $c$ is the speed of light in vacuum, $Z_i$ is the charge number for particle species $i$, and $e$ is the elementary charge. The acceleration efficiency $\eta$ is defined such that a higher value means more efficient acceleration. In this paper, we use $\eta = 0.1$, which is quite conservative \citep{Protheroe2004, Rieger2007} and in reasonable agreement with numerical simulations, see e.g. \citet{Caprioli2014}. The true value of $\eta$ is unknown and is most likely dependent on the dynamics of the acceleration. For completeness, the effect of varying the acceleration efficiency is discussed in Section \ref{Sec:Discussion}. 
Synchrotron energy loss time is dependent on the mass of the radiator. For particle species $i$, it is given by
%
\begin{equation}
	t'_{\textrm{sync},i} = \frac{6\pi}{Z_i^4 \sigma_{\rm T}} \frac{(m_i c^2)^2}{c} \left ( \frac{m_i}{m_e} \right )^2 \frac{1}{ (E/\Gamma) {B'}^2} \label{eq:t_sync}
\end{equation}
where $m_i$ is the mass of particle species $i$, $\sigma_{\rm T}$ is the Thompson cross section, and $m_e$ is the electron mass. At distance $r$ from the progenitor, the adiabatic cooling time scale, which is the same as the expansion time scale, is 
\begin{equation}
	t'_\textrm{ad} = \frac{r}{c\Gamma}. \label{eq:t_exp}
\end{equation}
The cooling time for protons by photo-hadronic interactions is $t'_{p\gamma} = (c K_{p\gamma} n'_\gamma \sigma_{p\gamma})^{-1}$, where $K_{p\gamma} \approx 0.2$ is the proton inelasticity \citep{Mucke1999}, $n'_\gamma$ is the photon density in the rest frame of the outflow and $\sigma_{p\gamma} \approx 10^{-28}$ cm$^2$ is the photo-hadronic cross section.\footnote{This value of the cross section is justified for very energetic protons. If anything, the cross section should be slightly larger \citep{CrossSectionPGamma2016,Hummer2010}, resulting in harsher results than those presented here.} The photon density can be estimated from the observed isotropic $\gamma$-luminosity as $L_\gamma = 4\pi r^2 c \Gamma n'_\gamma \left< \varepsilon \right>$, where $\left< \varepsilon \right>$ is the typical photon energy observed. Thus,
\begin{equation}\label{t_pg}
	t'_{p\gamma} = \frac{20 \pi r^2 \Gamma \left< \varepsilon \right>}{\sigma_{p\gamma}L_\gamma}.
\end{equation}
In addition to the condition that the acceleration time scale needs to be less than the smallest energy loss time scale, 
the Larmor radius of the particle should be smaller than the system size, of the order of $r/ \Gamma$. However, as pointed out in W95, this constraint is strictly weaker than the constraint $t'_{\rm acc} < t'_{\rm ad}$ for $\eta < 1$, and gives no additional information. If $\eta > 1$, the requirement on Larmor radius has to be considered.  

From now on, we consider a pure proton UHECR composition. Indeed, the metallicity in GRB jets is expected to be very low, since heavier elements are destroyed by the intense photon field \citep{Horiuchi2012}. This was thoroughly investigated in \citet{Zhang2018}. From Figure 6 in their paper it is evident that for a typical injection radius $r_0 \sim 10^7 - 10^8$ cm, iron and other heavy elements will be disintegrated by the high radiation field at the base of the jet. This is not necessarily true for low-luminosity GRBs, and we consider these events in Section \ref{Sec:LowLum}.
To achieve proton acceleration to energy $E$, one gets the following requirements on the comoving magnetic field: 
\begin{alignat}{4}
	t'_\textrm{acc}  &< t'_\textrm{sync} && \longrightarrow \ & B' & < B'_{\rm sync} && \equiv \frac{6\pi}{\sigma_{\textrm{T}}}\frac{(m_p c^2)^2}{ \left(m_e/m_p \right)^2} \frac{\eta e}{\left(E/\Gamma \right)^2} \label{Bsync}  \\
    && && && & \sim 4.1 \times 10^3 \eta_{-1} \left(\frac{\Gamma_2}{E_{20}}\right)^2 \textrm{ G}, \nonumber
\end{alignat}
\begin{alignat}{4}
	t'_\textrm{acc}  &< t'_\textrm{ad} && \longrightarrow \ & B' &> B'_{\rm ad} && \equiv \frac{E}{\eta e r} \label{Bad} \\
    && && && & \sim 3.3 \times 10^4 \frac{E_{20}}{\eta_{-1}r_{14}} \textrm{ G},\nonumber
\end{alignat}
\begin{alignat}{4}
	t'_\textrm{acc} & < t'_{p \gamma} && \longrightarrow \ & B' &> B'_{p\gamma} && \equiv \frac{ \sigma_{p\gamma}}{20 \pi e c}  \frac{E L_\gamma}{\eta \left<\varepsilon\right> r^2 \Gamma ^2} \label{Bpgamma} \\
     && && && & \sim 1.1 \times 10^3 \frac{E_{20}L_{\gamma, 51}}{\eta_{-1}\Gamma_2^2 \left<\varepsilon\right>_6 r_{14}^2} \textrm{ G}. \nonumber
\end{alignat}
%
%
In the equations above, we used the notation $Q = 10^xQ_x$, and we also introduced $m_p$ as the proton mass. 
In Figure \ref{Fig:MagneticFieldRestrictionSynchrotron} on the left, we plot the parameter space where UHECR acceleration is allowed, as a function of $B'$ and $r$ for three values of the outflow bulk Lorentz factor: $\Gamma$ = 100, 300 and 1000 from top to bottom. The observed luminosity is taken to be $L_\gamma = 10^{51}$ erg s$^{-1}$, $\eta = 0.1$, and the peak energy $\left< \varepsilon \right> = 300$ keV. The three limiting conditions $B'_{\rm sync}$,  $B'_{\rm ad}$, and  $B'_{p\gamma}$ are given by the dotted, dot-dashed and solid lines respectively. They are plotted for $\log(E/ \textrm{eV}) = 17$, 18, 19, 20, and 21 as indicated in the figures. The parameter space is bounded at low radii by the photospheric radius $r_{\rm ph} = L_{\rm tot}\sigma_{\rm T}/(8\pi m_p c^3\Gamma^3)$ where $L_{\rm tot}$ is the total isotropic luminosity of the burst, see e.g. \cite{Peer2015}, shown as the dashed vertical line. Indeed, it is unclear that particles can be accelerated below the photospheric radius to extreme relativistic velocities \citep{Budnik2010, Murase2013, Beloborodov2017}. On the top of Figure \ref{Fig:MagneticFieldRestrictionSynchrotron} and in forthcoming figures, there is an x-axis showing $\log (r/(2\Gamma^2 c))$, to give an indication of the expected variability time.

From Equations \ref{Bsync}, \ref{Bad} and \ref{Bpgamma}, we see that increasing $\eta$, i.e., faster particle acceleration, shifts the dotted horizontal synchrotron line to higher $B'$ values, while it shifts the dot-dashed and continuous lines to the left, resulting in larger parameter space available for particle acceleration. Changing the ratio of the luminosity to the peak energy $L_\gamma/\left< \varepsilon \right>$ to larger values shifts the solid line to the right, reducing the available parameter space. We also checked the possibility of correlating the $\gamma$-ray luminosity to the peak energy via the Yonetoku relation \citep{Yonetoku2010}, but the additional assumption does not change the  results presented below and is therefore not required for our analysis.

%
%

%
%
\section{Results}\label{Sec:Results}
\subsection{Synchrotron dominated prompt emission}\label{Sec:SynchrotronPrompt}
\begin{figure*}
    \includegraphics[width=0.45\columnwidth]{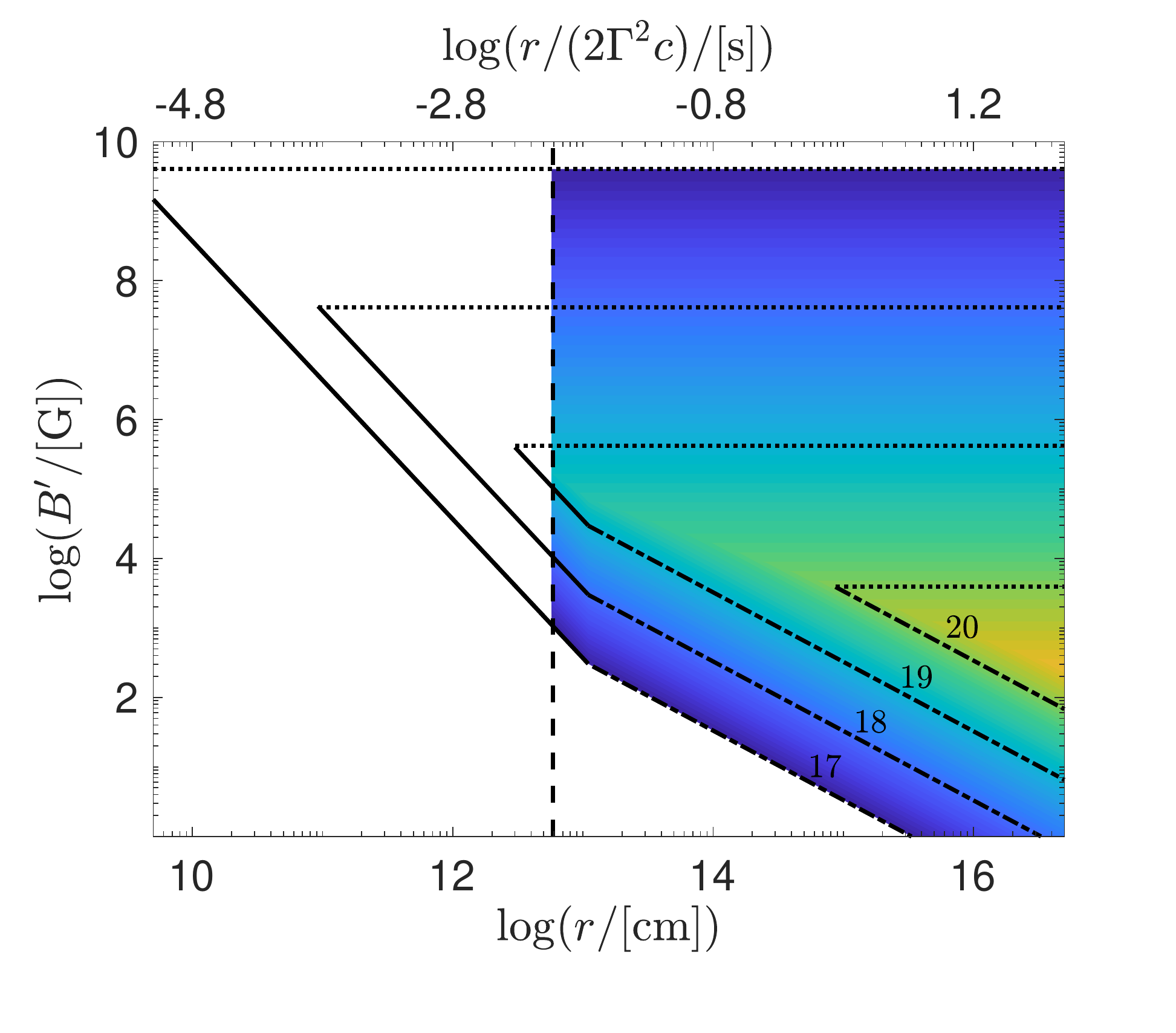}\hfill
    \includegraphics[width=0.45\columnwidth]{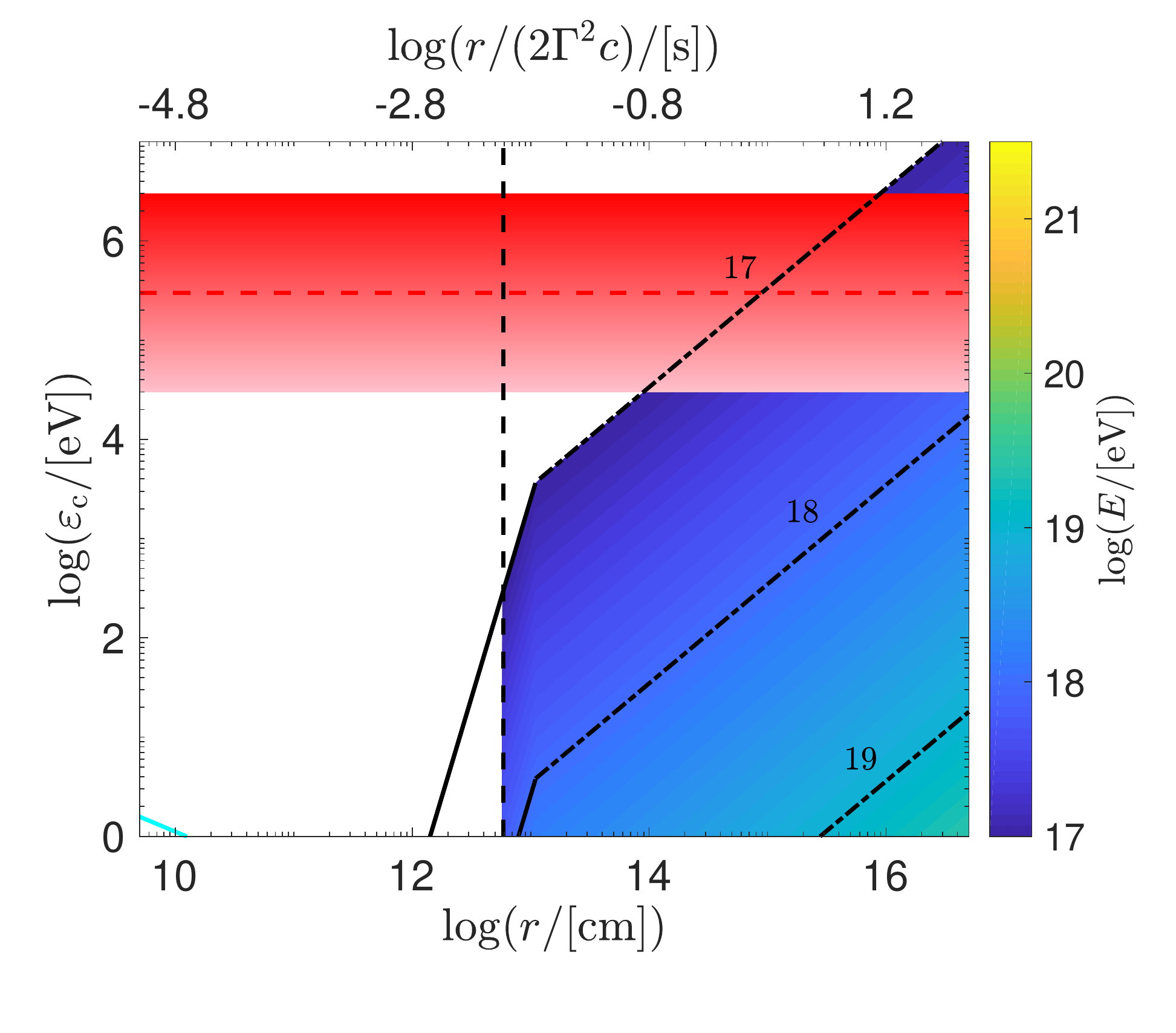}\hfill
	\includegraphics[width=0.45\columnwidth]{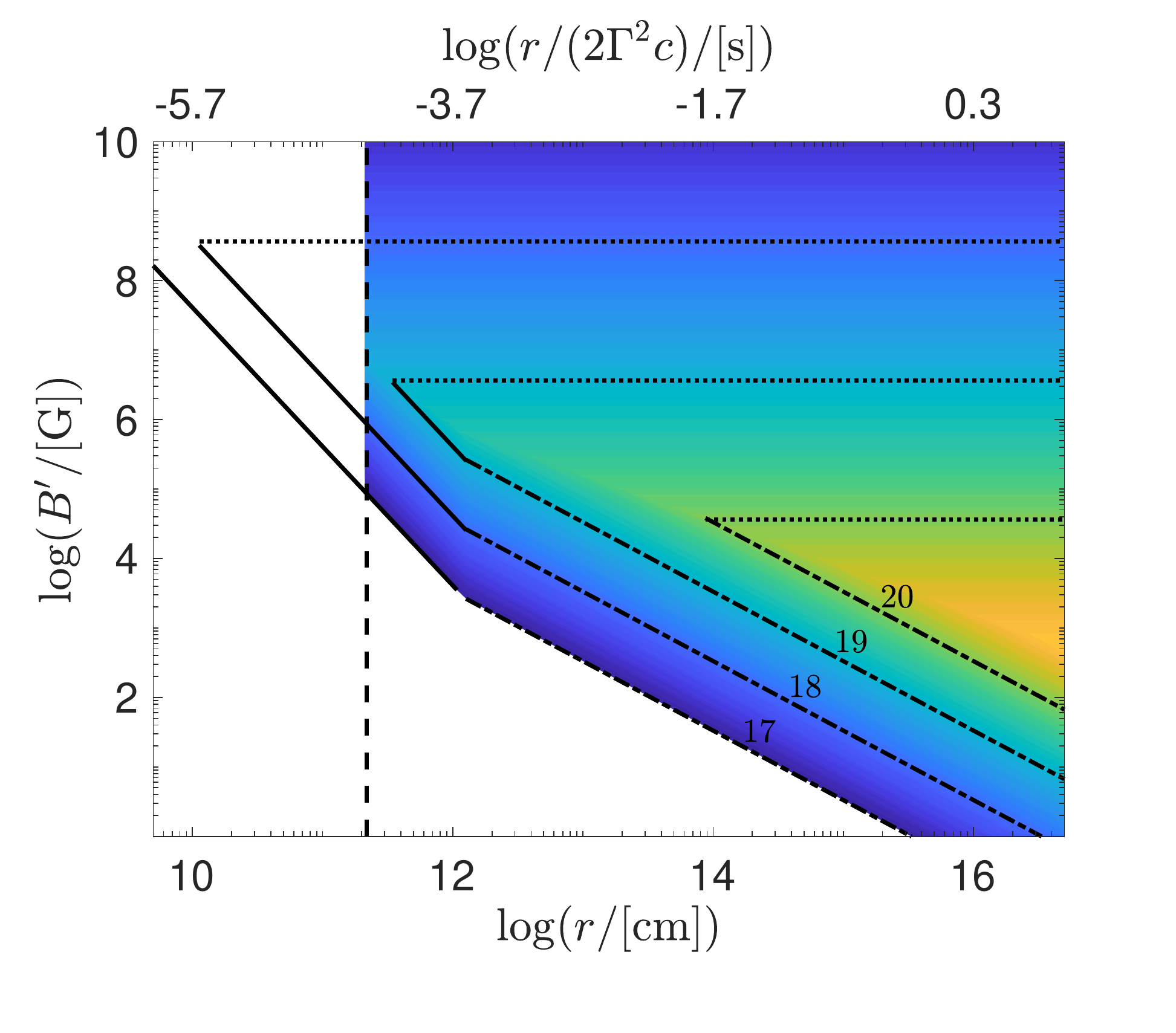}\hfill
    \includegraphics[width=0.45\columnwidth]{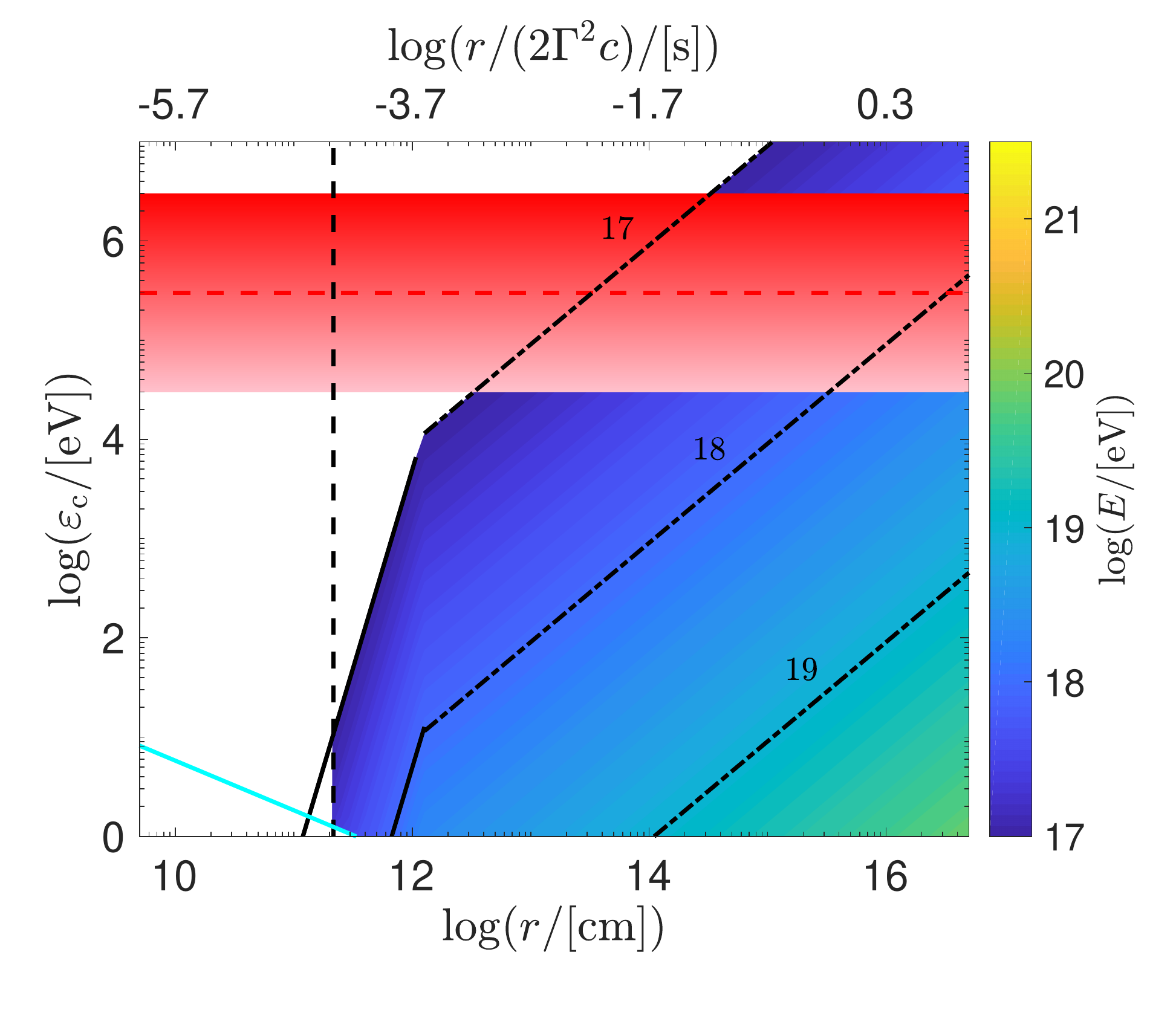}\hfill
 	\includegraphics[width=0.45\columnwidth]{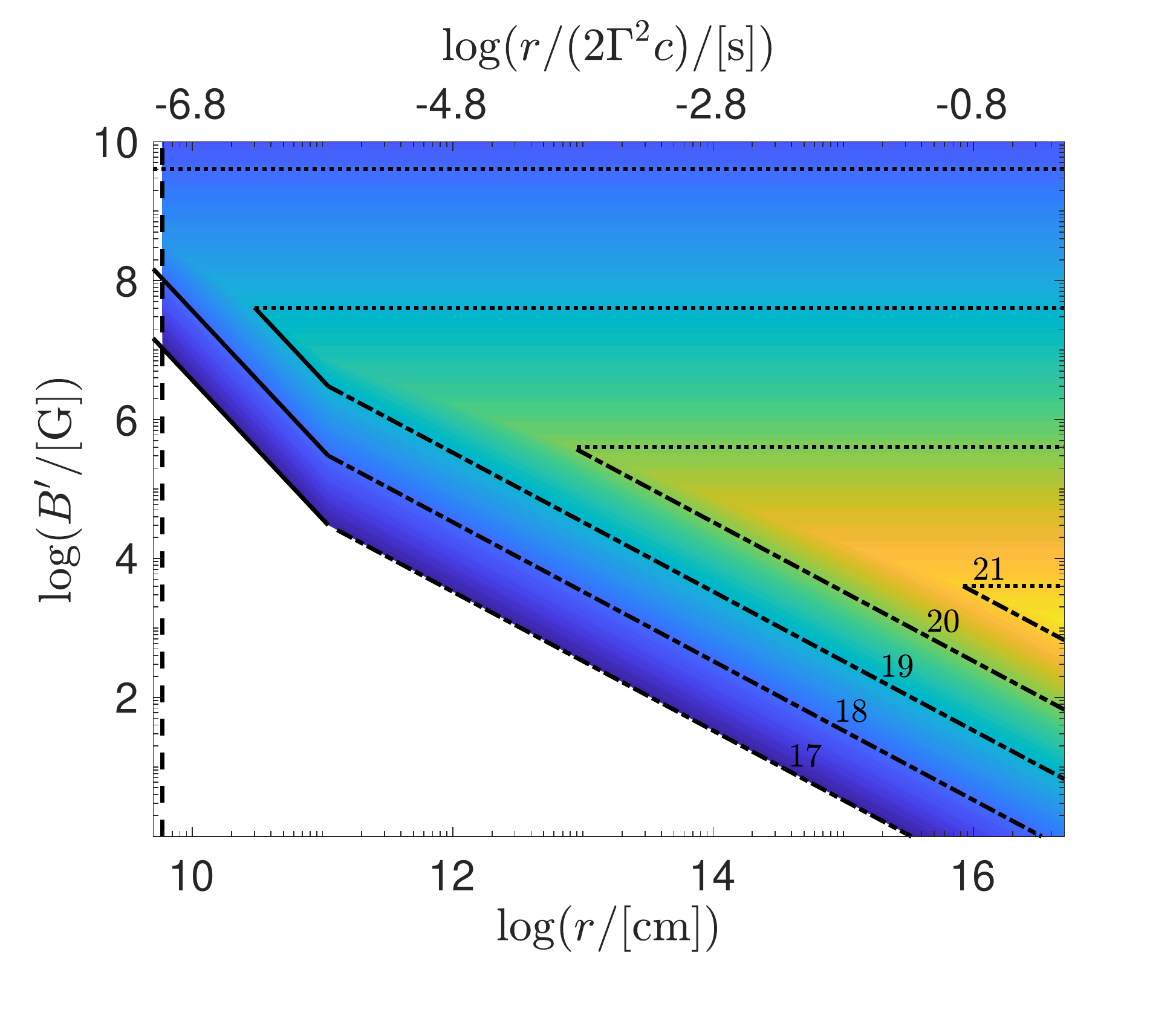}\hfill
    \includegraphics[width=0.45\columnwidth]{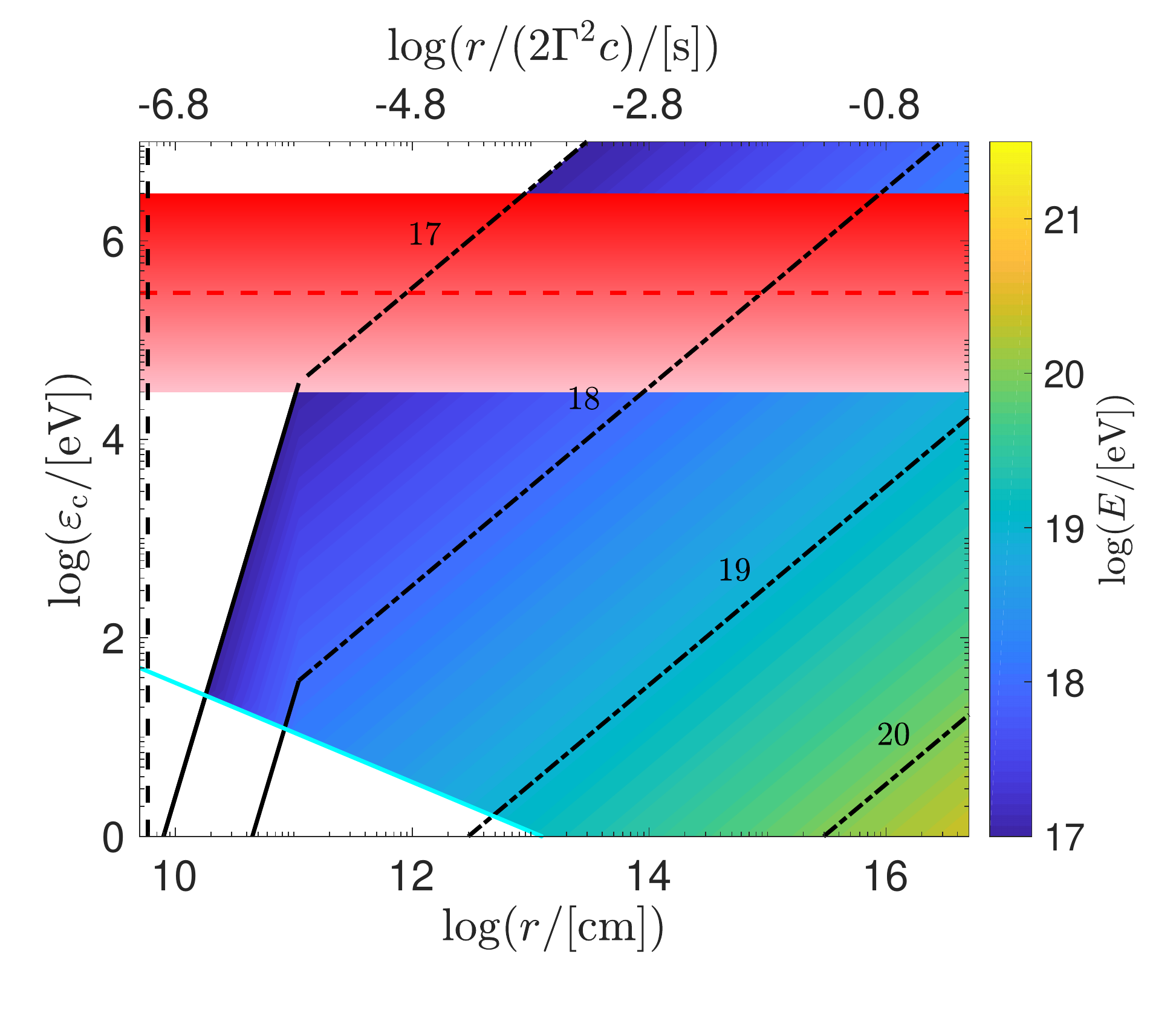}\hfill
    
    \caption{Allowed parameter space for $B'$ (left) and $\varepsilon_{\rm c}$ (right) as function of $r$ for bulk Lorentz factor $\Gamma$ = 100, 300, and 1000 from top to bottom. The color bar shows $\log(E)$ and the dashed vertical line shows the photosphere $r_{\rm ph}$. The dotted lines on the left (right) show $B'_{\rm sync}$ ($\varepsilon_{\rm c, sync}$), the dot-dashed lines show $B'_{\rm ad}$ ($\varepsilon_{\rm ad}$), and the solid lines show $B'_{p\gamma}$ ($\varepsilon_{\text{c}, p\gamma}$), all for integer values of $\log(E)$ as indicated in the plots. The pink-red gradient shows possible values for $\varepsilon_{\rm peak}^{\rm  obs}$, running from 30 keV to 3 MeV with a dashed red horizontal line indicating $\varepsilon_{\rm c} = 300$ keV. The solid, cyan line visible on the right indicates the parameter space neglected because it would result in $\gamma_{\rm c}' < 1$. Numerical values used are $L_{\gamma} = 10^{51}$ erg s$^{-1}$, $\eta = 0.1$, and $\left< \varepsilon \right> = 300$ keV.}
    \label{Fig:MagneticFieldRestrictionSynchrotron}
\end{figure*}

In this subsection we evaluate the allowed parameter space for the cooling break energy $\varepsilon_{\rm c}$, defined as the characteristic observed energy of synchrotron photons emitted by electrons with Lorentz factor $\gamma'_{\rm c}$. The position of the cooling break energy help us differentiate between if emission occurs in the FC or SC regime. We underline that in this section we ignore corrections for redshift, as UHECRs need to be generated sufficiently close to us or they would lose their energy through interactions with the CMB \citep{Waxman1995}. We note that including the redshift would only change the results by a factor of order unity.
A relativistic electron with Lorentz factor $\gamma'_e$ in an outflow with bulk Lorentz factor $\Gamma$ and magnetic field $B'$ radiates photons with the observed characteristic synchrotron energy
\begin{equation}\label{varepsilon}
	\varepsilon(\gamma_e') = 2\Gamma \frac{\hbar e}{m_e c}(\gamma'_e)^2 B',
\end{equation}
where $\hbar$ is the reduced Planck constant and the factor $2\Gamma$ transforms the energy from the outflow frame to the observer frame. 
By equating the synchrotron time scale for electrons with the expansion time scale, respectively given by Equations \eqref{eq:t_sync} and \eqref{eq:t_exp}, we obtain the cooling break Lorentz factor as
\begin{equation}\label{gamma_c}
	\gamma'_{\rm c} = \frac{6\pi m_e c^2}{\sigma_\textrm{T}} \frac{\Gamma}{r {B'}^2}.
\end{equation}
%
Equation \eqref{gamma_c} is only valid as long as $\gamma'_{\rm c} \geq 1$. For some of the parameter space on the left-hand side in Figure \ref{Fig:MagneticFieldRestrictionSynchrotron}, this requirement will not be met. From here on out, we limit our investigation to regions where Equation \eqref{gamma_c} is valid and the parameter space that is excluded because of this is indicated by a solid, cyan colored line in forthcoming figures. As evident from these figures, the excluded regions always lie where they are rejected because of the other arguments in the paper, so for our discussion, the restriction $\gamma'_{\rm c} \geq 1$ is of minor concern.

We assume that electrons are accelerated to a power-law of index $-p$, with $p>0$, between Lorentz factors $\gamma'_{\rm  m}$ and $\gamma'_{\rm max}$. 
If $\gamma'_{\rm c} < \gamma'_{\rm m}$, the electrons are FC and they quickly cool down to $\gamma'_{\rm c}$. If $\gamma'_{\rm m} < \gamma'_{\rm c}$ the electrons are SC as at least some of them cool mainly due to the adiabatic expansion of the ejecta. Several characteristics of the synchrotron spectrum depends on in which regime the emission takes place. 
One of the differences is the position of the peak energy in a $\nu F_\nu$-spectrum $\varepsilon_{\rm peak}^{\rm sync}$. 
As long as $p < 3$, the peak is given by $\varepsilon_{\rm peak}^{\rm sync} = \max(\varepsilon_{\rm c}, \varepsilon_{\rm m})$, where $\varepsilon_{\rm c} \equiv \varepsilon(\gamma'_{\rm c})$ and $\varepsilon_{\rm m} \equiv \varepsilon(\gamma'_{\rm m})$ \citep{Sari1998}.

In the FC regime, the power-law photon spectral index below $\varepsilon_{\rm peak}^{\rm sync}$ is $\alpha = -3/2$ between $\varepsilon_{\rm c}$ and $\varepsilon_{\rm m}$. The fact that 2/3 of GRBs have a harder $\alpha$ than what FC synchrotron emission can produce has lead to the synchrotron line of death problem \citep{Preece1998}. SC emission has a somewhat harder low-energy photon spectral index with $\alpha = -2/3$ between $\varepsilon_{\rm m}$ and $\varepsilon_{\rm c}$, but about a third of observed burst have an $\alpha$ that is harder still. 
This contradiction by observation lead to the suggestion of a \textit{marginally fast cooling} (MFC) regime \citep{Daigne2011}, where $\gamma'_{\rm c} \sim \gamma'_{\rm m}$. Such a scenario would produce $\varepsilon_{\rm c} \lesssim \varepsilon_{\rm m}$; the two breaks would be so close together that we fail to observe the $\alpha = -3/2$ segment of the spectrum and we would instead see a harder low-energy slope. In addition, MFC emission would be relatively efficient in converting the kinetic energy of the outflow into radiation, as opposed to SC emission. However, a MFC emission region requires fine tuning of the parameters \citep{Beniamini2018}.

From the discussion above we are motivated to evaluate the allowed parameter space for $\varepsilon_{\rm c}$, as this will give us information about the spectral shape of the synchrotron emission. By inserting $\gamma'_{\rm c}$ from Equation \eqref{gamma_c} into Equation \eqref{varepsilon} we get
\begin{equation}\label{eq:varepsilon_c}
	\varepsilon_{\rm c} = \frac{2\hbar e}{m_ec} \left(\frac{6\pi m_e c^2}{\sigma_T}\right)^2 \frac{\Gamma^3}{r^2 {B'}^3}
\end{equation}
The requirements on the magnetic field strength can now be translated into requirements on $\varepsilon_{\rm c}$ by inserting the magnetic fields given in Equations \eqref{Bsync} - \eqref{Bpgamma} into the equation above:
\begin{alignat}{2}
	\varepsilon_{\rm c} &> \varepsilon_{\rm c, sync} && \equiv \frac{ \hbar \sigma_T}{3 \pi e^2c^{9}} \frac{m_e^7}{m_p^{12}}\frac{1}{r^2} \left(\frac{E^2}{\Gamma\eta}\right)^3 \label{varepsilonsync}  \\
    && & \sim 1.9\times 10^{-2} \ \frac{E_{20}^6}{r_{14}^2\Gamma_2^3\eta_{-1}^3} \textrm{ eV},\nonumber
\end{alignat}
\begin{alignat}{2}
	\varepsilon_{\rm c} & < \varepsilon_{\rm c, ad} && \equiv 72\pi^2 \frac{\hbar e^4 m_e c^3 }{\sigma_T^2} \frac{r \Gamma^3 \eta^3}{E^3} \label{varepsilonad} \\
    && & \sim 3.4 \times 10^{-5} \ \frac{r_{14}\Gamma_2^3\eta_{-1}^3}{E_{20}^3} \textrm{ eV}, \nonumber 
\end{alignat}
\begin{alignat}{2}
	\varepsilon_{\rm c} & < \varepsilon_{\text{c}, p\gamma} && \equiv 72 \times 20^3 \pi^5 \frac{\hbar e^4 m_e c^6 }{\sigma_T^2 \sigma_{p\gamma}^3} r^4 \left (\frac{\left<\varepsilon\right> \eta \Gamma^3}{E L_\gamma} \right )^3  \label{varepsilonpgamma} \\
    && & \sim 0.92 \ \frac{ r_{14}^4 \left<\varepsilon\right>_6^3 \eta_{-1}^3 \Gamma_2^9}{E_{20}^3L_{\gamma, 51}^3} \textrm{ eV} \nonumber.
\end{alignat}
%
%
In the right-hand panels in Figure \ref{Fig:MagneticFieldRestrictionSynchrotron} we plot the allowed parameter space for $\varepsilon_{\rm c}$, translated from the allowed parameter space for $B'$ shown on the left. The color coding and line coding are similar  to that on the left. 

Assuming the observed prompt emission is dominated by synchrotron radiation, the peak of the synchrotron spectrum $\varepsilon_{\rm peak}^{\rm sync}$ coincides with the peak of the observed spectrum $\varepsilon_{\rm peak}^{\rm obs}$. Thus, one can differentiate the regimes depending on the value of $\varepsilon_{\rm c}$ compared to the observed $\nu F_\nu$-peak:
\begin{alignat}{3}
	\varepsilon_{\rm c} &< \varepsilon_{\rm peak}^{\rm obs} && \quad \longrightarrow  \quad &&\rm  FC, \nonumber \\
    \varepsilon_{\rm c} &\sim \varepsilon_{\rm peak}^{\rm obs} &&\quad  \longrightarrow \quad &&\textrm{MFC or SC}. \nonumber
\end{alignat}
Experimentally, the observed peak energy is around 300 keV (average values are $\varepsilon_{\rm peak}^{\rm obs} \sim 200$ keV for long GRBs and $\varepsilon_{\rm peak}^{\rm obs} \sim 400$ keV for short GRBs \citep{Goldstein2012,Yu2016}). Variations of the peak energy are represented by the pink-red gradient showing 30 keV to 3 MeV in Figure \ref{Fig:MagneticFieldRestrictionSynchrotron}. From the figure, we see that the acceleration of UHECR implies the emission to be extremely FC for most of the parameter space. Indeed $\varepsilon_{\rm c} \ll 300$ keV, except for the lowest UHECR energies. 
As a consequence, the low-energy slope of GRB spectra should be of index $\alpha = -3/2$, in contradiction with observations. 
As pointed out in \citet{Nakar2009, Daigne2011}, synchrotron self-Compton (SSC) could have a significant effect on the observed spectrum from astrophysical sources. If the Compton parameter $Y$ is large, $Y \geq 1$, then SSC would harden the spectrum. \citet{Beniamini2013} thoroughly investigated what restrictions on the emitting region are necessary for a FC synchrotron emission model to match observations. As can be seen in Figures 3 and 4 of their paper, the parameter space for $B'$ that we obtain for the highest energy UHECR could indeed be compatible with a FC emission scenario as long as $\Gamma \gtrsim 200$ and the emission happens sufficiently far out ($r > 10^{15}$ cm). SSC emission from GRBs is expected to be at least partially in the Fermi LAT band (20 MeV - 300 GeV). In a recent paper, the Fermi LAT collaboration reported that they have not yet seen any statistically significant emission in excess of what is expected from extrapolation of the synchrotron emission from lower energies \citep{FermiSwift2018}. If a SSC component is present, then it is either weak enough to be outshined by the high energy synchrotron tail or the SSC peak resides outside of the LAT sensitivity range \citep{Beniamini2013, FermiSwift2018}. Both scenarios pose problems. If the SSC emission is weaker than the extended synchrotron emission, this would imply a low Compton parameter $Y$ and the fraction of upscattered photons would not be sufficient to harden the spectrum to match observations. Furthermore, suppression of the LAT flux due to Klein-Nishina effects should be negligible in a FC synchrotron scenario as the Klein-Nishina frequency should lie above the LAT band \citep{Beniamini2013}. A strong SSC component above the LAT is also disfavored by recent results from the HAWC collaboration, showing that no very high energy emission ($\sim 300$ GeV) from GRBs have been detected by HAWC \citep{HAWC2018}. However, we note that these results are only marginally constraining.

\subsection{Photospherically dominated prompt emission}\label{Sec:PhotosphericEmission}
We next consider hybrid models, where we assume that in addition to a synchrotron component the prompt $\gamma$-ray photons are produced at the photosphere. In particular, we assume the observed sub-MeV peak to be thermal; if this is not the case we refer back to the discussion in Section \ref{Sec:SynchrotronPrompt}. 
Below the photosphere, the charged particles are too tightly bound to the photon field to be efficiently accelerated \citep{Budnik2010,Murase2013,Beloborodov2017}. However, the acceleration of UHECR could still occur above the photosphere, provided that the synchrotron flux from the co-accelerated electrons in the UHECR acceleration region do not outshine the photospheric emission. In this subsection we add an additional constraint, which is that the magnetic luminosity $L_{\rm mag} = B'^2/(8\pi)\times 4\pi r^2c\Gamma^2$ cannot be larger than the total luminosity of the burst $L_{\rm tot}$. How the value of $L_{\rm tot}$ influence the results is discussed in Section \ref{Sec:Discussion}. This leads to an upper limit on the magnetic field in complement to the limits given in Equations \eqref{Bsync} - \eqref{Bpgamma}:
\begin{equation}\label{Blum}
	B' < B'_{\rm lum} \equiv \left(\frac{2L_{\rm tot}}{c r^2\Gamma^2}\right)^{1/2}.
\end{equation}

In order to characterize the maximum proton energy, we use two constraints on the spectral (per frequency) flux of the synchrotron component. Firstly, for the observed sub-MeV peak to be mainly thermal, the synchrotron component is not allowed to outshine the photosphere around these energies. 
If we denote the synchrotron flux per frequency as $F_{\nu}^{\rm sync}$, then we use the condition $F_{\nu}^{\rm sync}(\varepsilon_{\rm peak}^{\rm obs}) < 0.2 F_{\nu}^{\rm obs}(\varepsilon_{\rm peak}^{\rm obs}) \equiv F^{\rm lim}_{\nu, \rm peak}$. We note that the factor of 0.2 is somewhat arbitrary, but the limit is conservative as the softer synchrotron flux should not overshoot the thermal flux at energies below the peak (around a few ten keV). It furthermore influences the final results only weakly. The observed peak flux differs for long and short GRBs, where the former have $F_{\nu}^{\rm obs}(\varepsilon_{\rm peak}^{\rm obs}) \approx 1$ mJy and the latter $F_{\nu}^{\rm obs}(\varepsilon_{\rm peak}^{\rm obs}) \approx 8$ mJy \citep{Ghirlanda2009}. In this work we use $F_{\nu}^{\rm obs}(\varepsilon_{\rm peak}^{\rm obs}) = 10$ mJy, which gives the limit $ F^{\rm lim}_{\nu, \rm peak} = 2$ mJy for the synchrotron component. The observed flux is redshift dependent and we use $z = 1$. 

Secondly, there are harsh observational constraints on the prompt flux in the optical band \citep{Greiner1996, Yost2007, Klotz2009}. 
Detections and upper limits place the optical component at less than 10 mJy after correcting for galactic extinction. We note that there are a few exceptions, most notably GRB 061007 that reached a peak flux of 500 mJy about 100 seconds after trigger. We use the condition $F_{\nu}^{\rm sync}(\varepsilon_{\rm opt}) < F^{\rm lim}_{\nu, \rm opt} = 100$ mJy, insuring at least an order of magnitude in leeway compared to the generic observed burst. To use conservative values for the fluxes also serves as a precaution for variations in the redshift.

\begin{figure*}
\begin{centering}
    \includegraphics[width=0.32\columnwidth]{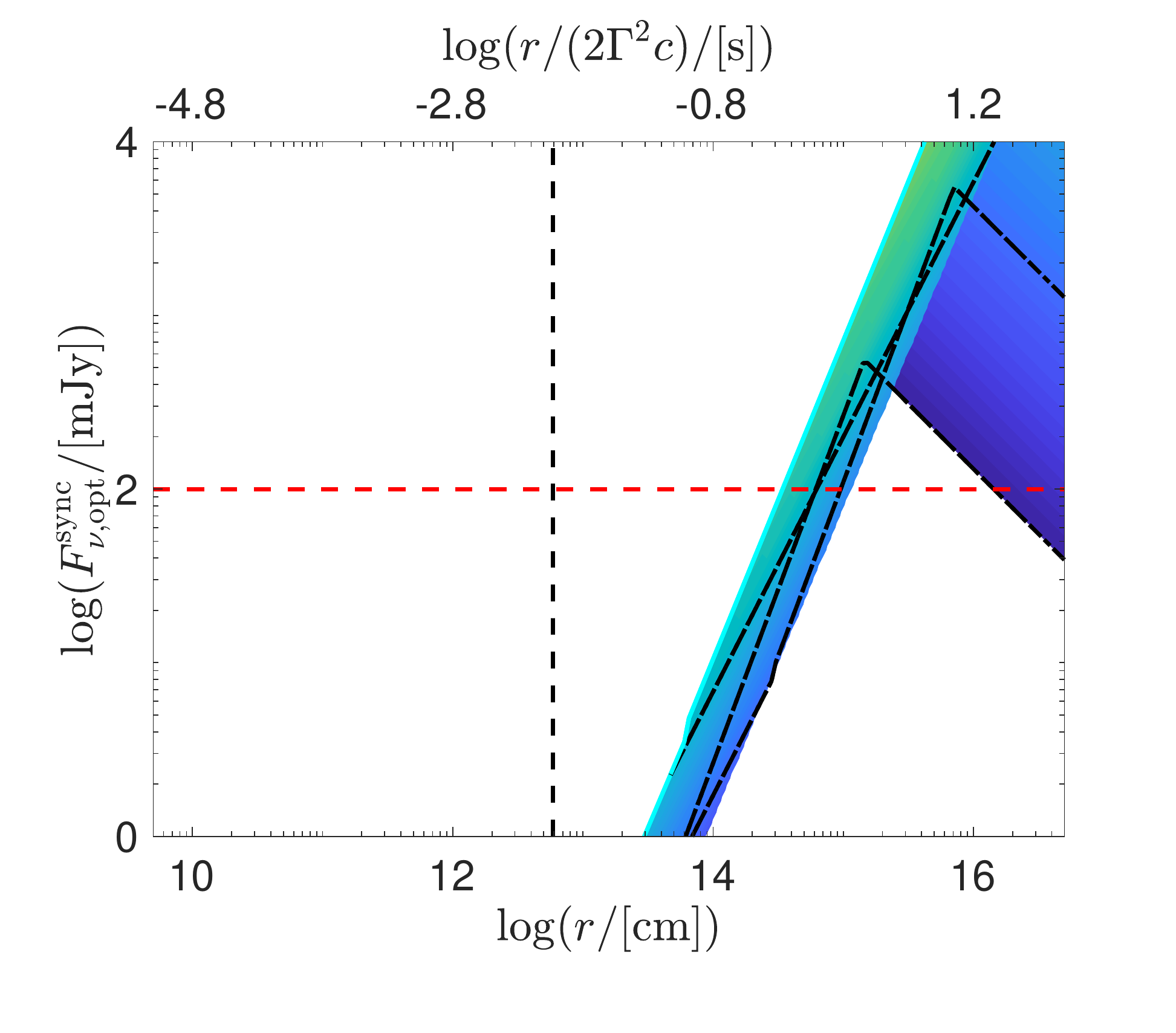}
    \includegraphics[width=0.32\columnwidth]{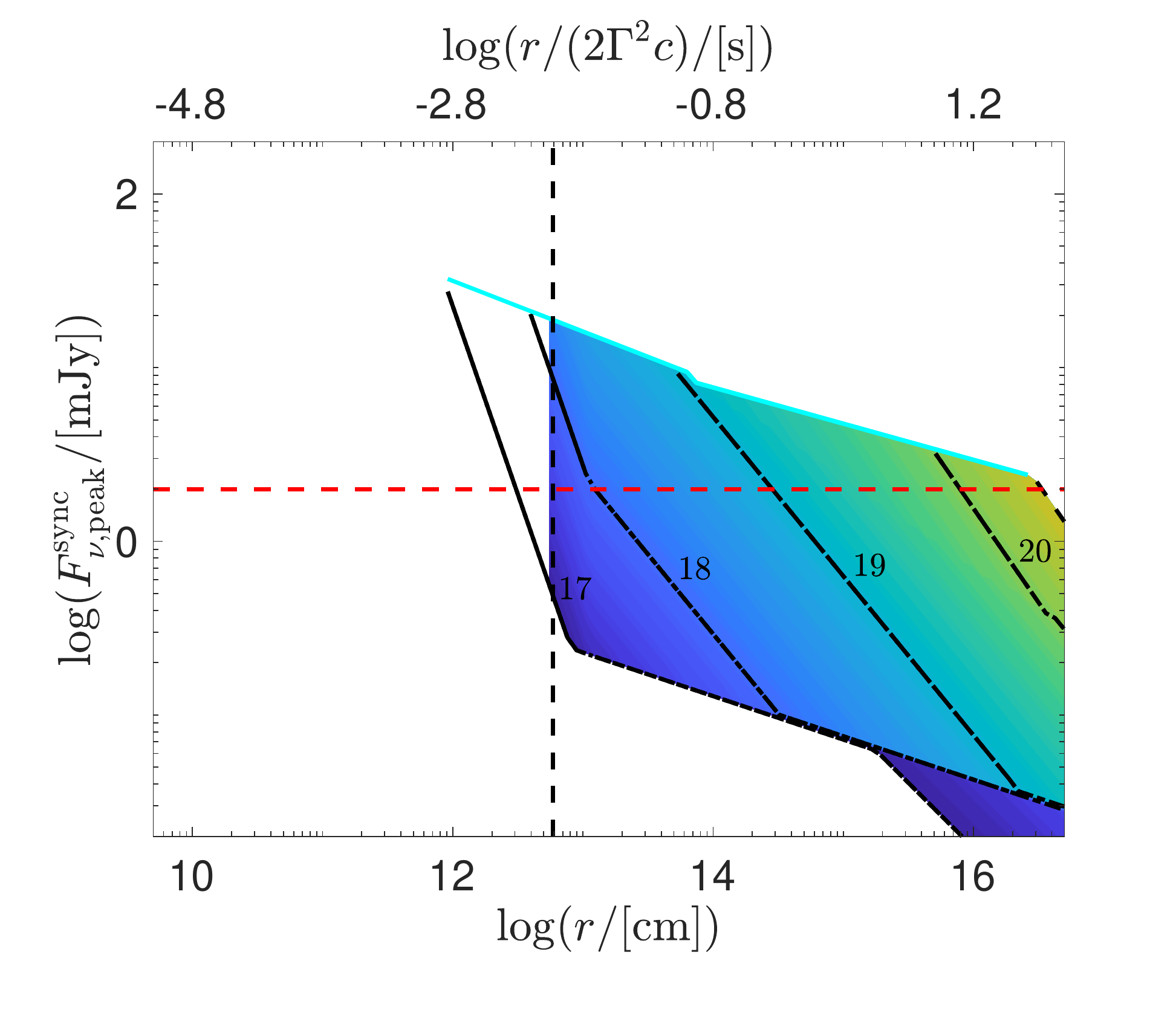}
    \includegraphics[width=0.32\columnwidth]{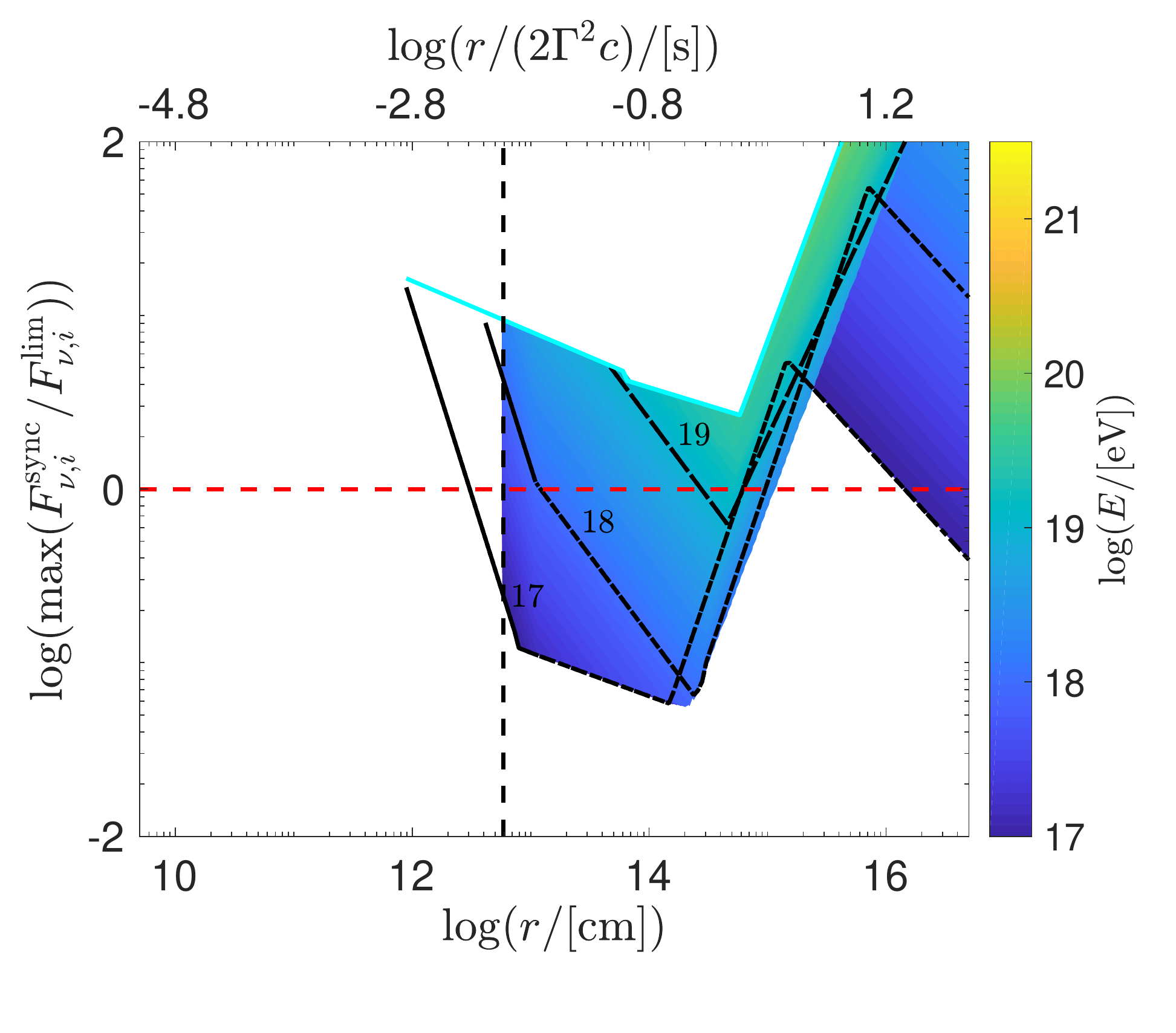}
    \includegraphics[width=0.32\columnwidth]{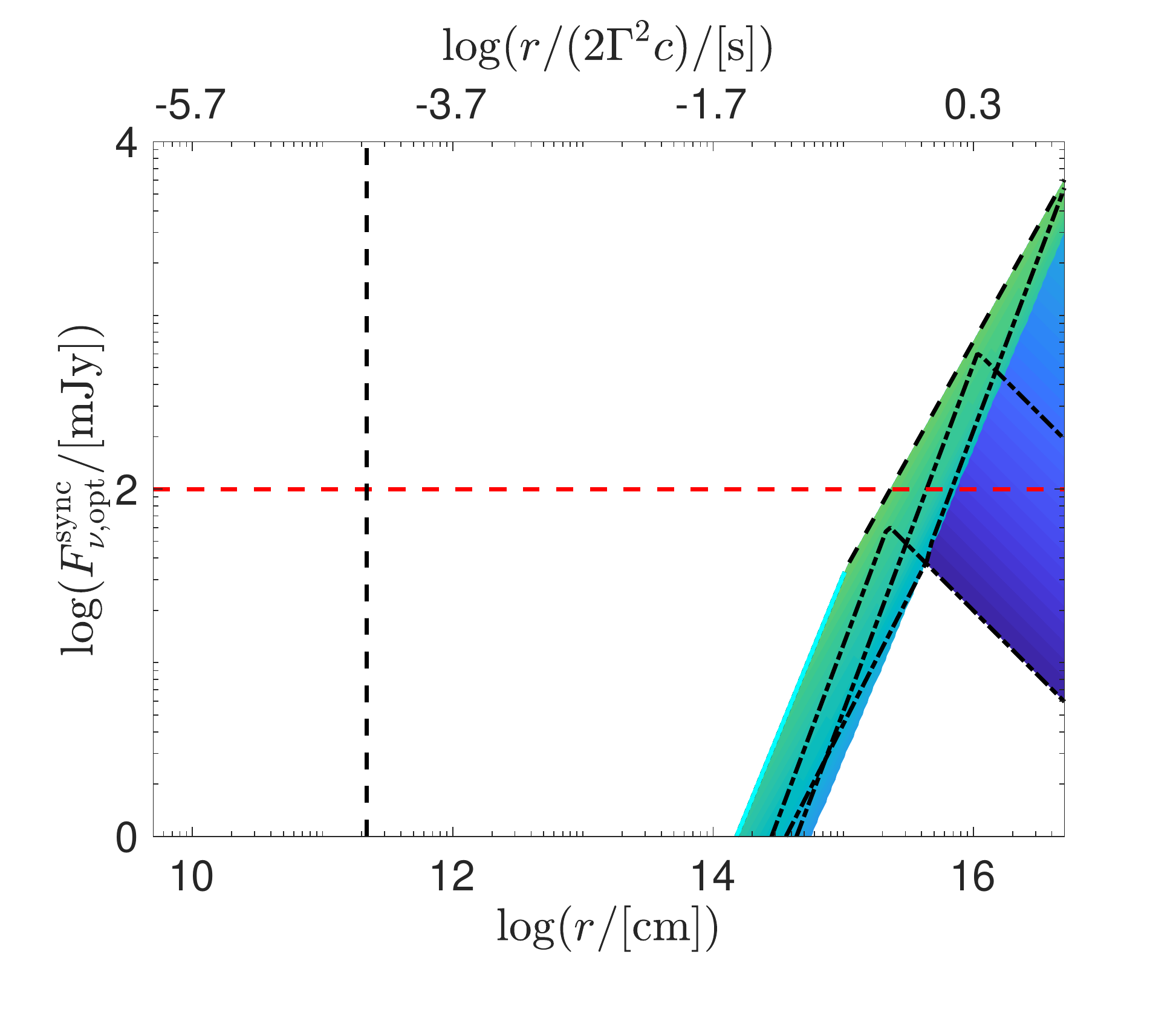}
    \includegraphics[width=0.32\columnwidth]{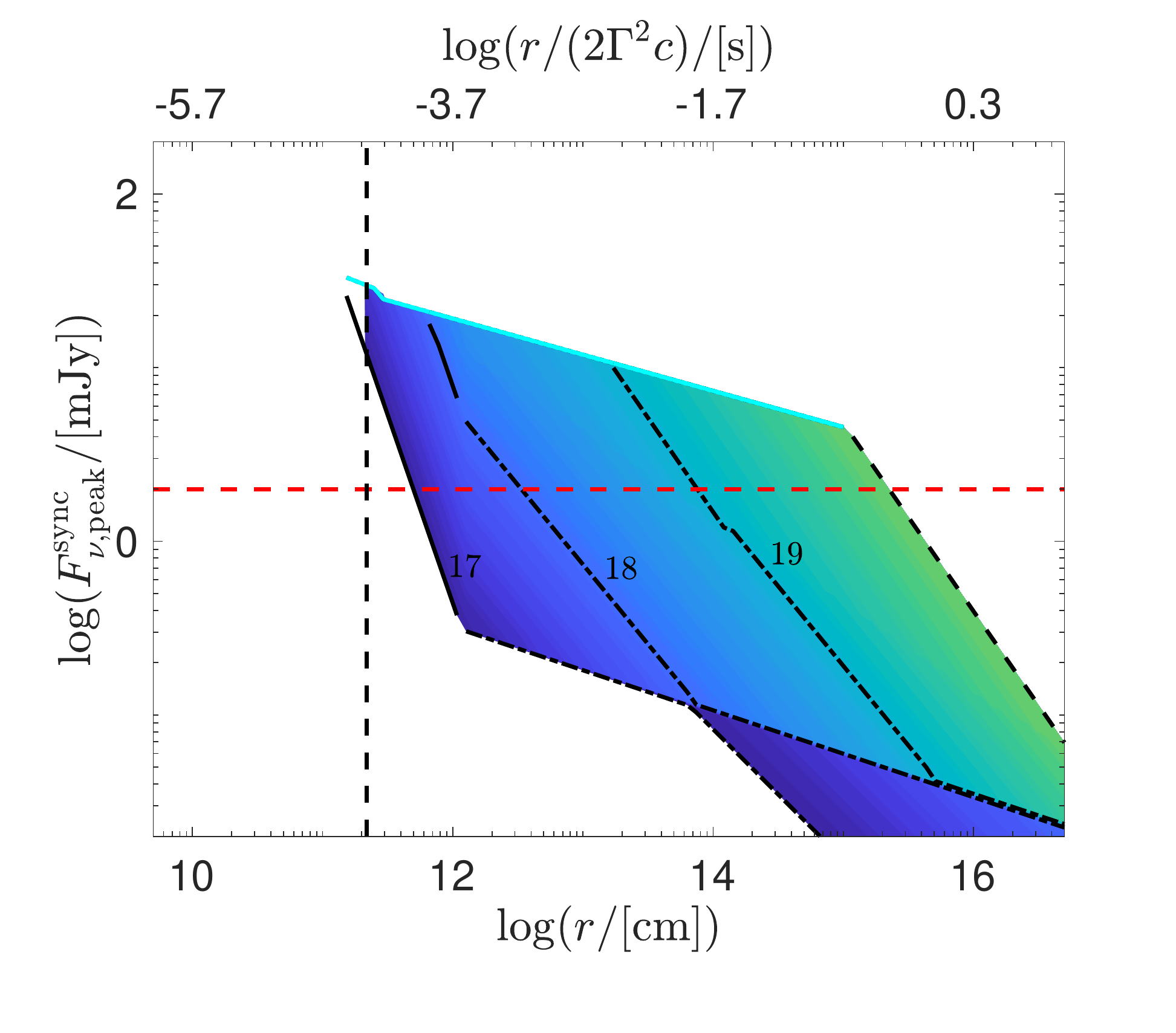}
    \includegraphics[width=0.32\columnwidth]{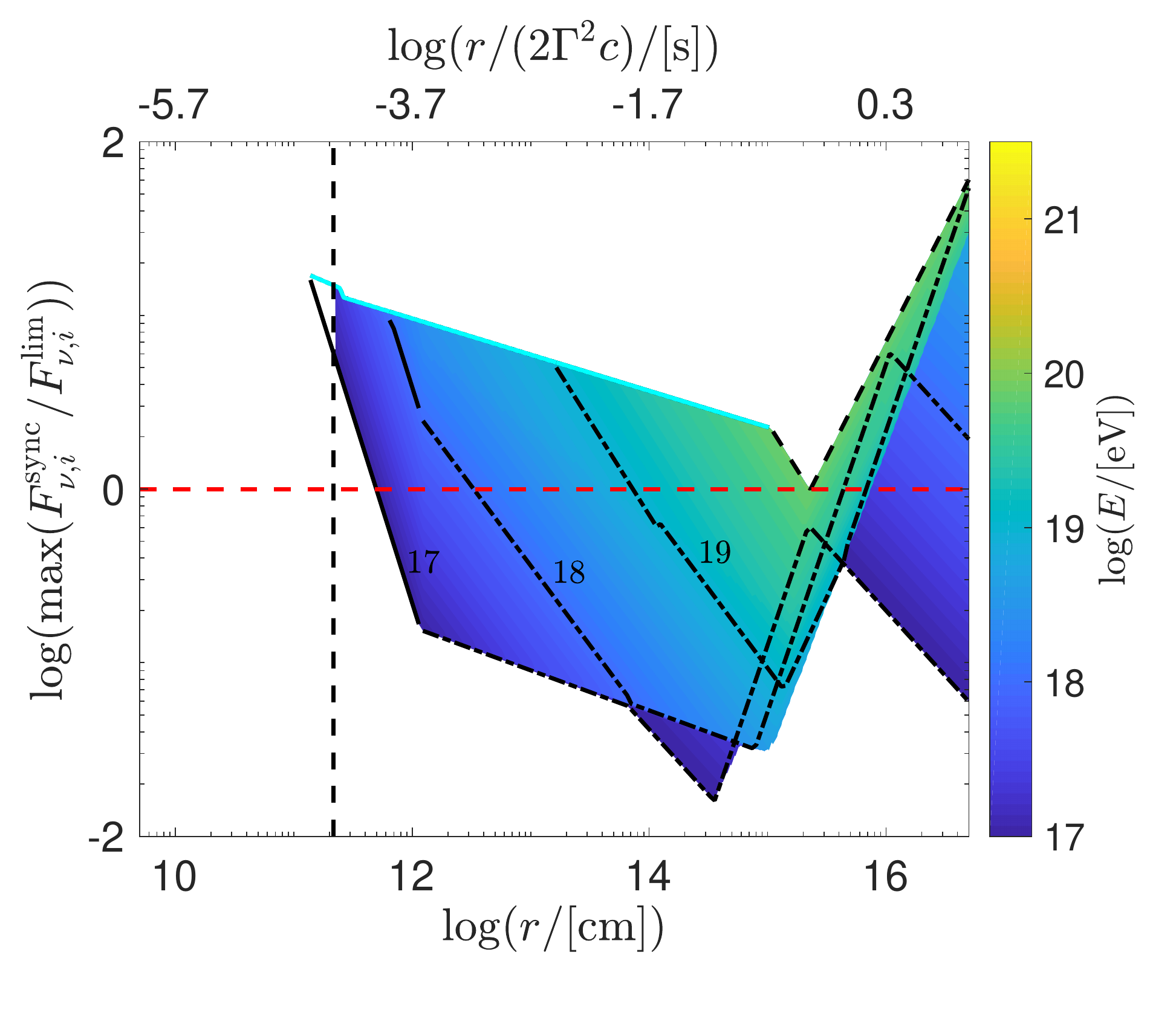}
    \includegraphics[width=0.32\columnwidth]{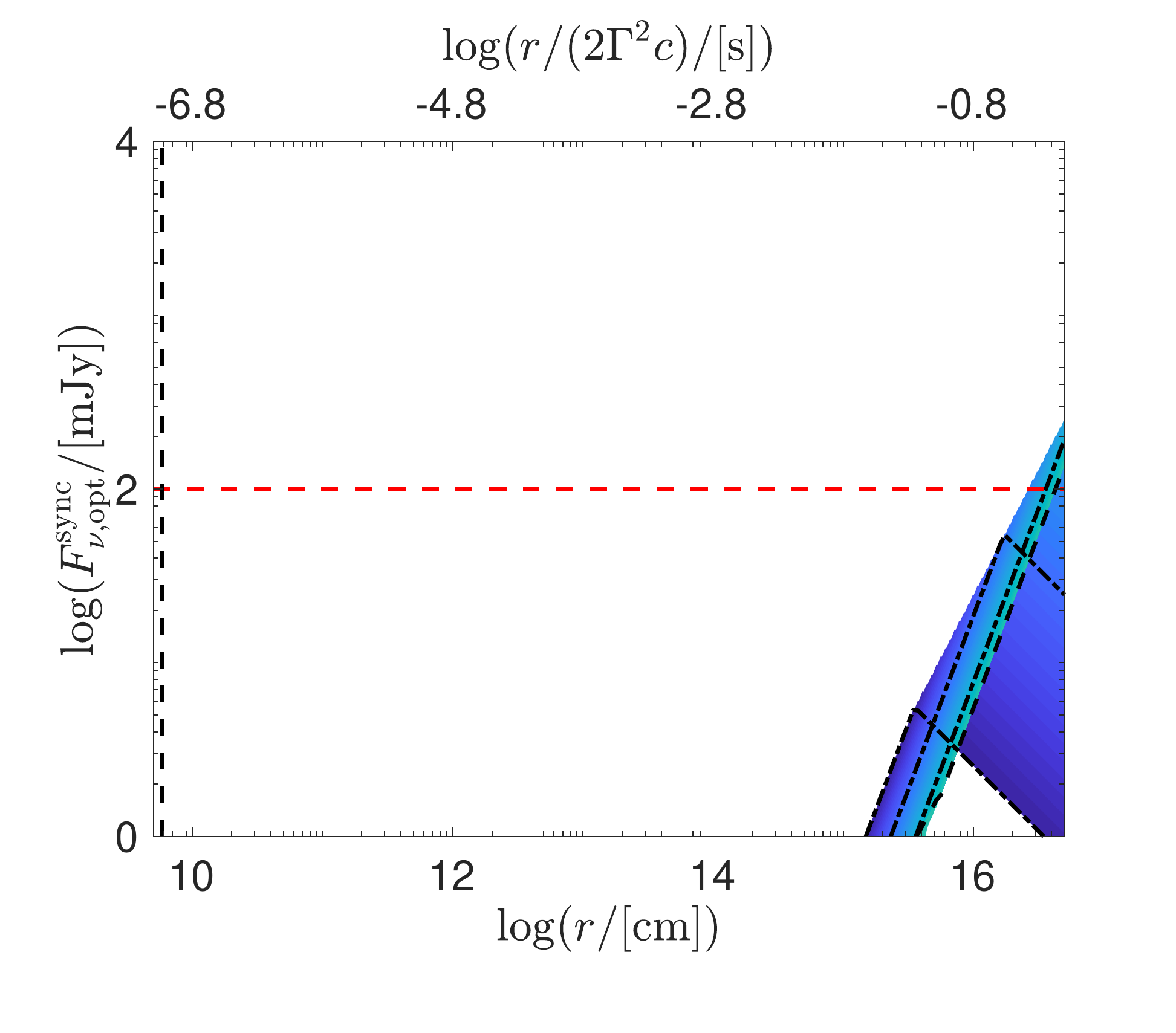}
    \includegraphics[width=0.32\columnwidth]{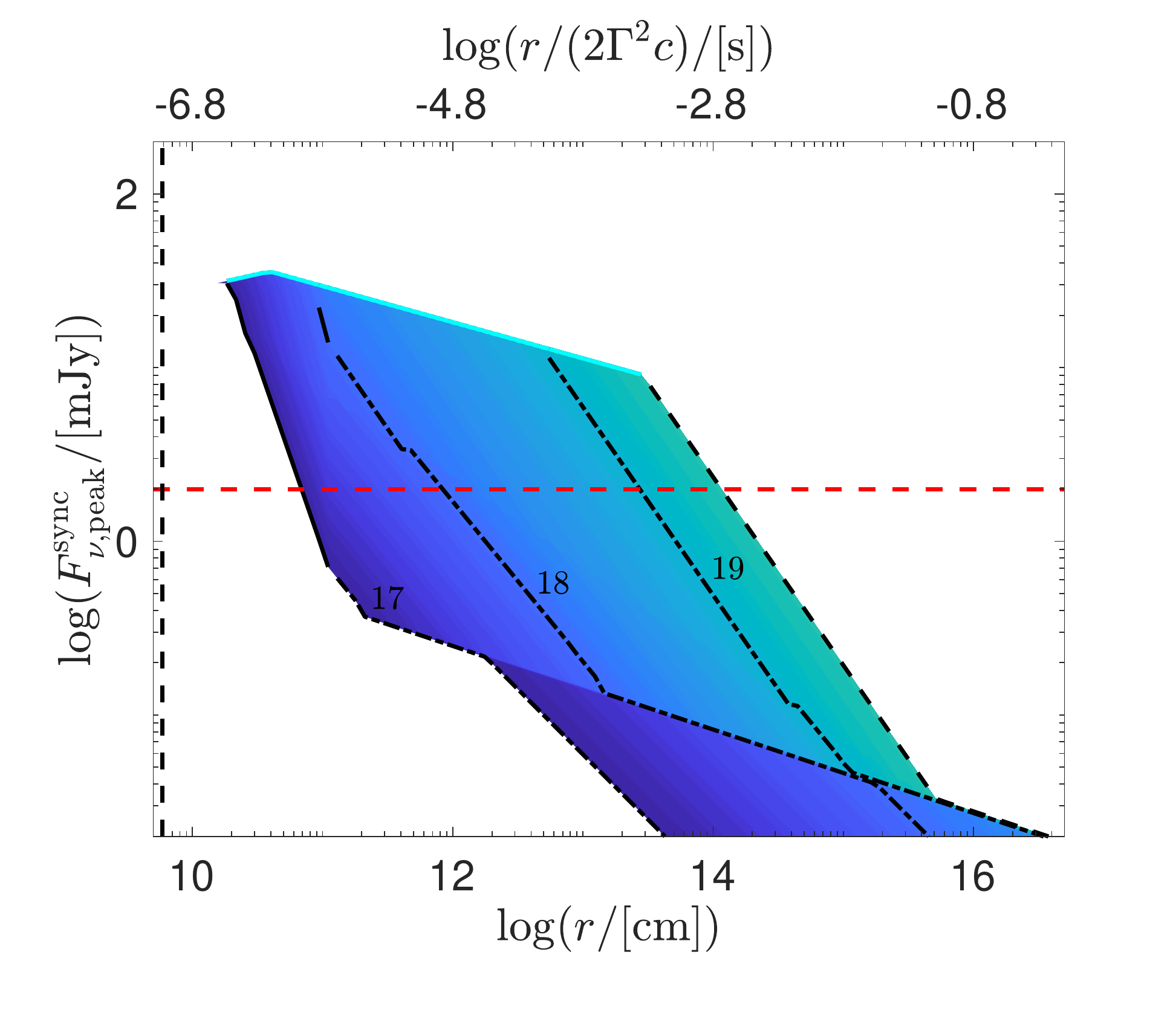}
    \includegraphics[width=0.32\columnwidth]{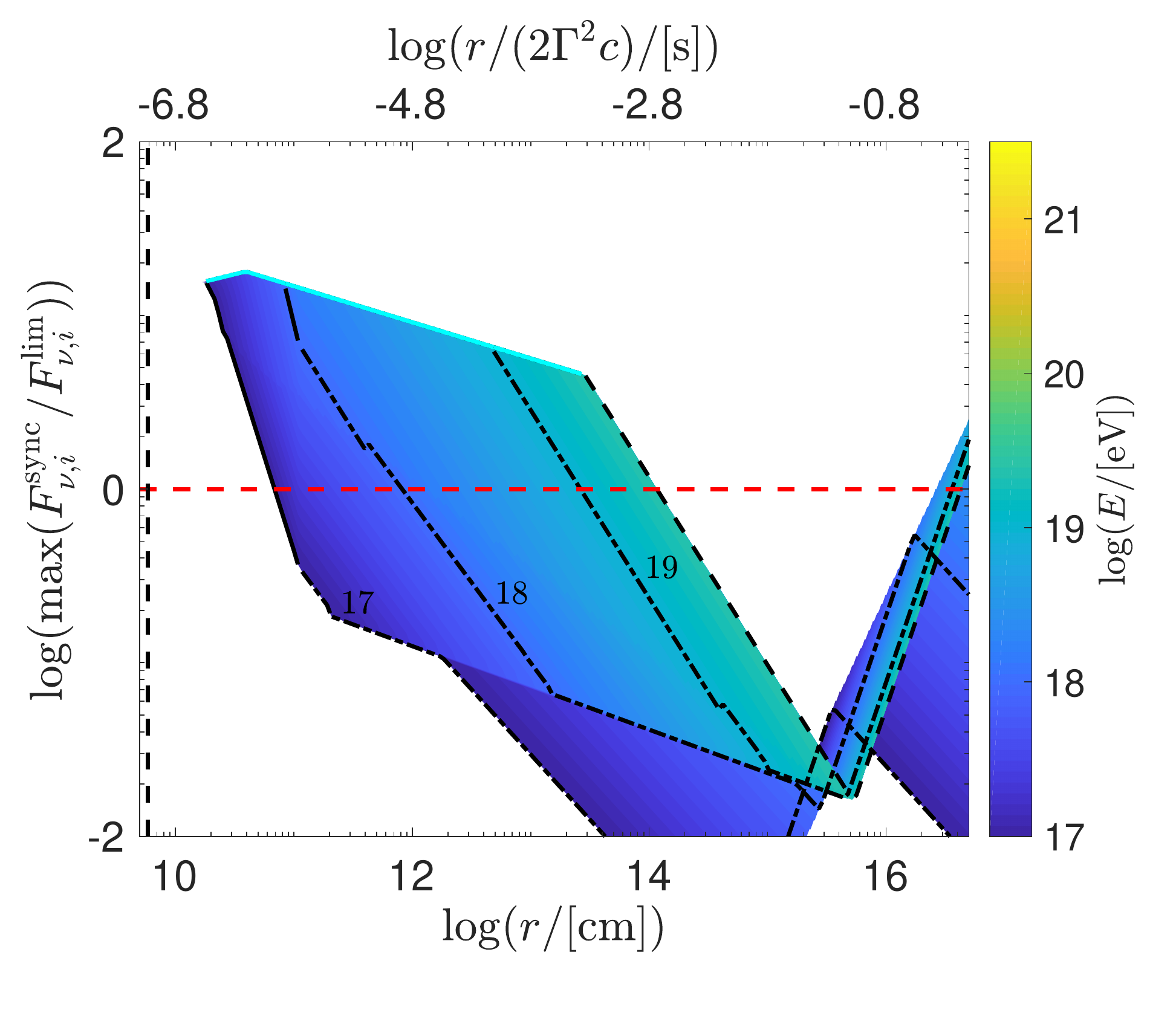}
\caption{Expected synchrotron fluxes at optical (left), 300 keV (middle) and maximum of the two when normalized to their respective observational limit (right), for proton acceleration in a high-luminosity GRB with $\Gamma = 100$, 300, and 1000 (top to bottom) as a function of $r$. The red dashed line shows our spectral flux limits $F_{\nu}^{\rm lim}$, so everything above this results in too high synchrotron fluxes. Close to the progenitor, the optical flux is completely absorbed due to synchrotron self-absorption, while the peak flux is high. For larger $r$ the peak flux is lower, mainly due to the decrease in $\varepsilon_{\rm m}$. The optical flux increases rapidly when $\varepsilon_{\rm SSA}$ decreases towards the optical band, peaking at $\varepsilon_{\rm SSA} = \varepsilon_{\rm opt}$. The dotted lines show limitations by synchrotron emission (Eq. \eqref{Bsync}), the dot-dashed lines show restrictions from adiabatic cooling (Eq. \eqref{Bad}), and the solid lines show restrictions from photo-hadronic interactions (Eq. \eqref{Bpgamma}), all for integer values of $\log(E)$ as indicated in the plots. Additionally, the dashed line shows restrictions from the magnetic luminosity (Eq. \eqref{Blum}). The solid, cyan line indicates the parameter space neglected because it would result in $\gamma_{\rm c}' < 1$. Total luminosity, spectral flux limits \added{and the redshift} are $L_{\rm tot} = 10^{52}$ erg s$^{-1}$, $F_{\nu, \rm opt}^{\rm lim} = 100$ mJy, \replaced{$F_{\nu, \rm peak}^{\rm lim} = 1$ mJy}{$F_{\nu, \rm peak}^{\rm lim} = 2$ mJy}, \added{and $z = 1$}. The vertical dashed line indicates the photosphere. Other numerical values used are given in Table \ref{tab:PhotosphereNumerics}.} 
\label{Fig:FluxLimits}
\end{centering}
\end{figure*}

We describe how we obtain $F_{\nu}^{\rm sync}(\varepsilon_{\rm peak}^{\rm obs})$ and $F_{\nu}^{\rm sync}(\varepsilon_{\rm opt})$ in Appendix \ref{App:SSA}. It is quite cumbersome, more so as synchrotron self-absorption (SSA) can not be ignored at low radii. Even though the shape of the synchrotron spectrum is well defined, the flux at a specific energy depends on the positions of $\varepsilon_{\rm c}$, $\varepsilon_{\rm m}$, and $\varepsilon_{\rm SSA}$, relative to each other and relative to the energy in question. 
This leads to 24 possible permutations in both cases, each with a corresponding flux profile (see Table \ref{tab:FluxCases} for a complete list). For every triplet ($B', r, \Gamma$), which is equivalent to the triplet ($E, r, \Gamma$), 
we calculate $\varepsilon_{\rm c}$, $\varepsilon_{\rm m}$, and $\varepsilon_{\rm SSA}$ as well as the peak of the synchrotron spectral flux spectrum $F_{\nu, \rm max}^{\rm sync}$. This completely characterizes the flux profile and the synchrotron fluxes in the optical band and around the observed peak can be estimated. The flux depends on the fraction of available electrons that are accelerated $\xi_a$ and on the fractional energy given to electrons $\epsilon_e$. We use $\xi_a = 1$ and $\epsilon_e = 0.1$ as our fiducial parameters. The value of $\epsilon_e$ is widely used and well motivated, see \textit{e.g.} \citet{Wijers1999,Panaitescu2000,Santana2014,Beniamini2017}. The parameter value $\xi_a \sim 1$ is less physically motivated but widely used, see e.g., \citet{Sari1998,Eichler2005,Santana2014}. The effect of varying these parameters is discussed in Section \ref{Sec:Discussion}.
The result is shown in Figure \ref{Fig:FluxLimits}. As in Figure \ref{Fig:MagneticFieldRestrictionSynchrotron}, the results are shown for $\Gamma = $100, 300, and 1000 from top to bottom, and the line coding is similar with the addition of the black dashed line showing the constraint from the magnetic luminosity (Equation \eqref{Blum}). The left column shows $F_{\nu}^{\rm sync}(\varepsilon_{\rm opt})$ and the middle column shows $F_{\nu}^{\rm sync}(\varepsilon_{\rm peak}^{\rm obs})$. The dashed red line shows $F_{\nu}^{\rm lim}$ for each case, meaning that everything above it predicts a synchrotron flux that is too high. The right column shows the allowed parameter space when both constraints normalized to their respective observational limit, are taken into account. 
For small values of $r$, the peak limit is the most constraining. When $r$ increases, $\varepsilon_{\rm m}$ decreases resulting in a longer soft high-energy tail, lowering the peak flux. Moreover, $\varepsilon_{\rm SSA}$ also decreases, shifting $F_{\nu, \rm max}^{\rm sync}$ to lower energies, further reducing the sub-MeV peak flux. Meanwhile, the optical flux is completely absorbed close to the progenitor. Once $\varepsilon_{\rm SSA}$ approaches $\varepsilon_{\rm opt}$, the optical flux rises dramatically, with peak at $\varepsilon_{\rm SSA} = \varepsilon_{\rm opt}$.

From Figure \ref{Fig:FluxLimits} it is evident that the constraints on the flux, together with the limit on the magnetic luminosity, rules out all acceleration above $10^{20}$ eV. 
The highest reachable energy is $\lesssim 10^{20}$ eV, obtainable when $\Gamma = 300$ at $r \sim 10^{15}$ cm. Comparing with the middle subfigure on the left in Figure \ref{Fig:MagneticFieldRestrictionSynchrotron}, we see that this require a magnetic field strength of $B' \sim 10^3$ G.

Our derivation is agnostic to the dynamics of the flow, whether it is thermal pressure or magnetic reconnection responsible for its acceleration. In both cases, several models combining synchrotron radiation with photospheric emission have been proposed to overcome the line of death problem. For instance \citet{Beniamini&Giannios2017} considered gradual dissipation of magnetic energy in the jet through reconnection. They considered the peak of the observed spectra to originate from a strong photospheric component overlaid on a non-thermal synchrotron emission of electrons accelerated above the photopsheric radius by magnetic reconnection. The model predicts most of the energy to be dissipated by magnetic reconnection around the saturation radius $r_s\sim 10^{13}$ cm. 
Investigation of the left side of Figure \ref{Fig:MagneticFieldRestrictionSynchrotron} shows that the highest UHECR cannot be obtained so close to the progenitor. Furthermore, for all models that consider the observed peak to be photospheric, Figure \ref{Fig:FluxLimits} still applies. 


\section{Low-luminosity GRBs as UHECR sources}\label{Sec:LowLum}
In a recent paper, \citet{Zhang2018} thoroughly discussed low-luminosity GRBs (llGRBs) with isotropic radiation luminosity $L < 10^{49}$ erg s$^{-1}$ as sources for UHECR. We note that their existence as a separate transient class is still highly debated \added{\citep{Guetta2007,Bromberg2011, Dereli2017}}. These transients escape several problems present in the standard UHECR-GRB picture. For one, the jets in these events could be characterized by a higher baryonic composition than a typical GRB; their weaker radiation field means that heavier nuclei have a lower risk of photodisintegration. While the Telescope Array seems to favor an intermediate-mass nuclei composition of UHECR that does not change with energy \citep{TelescopeArray2018}, the Pierre Auger Observatory has seen indications that above $\sim 10^{18.5}$ eV, the cosmic ray composition starts to shift towards higher-mass nuclei with increasing energy \citep{PierreAugerICRC2017}. We therefore consider both protons and completely stripped iron in this section. If llGRBs are not a separate transient class however, then the photodisintegration problem remains as the initial radiation luminosity at the base of the jet would not be lower than for a canonical GRB. 
Secondly, IceCube has put strong upper limits on the GRB contribution to the high energy CR flux through stacking analysis 
\citep{IceCubeCollaboration2015GRBLimit}. 
Because llGRBs are much fainter they could contribute significantly to the diffuse neutrino flux observed by IceCube without being detected \citep{Murase2006}. Lastly, llGRBs are also expected to be more common than standard GRBs in our local universe. While the local rates estimates of llGRBs relies only on a few events leading to large uncertainties \citep{Guetta2007}, they are generally believed to be $\sim $ 100 times more common than long GRBs \citep{Sun2015}. 

Of the limits given in Equations \eqref{Bsync} - \eqref{Bpgamma}, the one in Equation \eqref{Bpgamma} is the only limit on the comoving magnetic field that changes with luminosity. Thus, the allowed parameter spaces shown in Figure \ref{Fig:MagneticFieldRestrictionSynchrotron} are mostly valid for protons in llGRBs as well. The consequence of lowering the luminosity is to shift the solid black lines to the left, only increasing the allowed parameter space for the lowest energy UHECRs. The photospheric radius also decreases but this does not alter the results presented. Iron acceleration to the highest energies is problematic as well, as this requires very large Lorentz factors of $\Gamma > 1000$, and llGRBs are believed to have slower outflows than canonical GRBs. High-energy acceleration with such large Lorentz factors are also incompatible with the constraint given in Equation \eqref{Blum}. Thus, neither protons nor iron can reach energies higher than a few $10^{19}$ eV in a llGRB in the synchrotron model without being in a deep fast cooling regime.

\begin{figure*}
\begin{centering}
    \includegraphics[width=0.32\columnwidth]{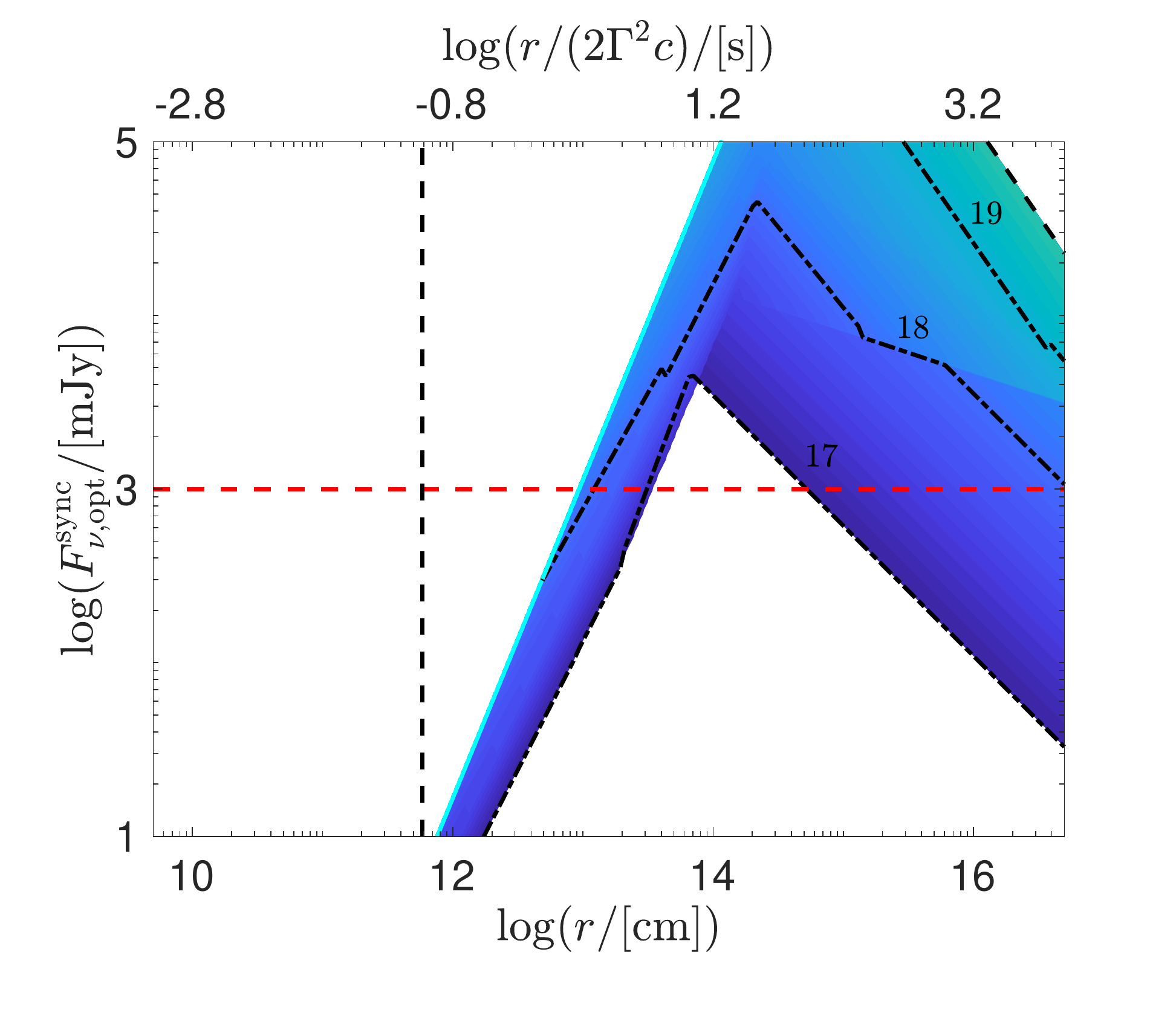}
    \includegraphics[width=0.32\columnwidth]{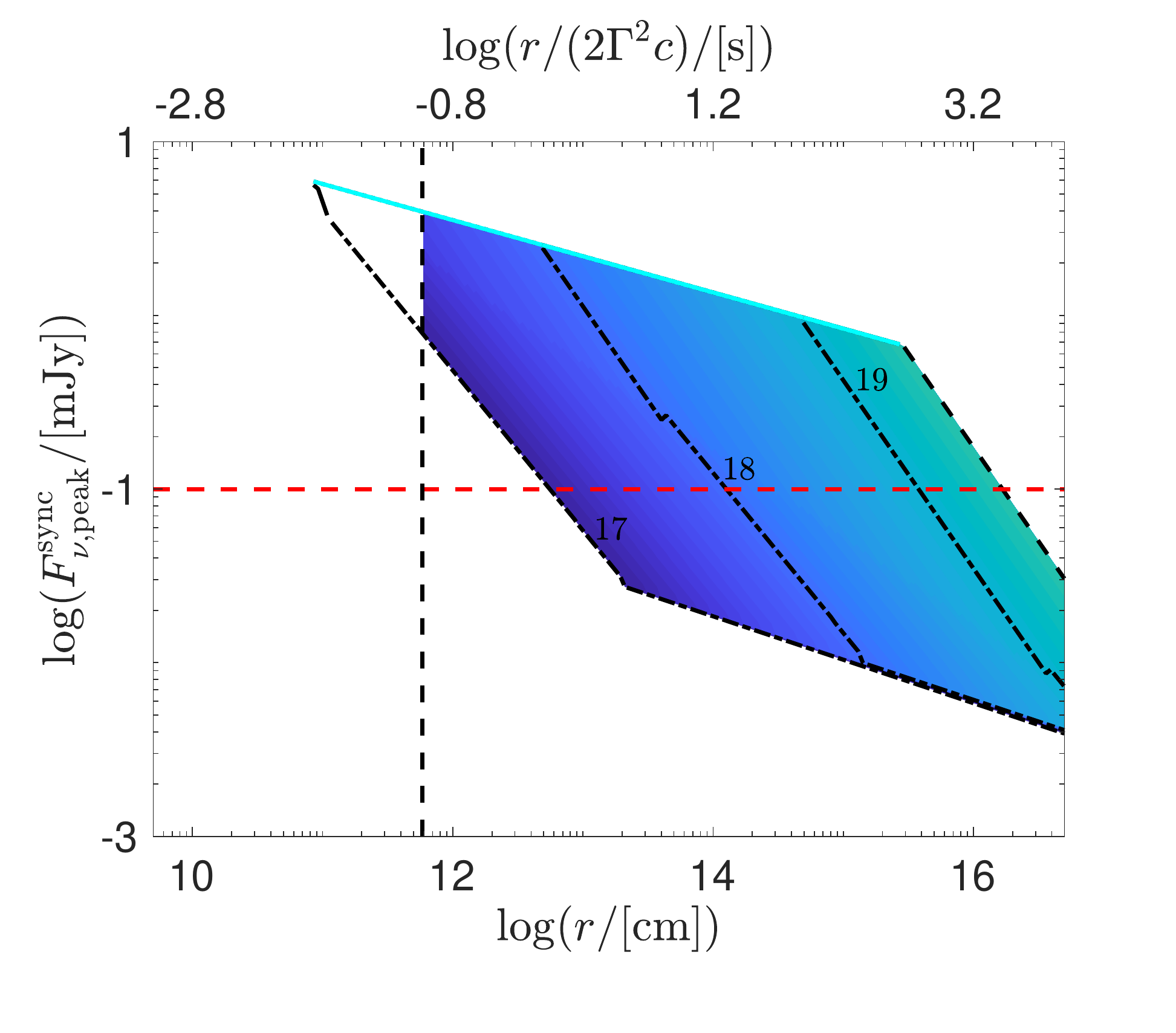}
    \includegraphics[width=0.32\columnwidth]{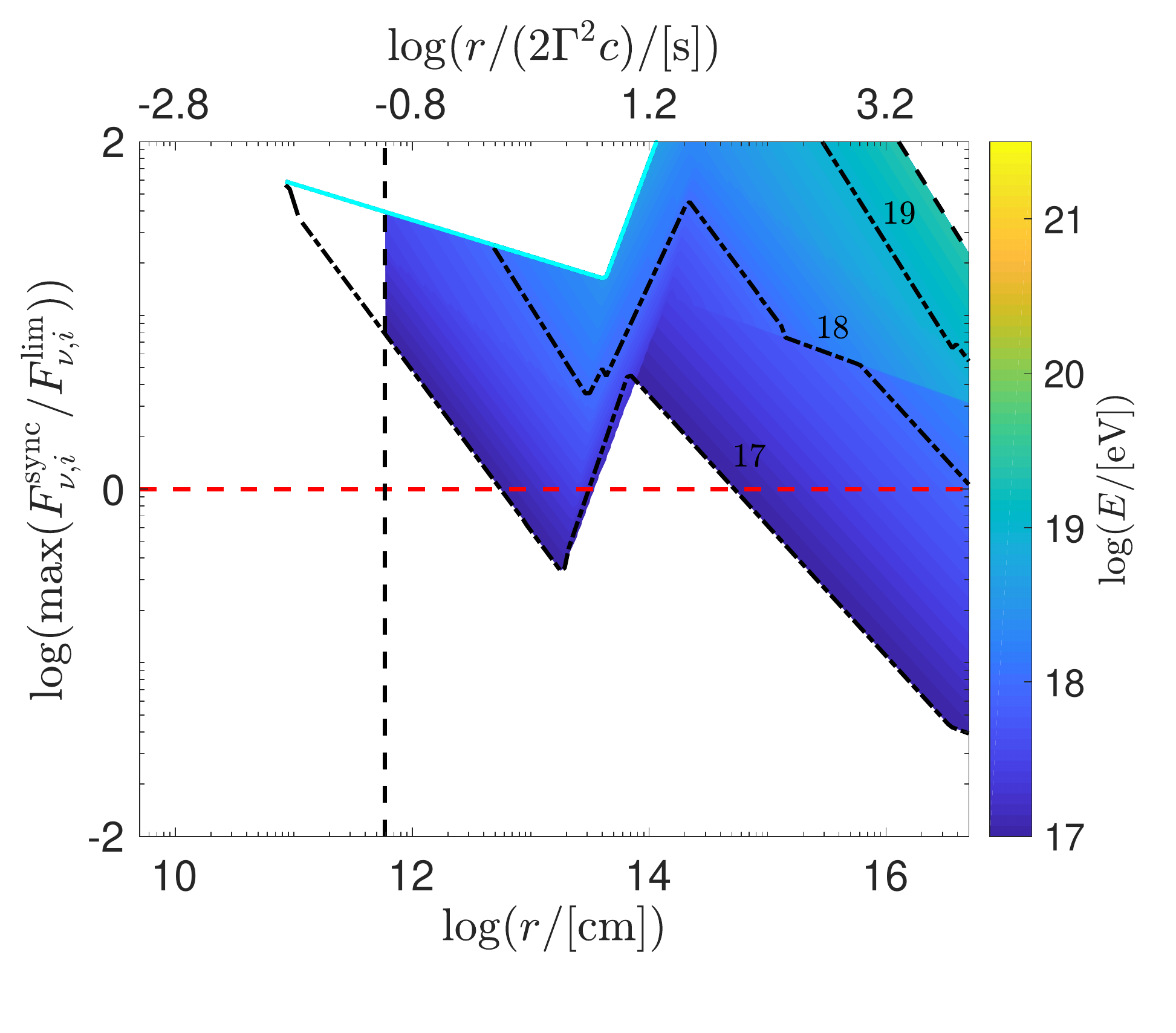}
    \includegraphics[width=0.32\columnwidth]{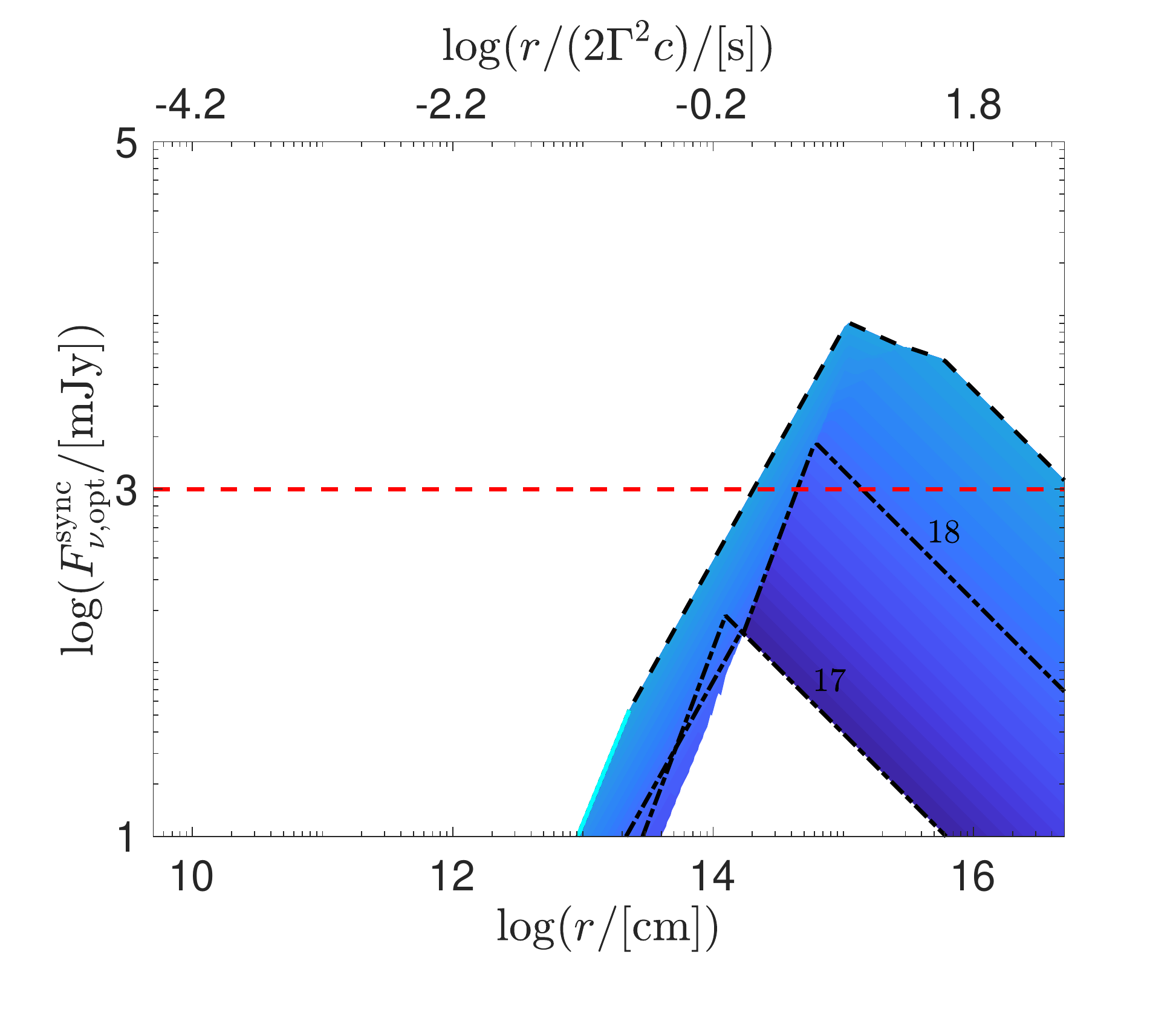}
    \includegraphics[width=0.32\columnwidth]{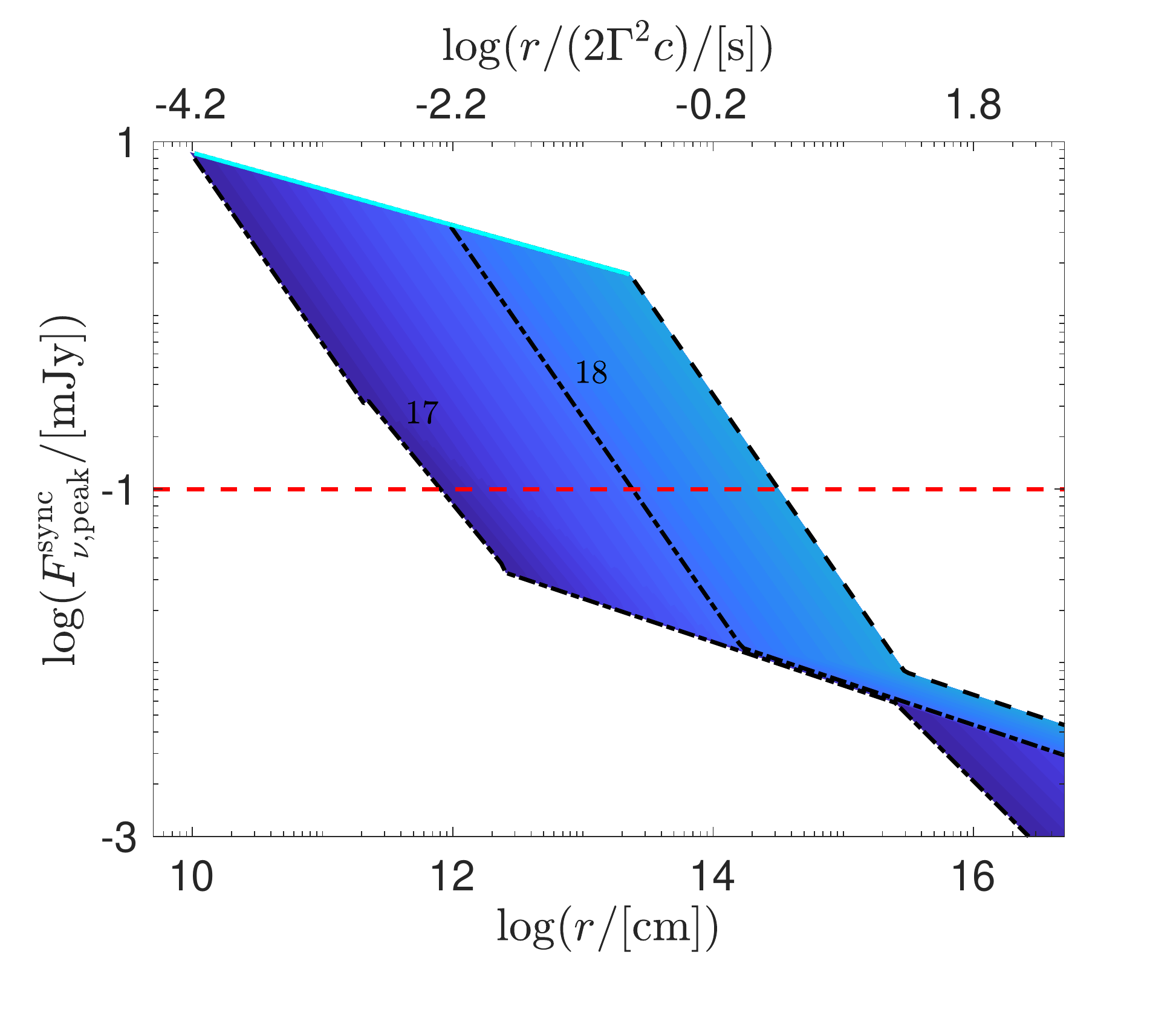}
    \includegraphics[width=0.32\columnwidth]{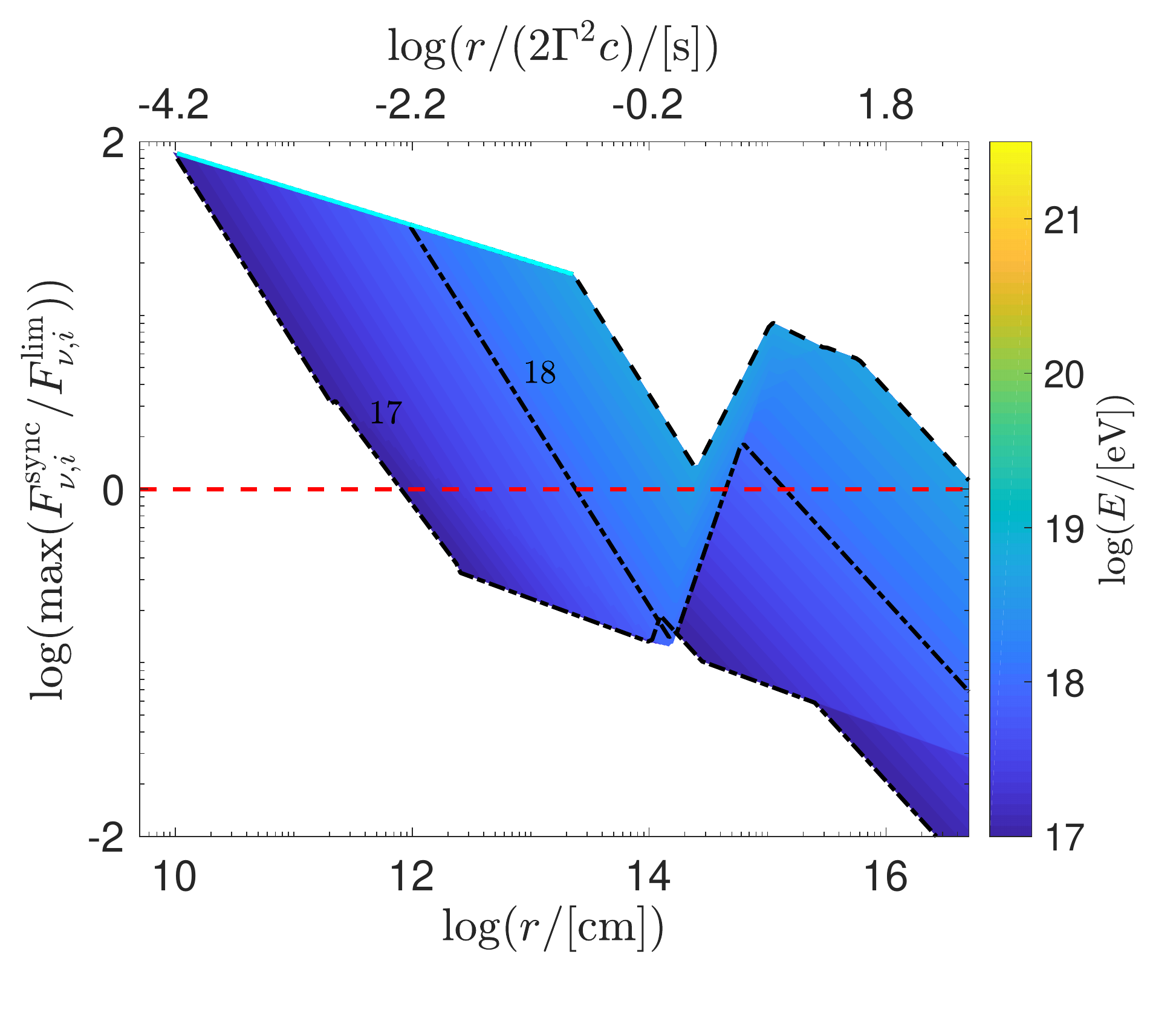}
    \includegraphics[width=0.32\columnwidth]{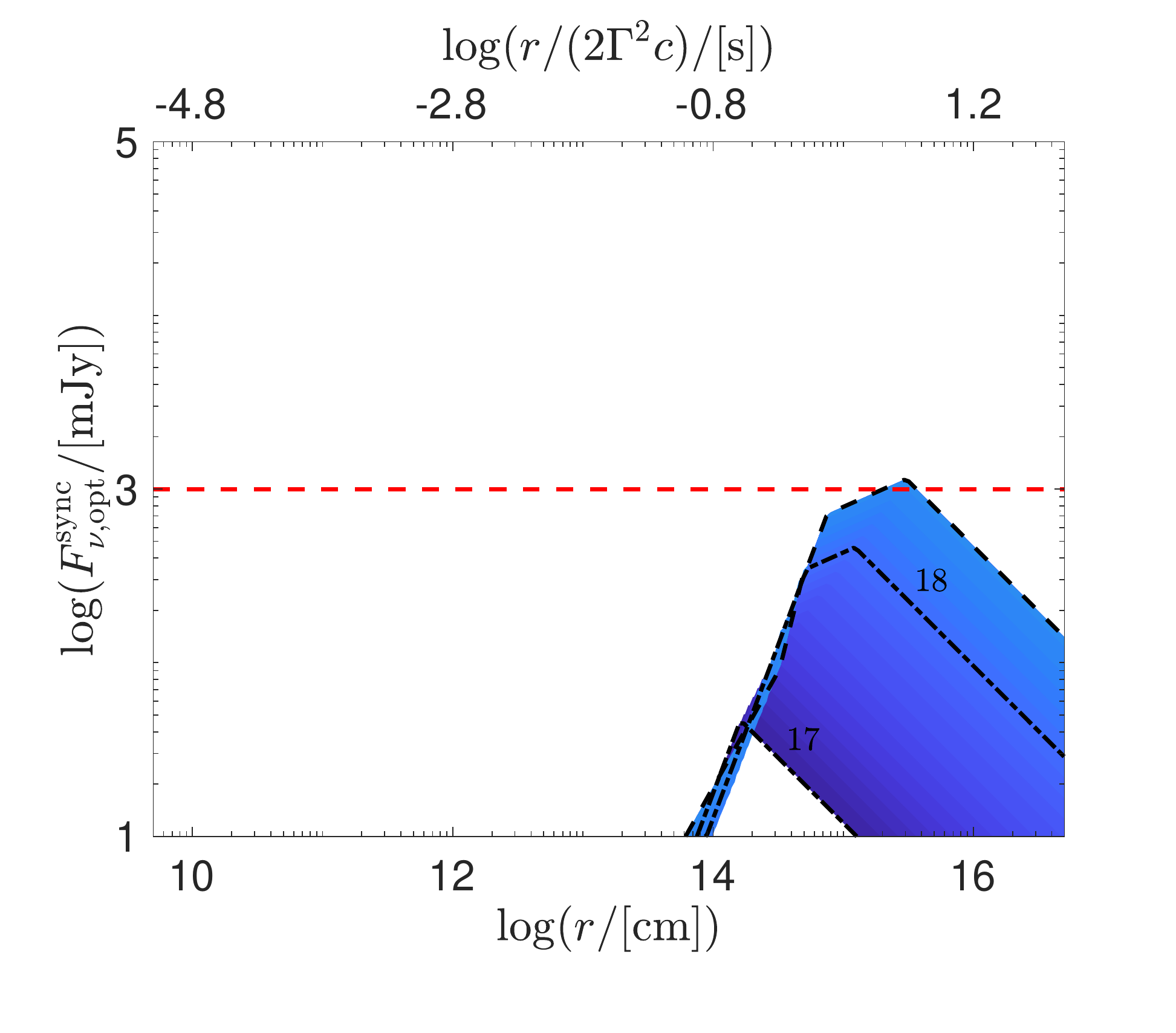}
    \includegraphics[width=0.32\columnwidth]{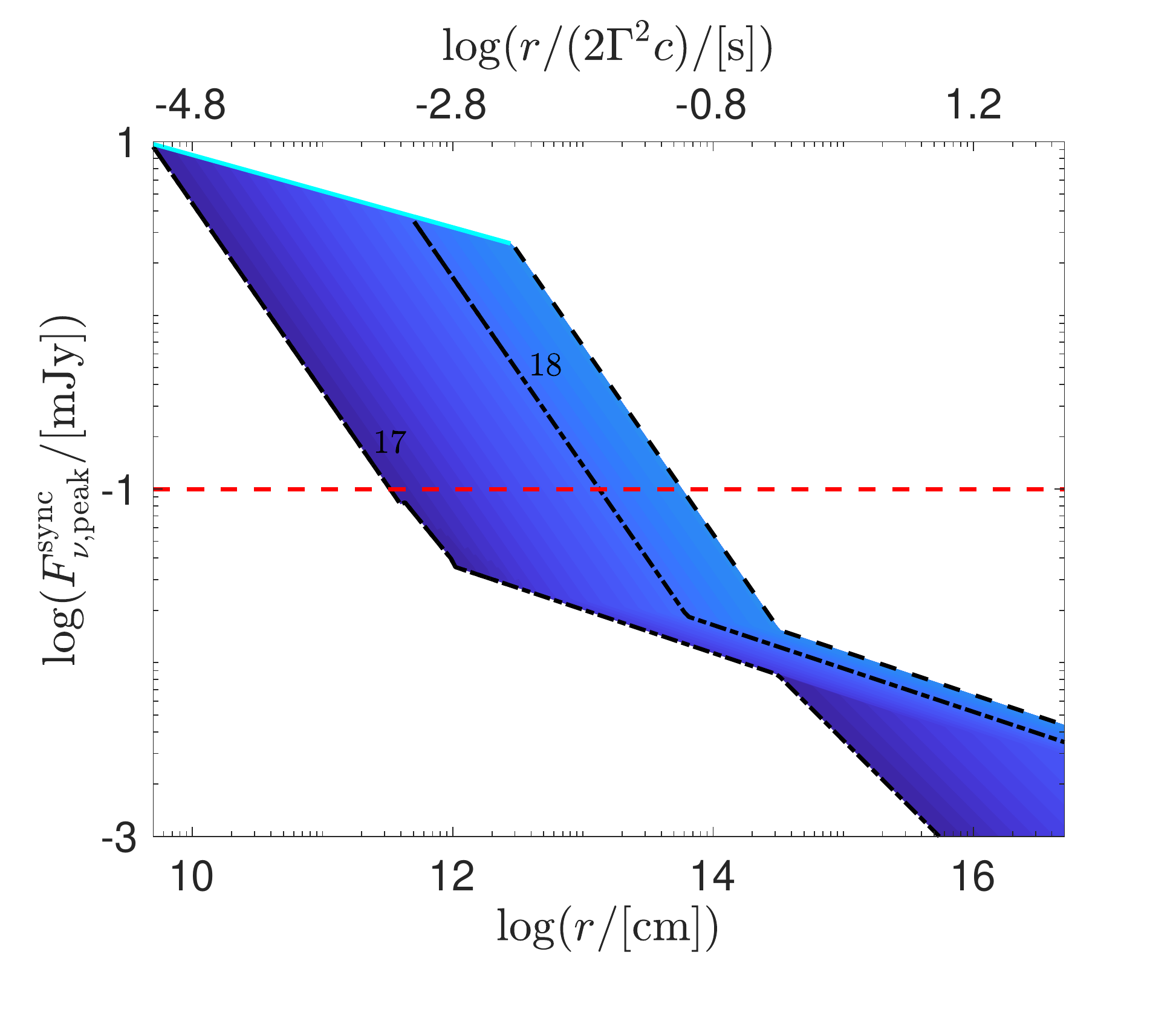}
    \includegraphics[width=0.32\columnwidth]{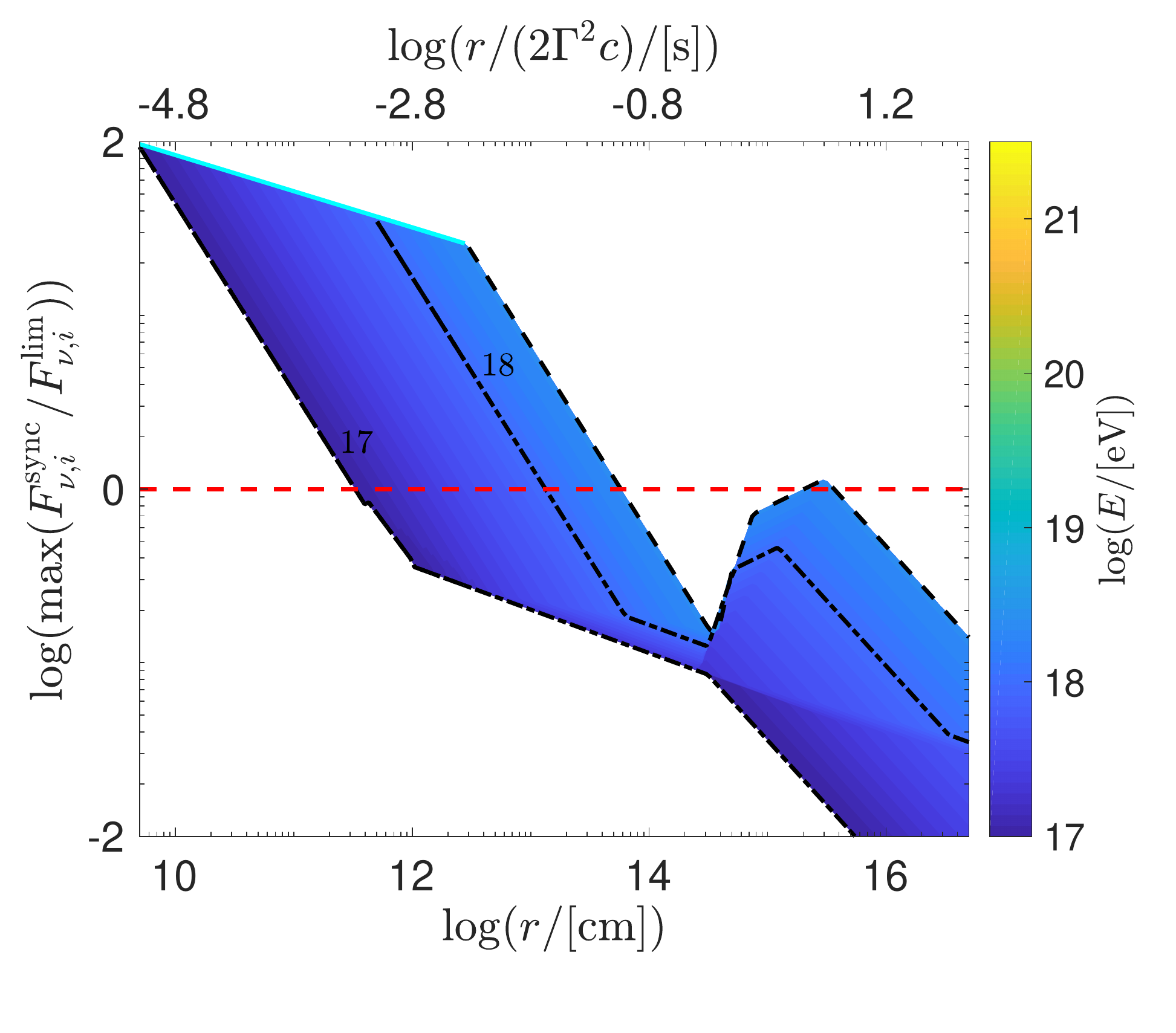}
    \includegraphics[width=0.32\columnwidth]{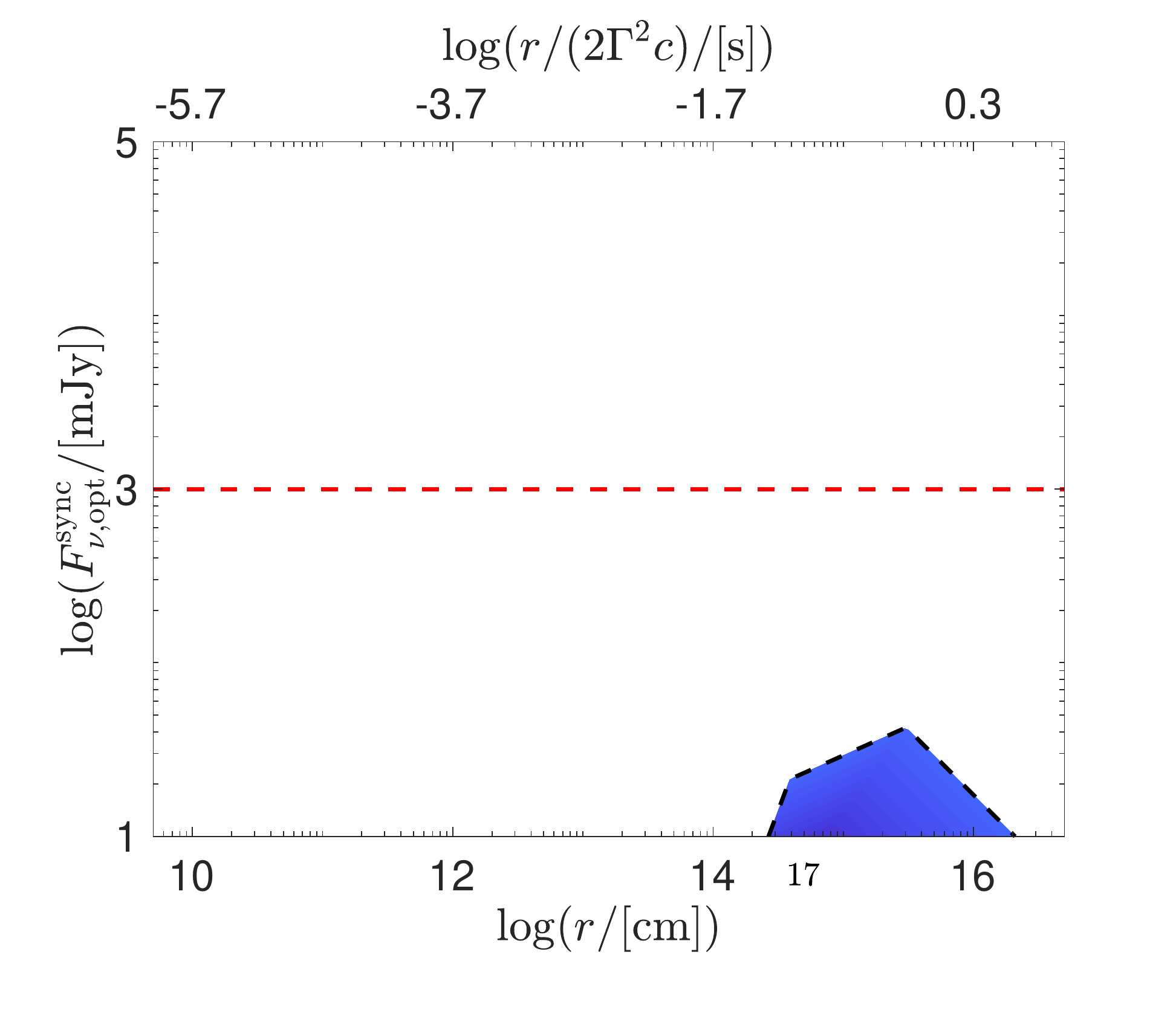}
    \includegraphics[width=0.32\columnwidth]{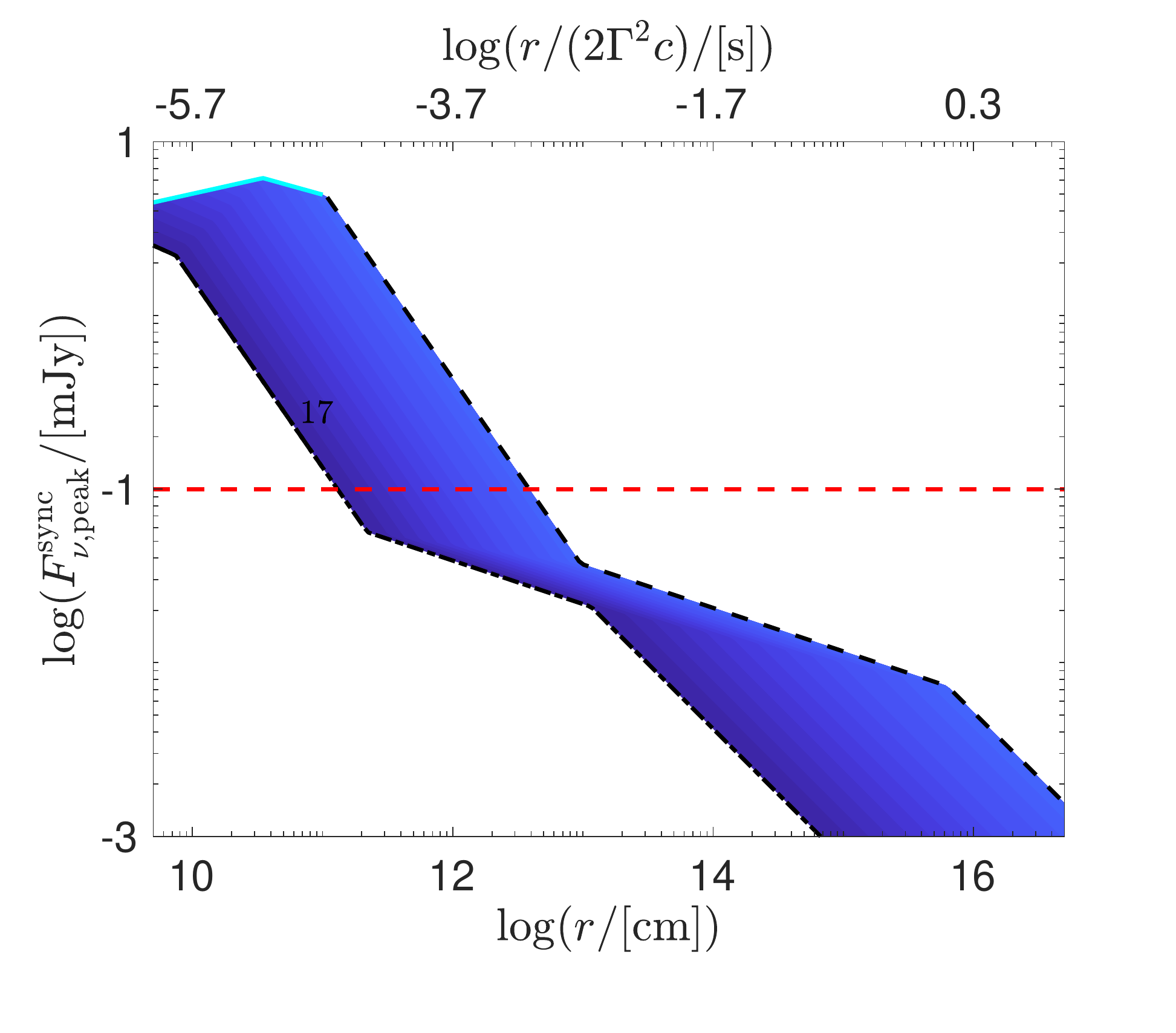}
    \includegraphics[width=0.32\columnwidth]{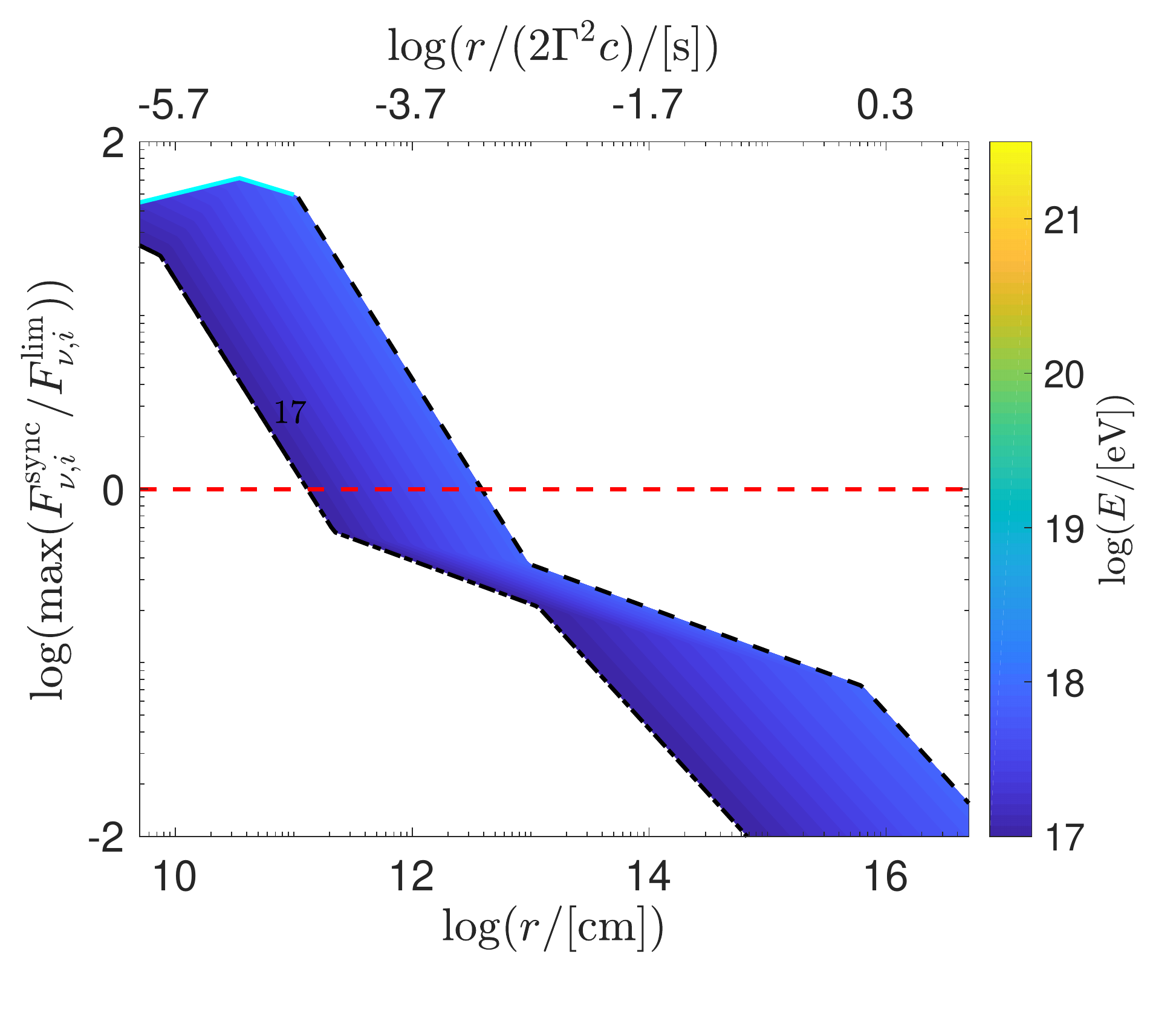}
\caption{Equivalent to Figure \ref{Fig:FluxLimits}, but for a low-luminosity GRB with $L_{\rm tot} = 10^{48}$ erg s$^{-1}$, $\varepsilon_{\rm ll, peak}=100$ keV, and $z_{\rm ll} = 0.05$. Plotted for $\Gamma = 10$, 50, 100, and 300 from top to bottom. Other numerical values used are given in Table \ref{tab:PhotosphereNumerics}.}
\label{Fig:FluxLimitsLL}
\end{centering}
\end{figure*}

\begin{figure*}
\begin{centering}
    \includegraphics[width=0.32\columnwidth]{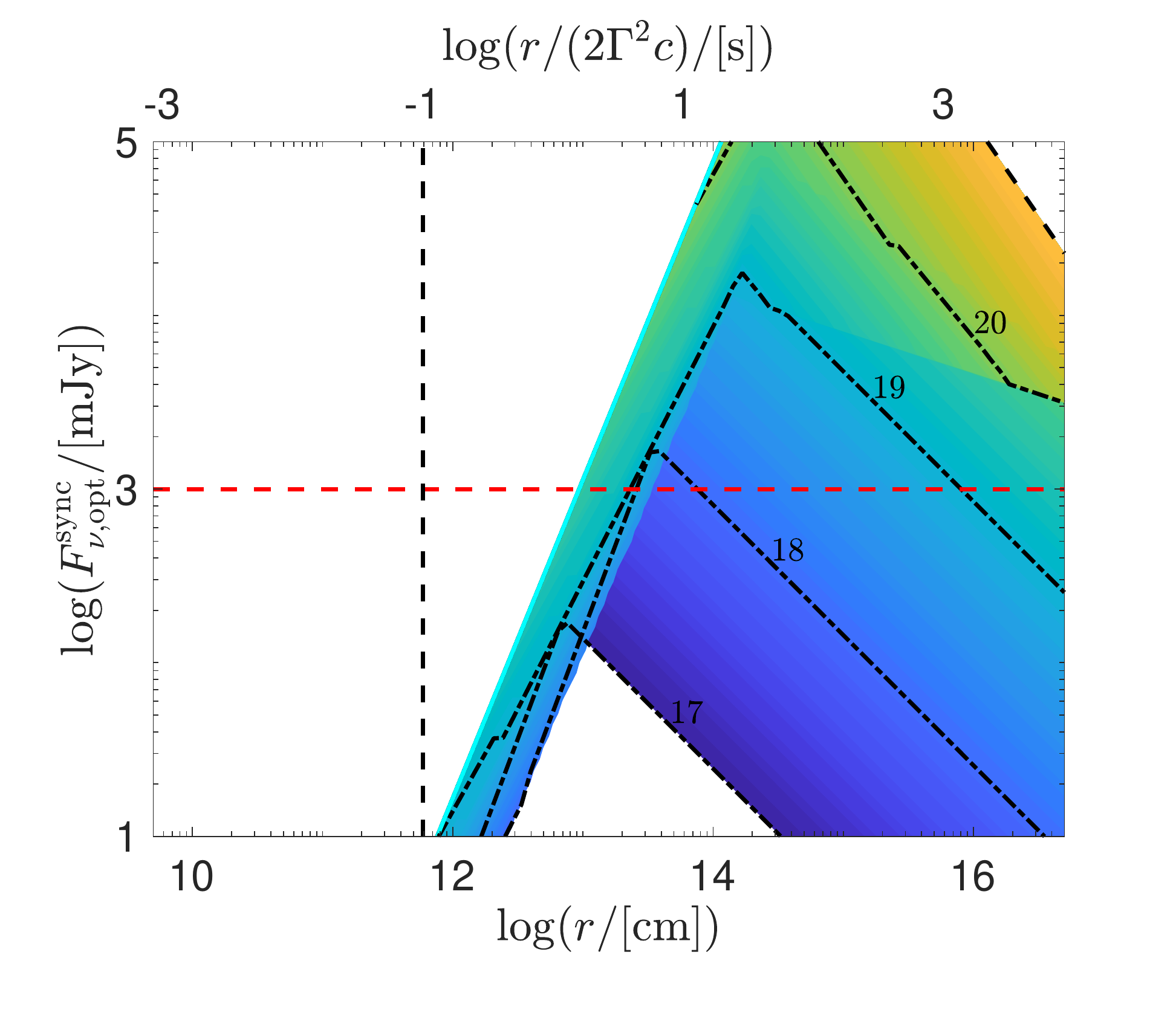}
    \includegraphics[width=0.32\columnwidth]{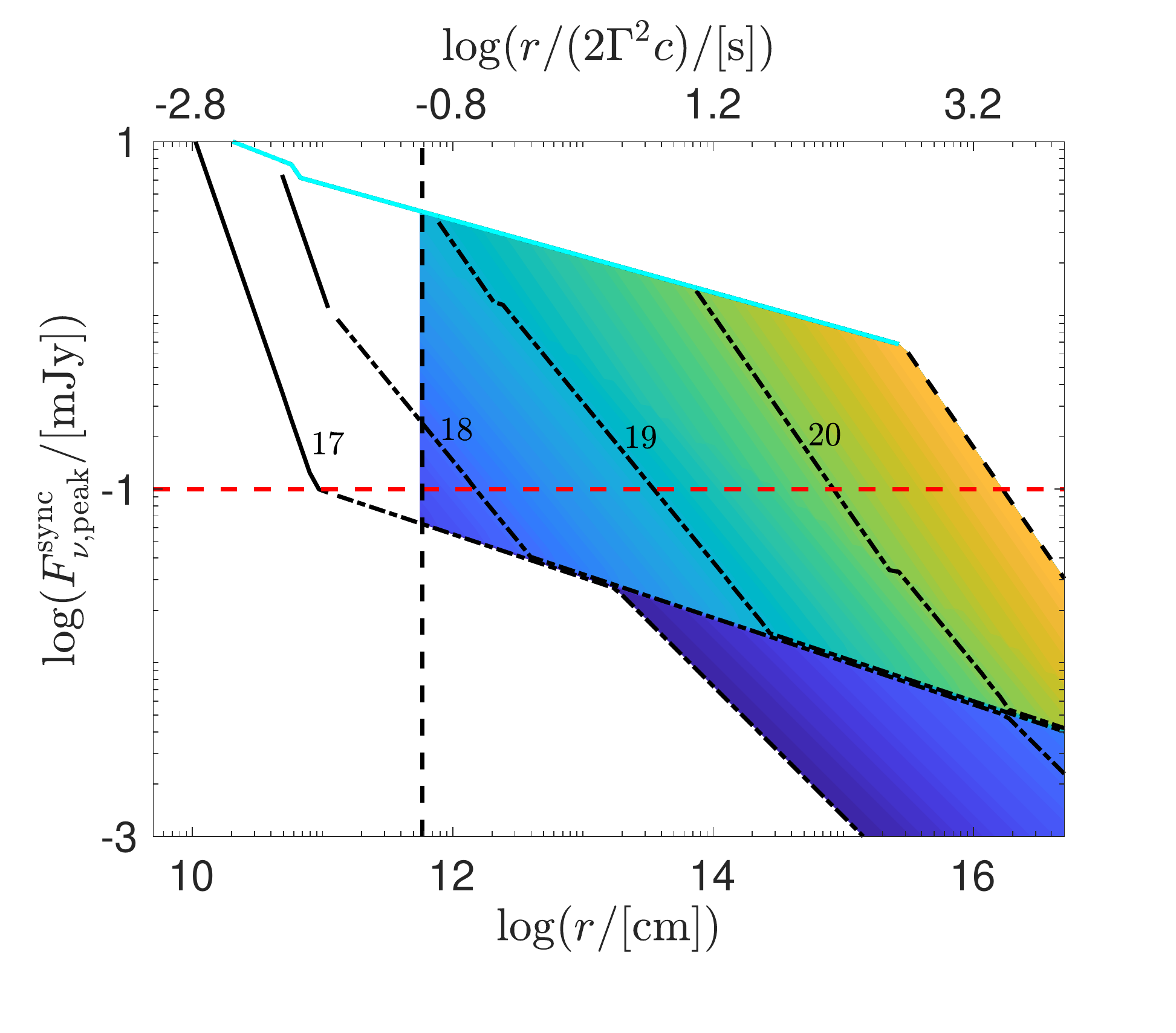}
    \includegraphics[width=0.32\columnwidth]{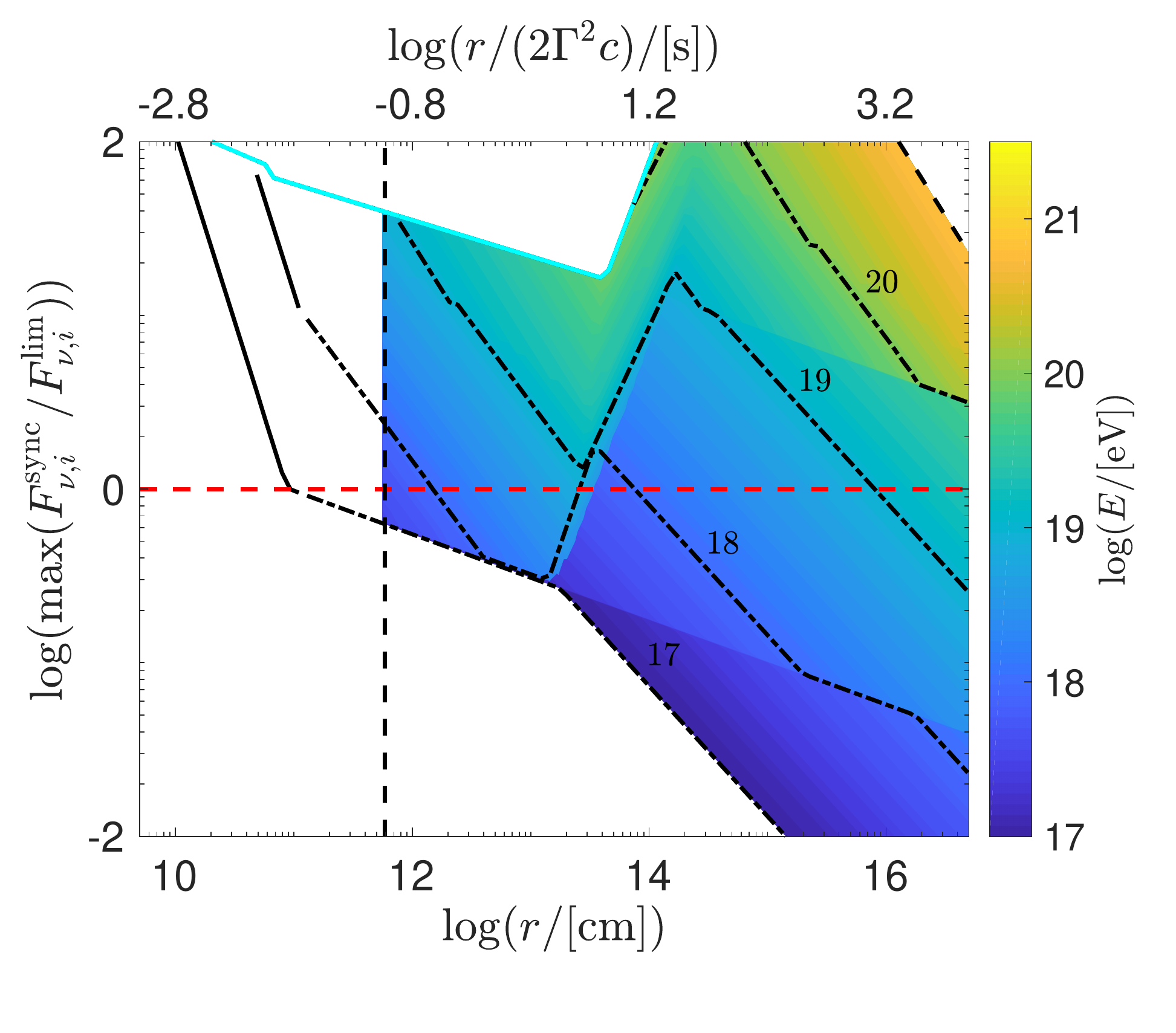}
    \includegraphics[width=0.32\columnwidth]{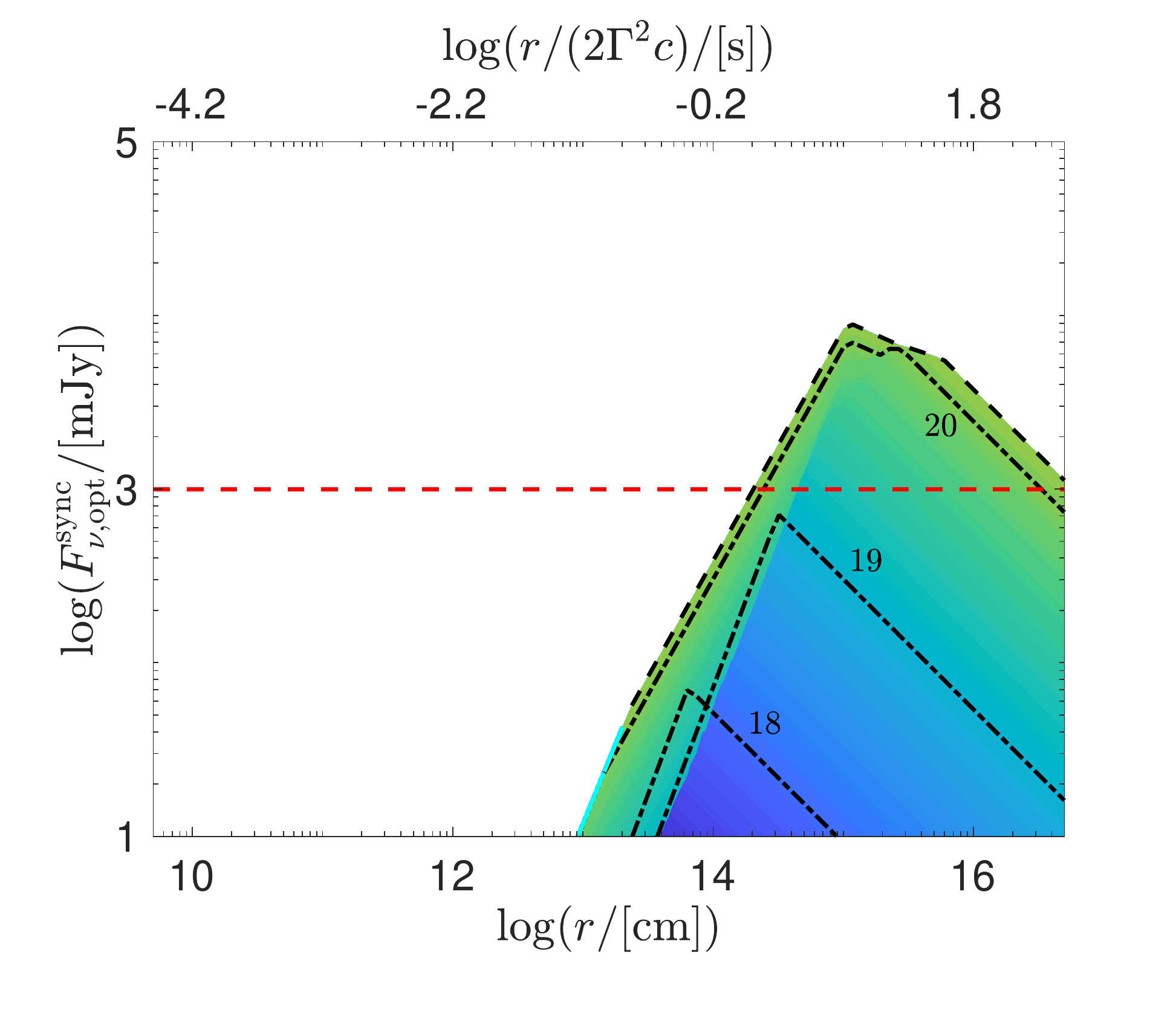}
    \includegraphics[width=0.32\columnwidth]{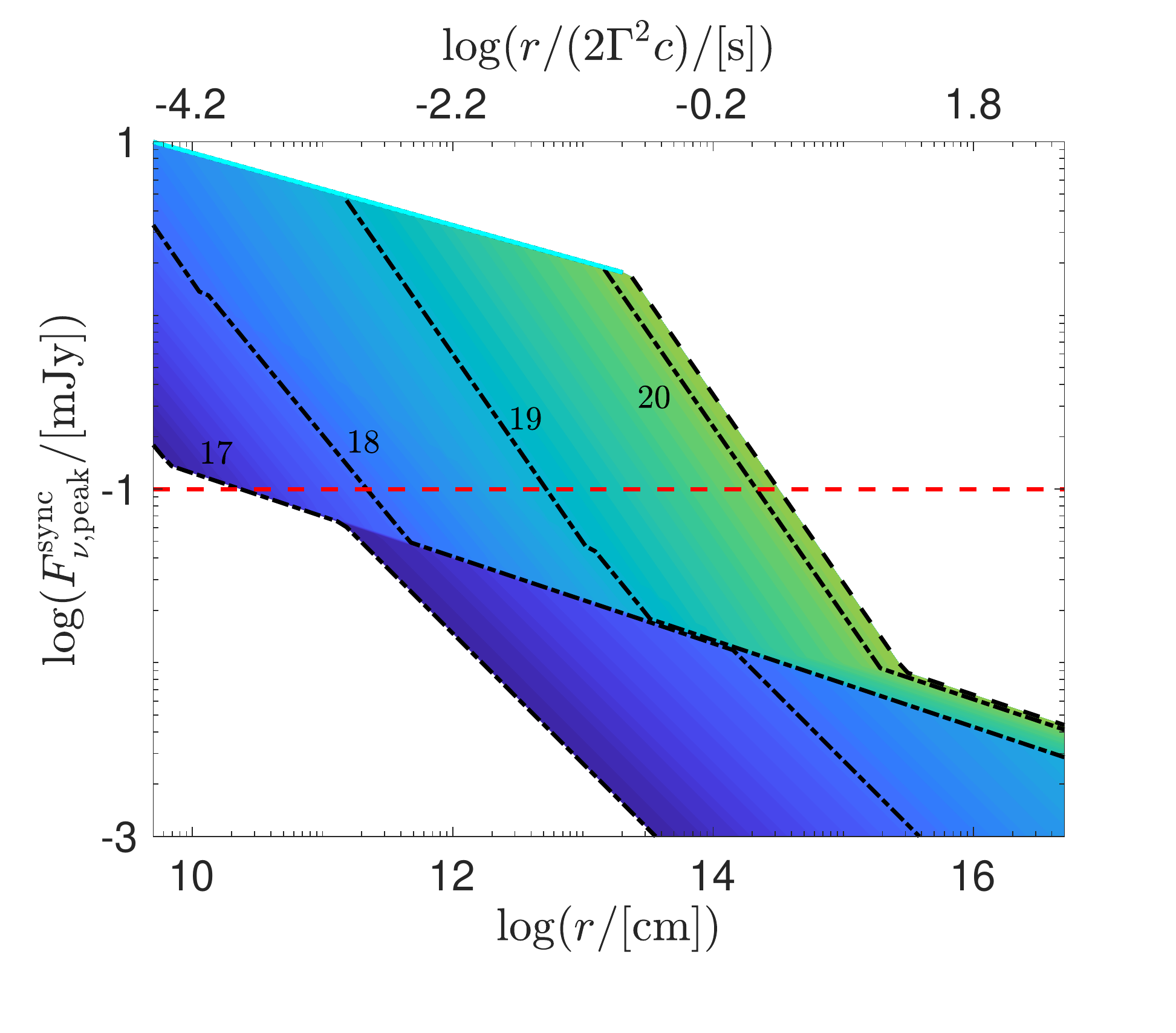}
    \includegraphics[width=0.32\columnwidth]{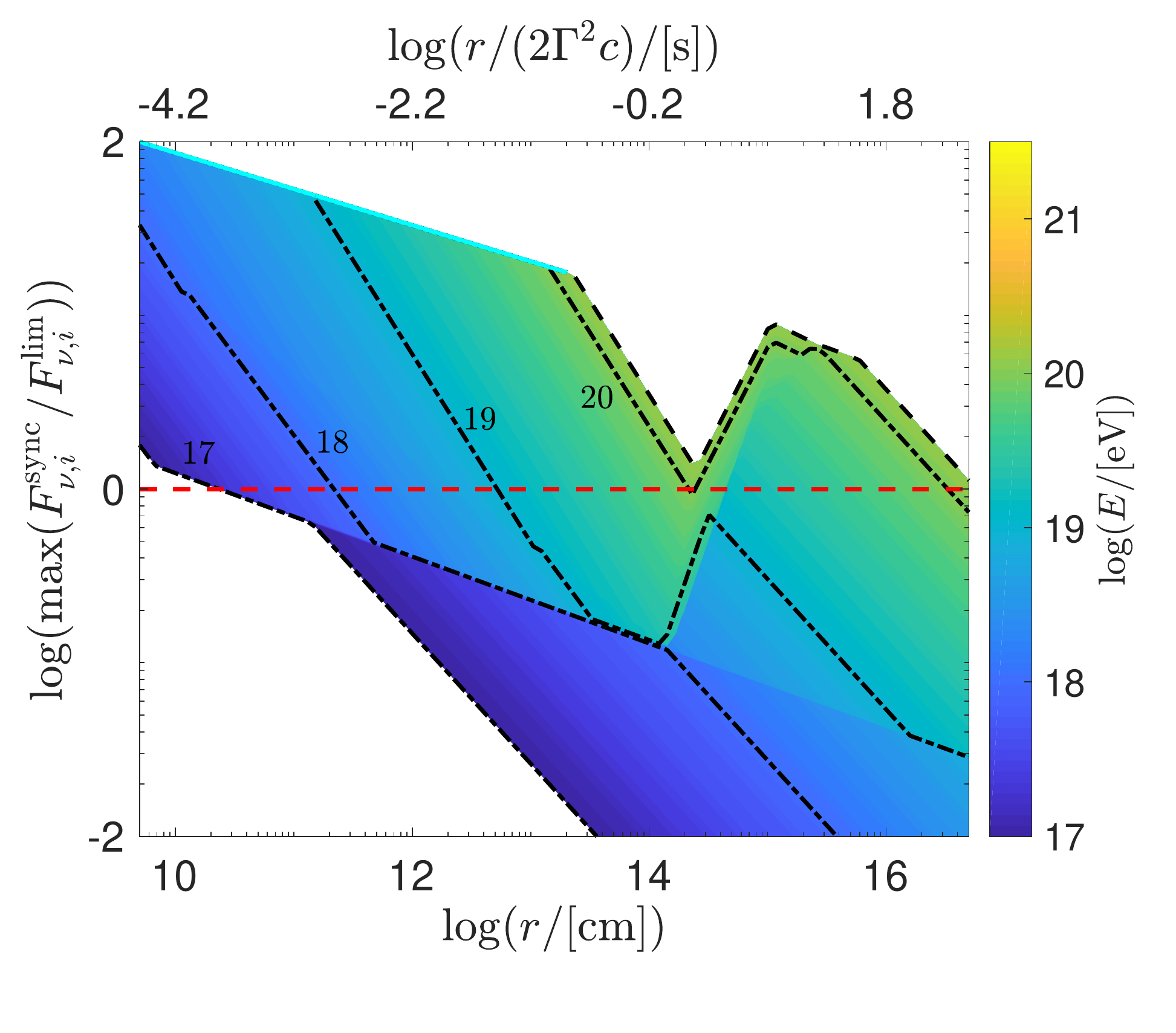}
    \includegraphics[width=0.32\columnwidth]{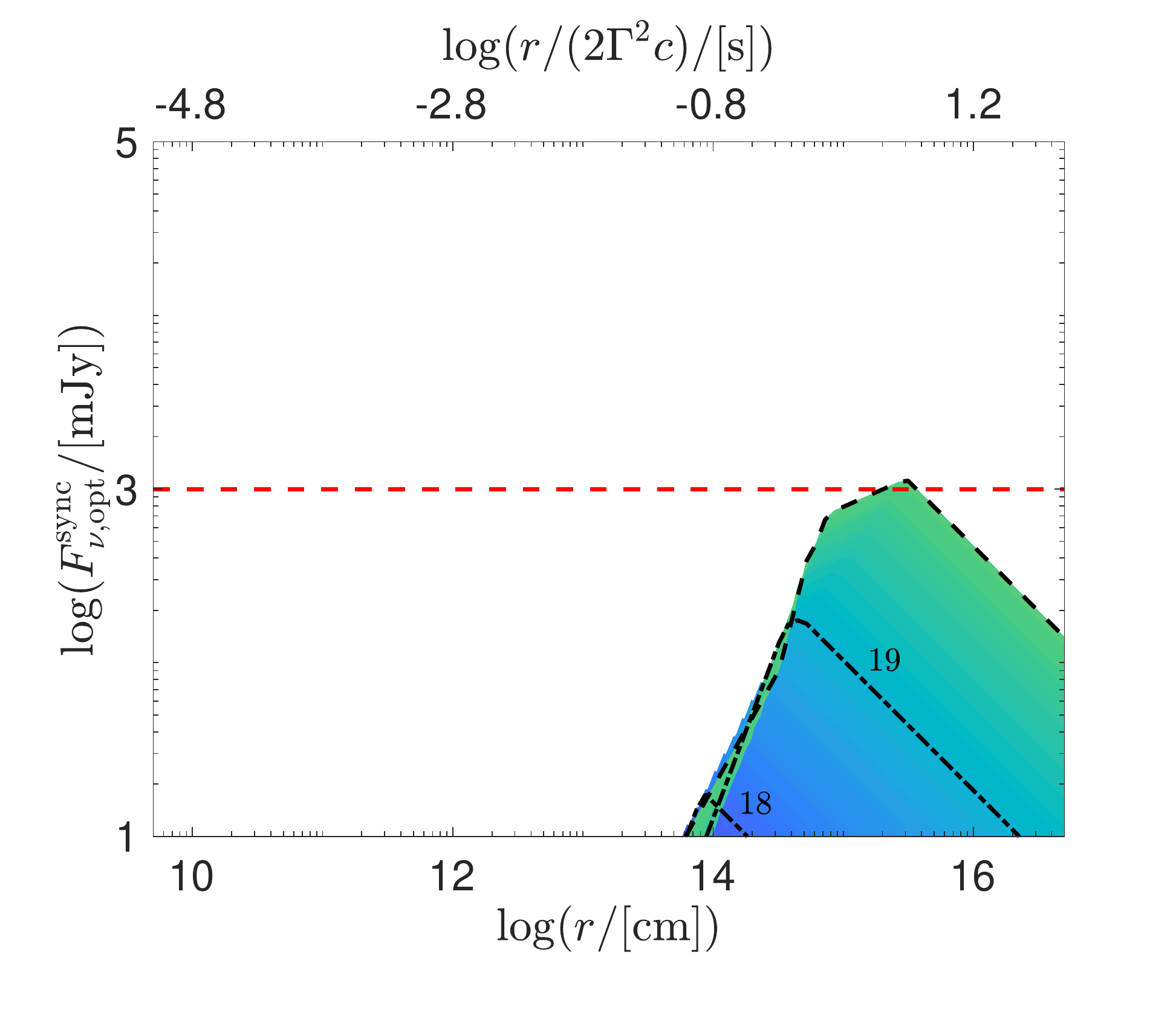}
    \includegraphics[width=0.32\columnwidth]{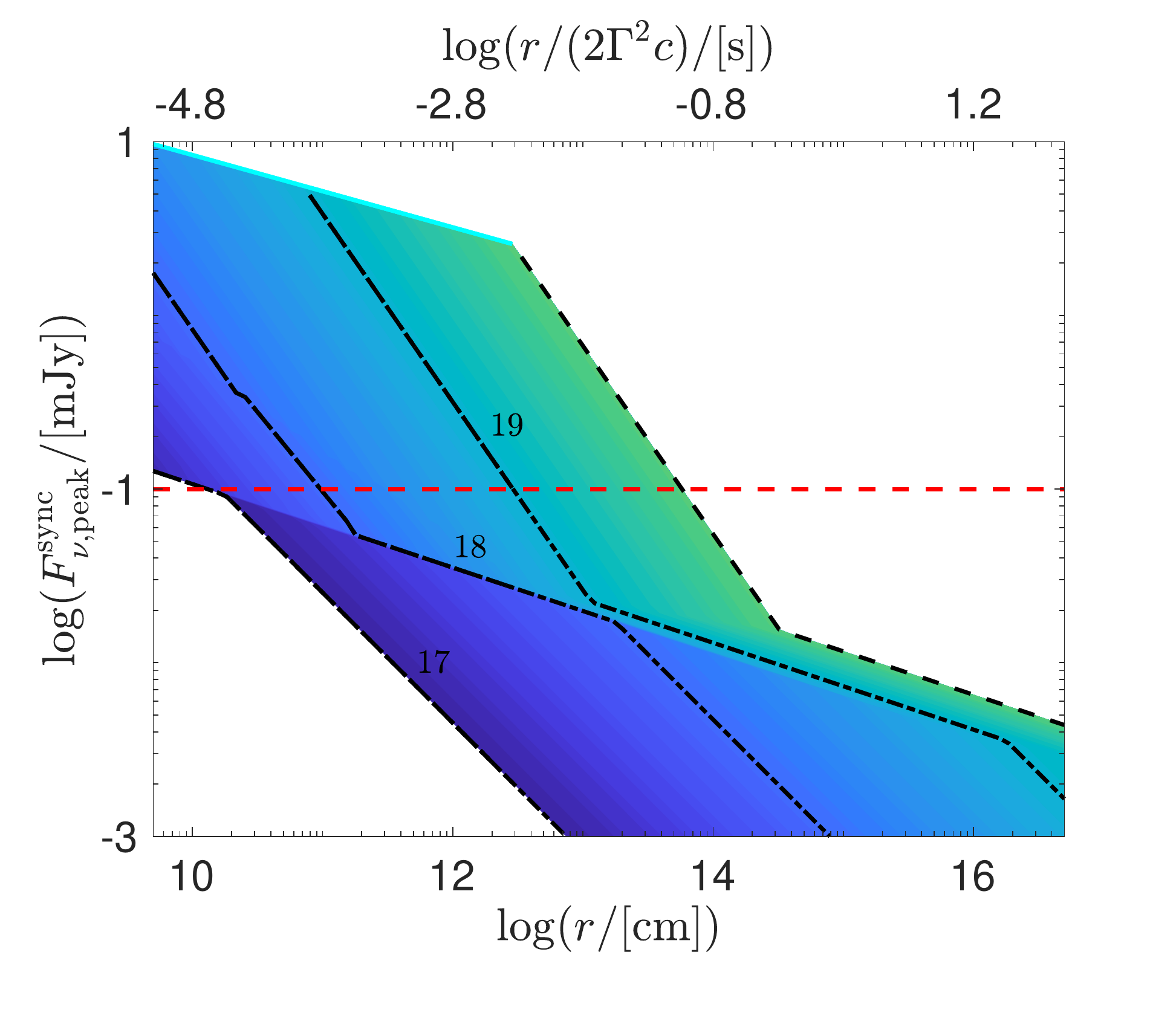}
    \includegraphics[width=0.32\columnwidth]{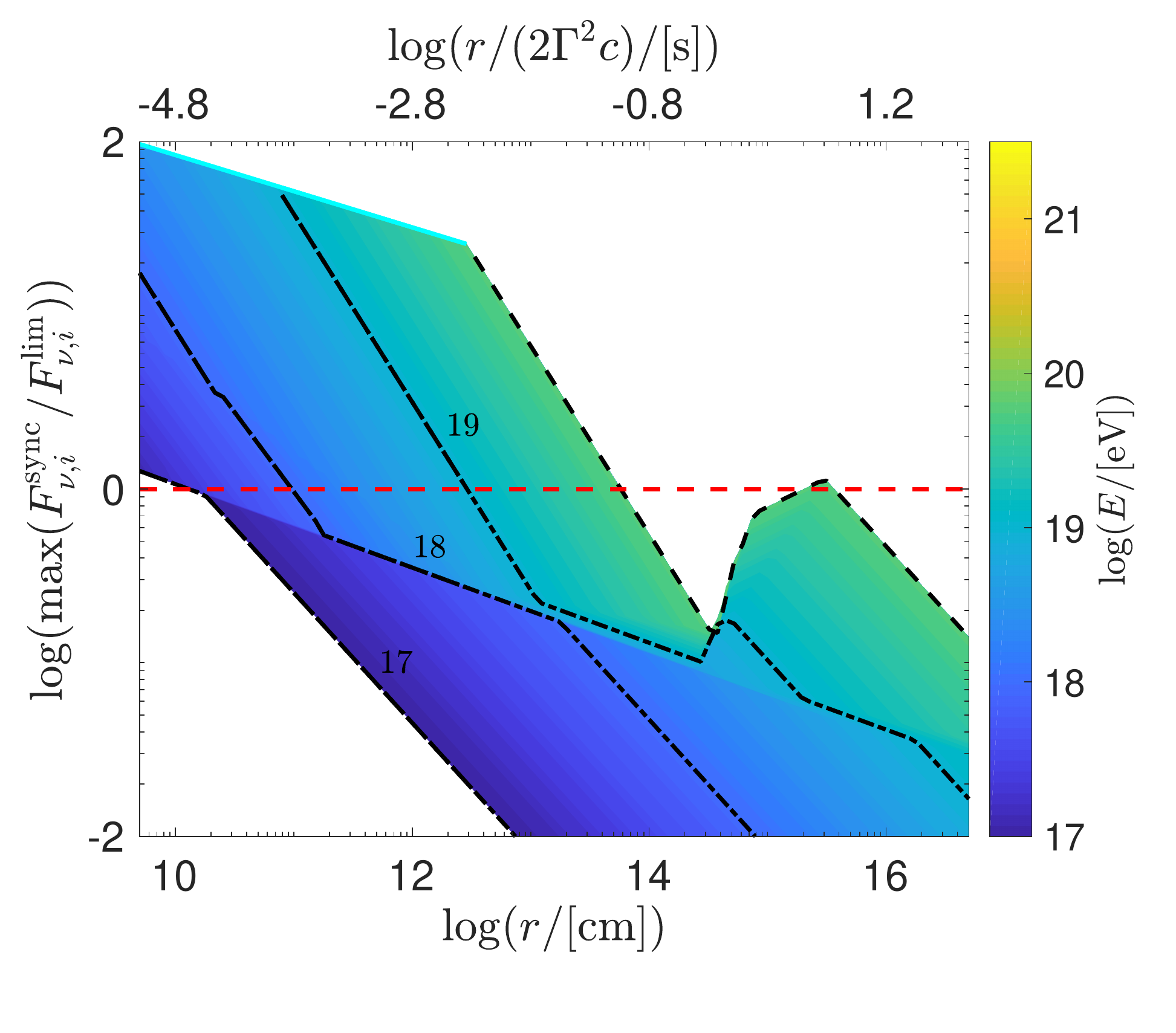}
    \includegraphics[width=0.32\columnwidth]{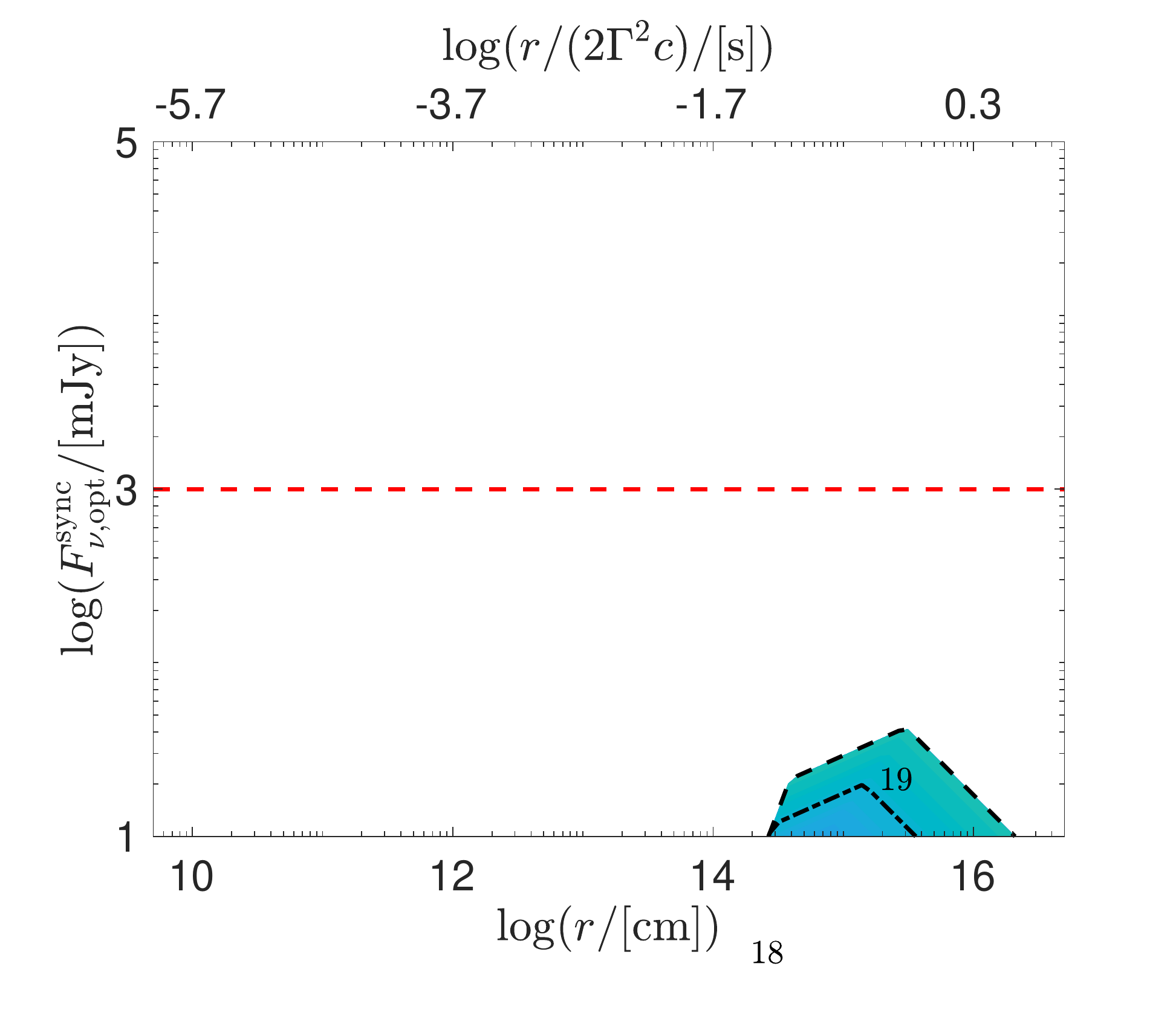}
    \includegraphics[width=0.32\columnwidth]{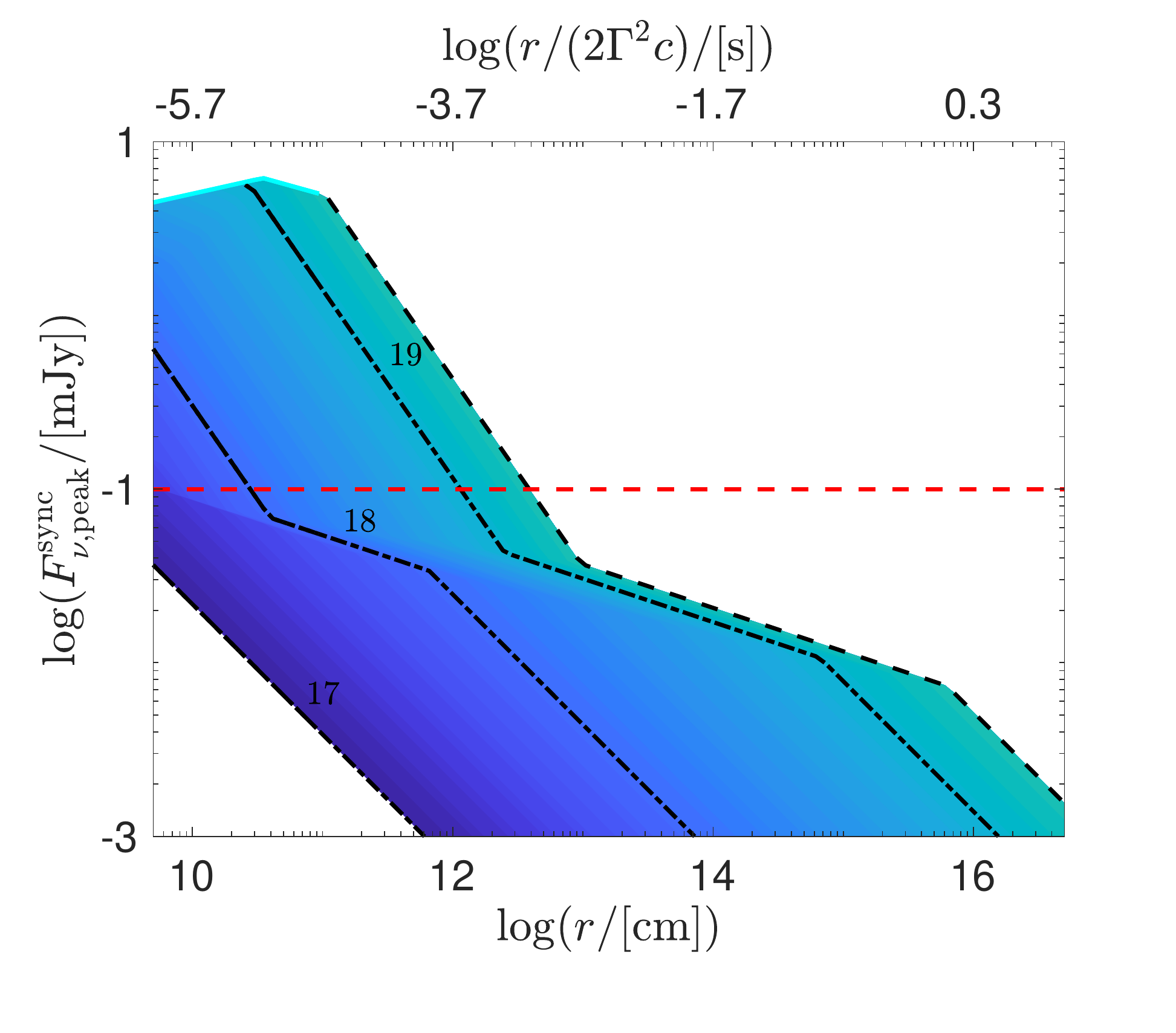}
    \includegraphics[width=0.32\columnwidth]{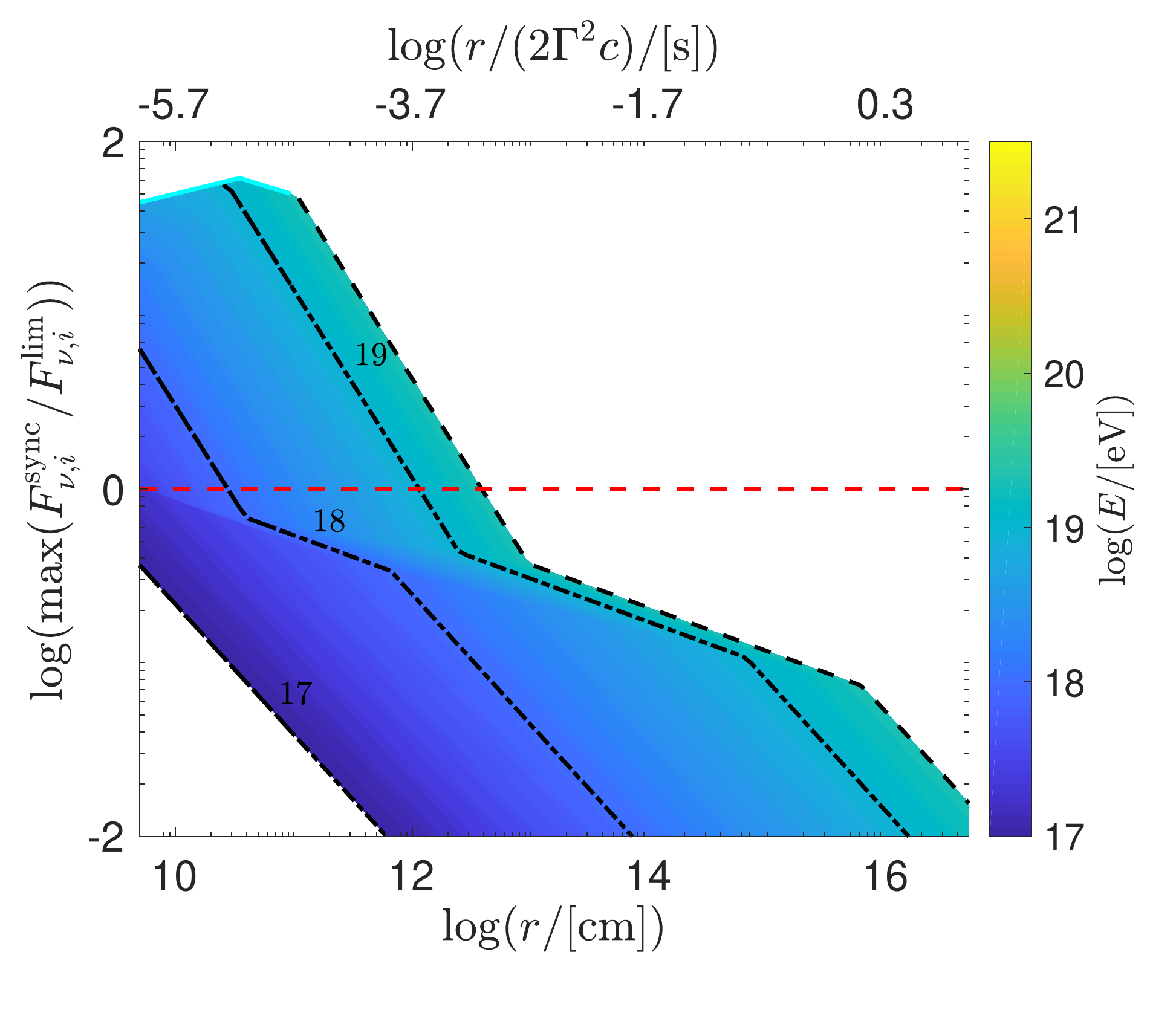}
\caption{Equivalent to Figure \ref{Fig:FluxLimits}, but for completely stripped iron in a low-luminosity GRB with $L_{\rm tot} = 10^{48}$ erg s$^{-1}$, $\varepsilon_{\rm ll, peak}=100$ keV, and $z_{\rm ll} = 0.05$. Plotted for $\Gamma = 10$, 50, 100, and 300 from top to bottom. Other numerical values used are given in Table \ref{tab:PhotosphereNumerics}.}
\label{Fig:FluxLimitsIronLL}
\end{centering}
\end{figure*}

In Figures \ref{Fig:FluxLimitsLL} and \ref{Fig:FluxLimitsIronLL} we show plots similar to Figure \ref{Fig:FluxLimits}, but for a llGRB with $L_{\rm tot} = 10^{48}$ erg s$^{-1}$. Figure \ref{Fig:FluxLimitsLL} show the results for protons and Figure \ref{Fig:FluxLimitsIronLL} for completely stripped iron. The bulk outflow Lorentz factors have been modified as llGRBs are believed to have lower outflow velocities. The plots are done with $\Gamma = 10$, 50, 100, and 300 from top to bottom. 
Even though their luminosities are several orders of magnitude below generic GRB luminosities, llGRB peak energies are only lower by a factor of a few, and their peak fluxes are lower by a factor of $\gtrsim 10$ \citep{Sun2015}. For the figures we use $\varepsilon_{\rm ll, peak}^{\rm obs} = 100$ keV and $F^{\rm lim}_{\rm ll,\nu, peak} = 0.1$ mJy. While these values are too high for some llGRBs, lowering either of them only reduces the allowed parameter space. A typical prompt optical flux from llGRBs is unknown, but untargeted optical transient surveys, see e.g. \citet{Kehoe2002, Rykoff2005, Rau2006}, have so far been unsuccessful in their searches for unknown optical GRB transients. 
As llGRBs are predicted to be quite common in the local universe, this might suggest low fluxes in optical. Furthermore, the non-detection by UVOT of XRF 100316D/SN 2010bh $\sim 150$ s after trigger \citep{Starling2011, Fan2011} and the low optical fluxes detected for GRB 060218 $\sim 150$ s after trigger \citep{GCN-4779, Ghisellini2007} together with very long prompt emission for both of $>1000$ seconds, suggests that a typical value for the optical flux in a llGRB is in the order of 1 mJy or even lower.
However, due to this uncertainty we set the limit on optical flux to $F^{\rm lim}_{\rm ll, \nu, opt} = 1$ Jy, which is one order of magnitude larger than the limit used in the canonical GRB case. 
As they are much fainter, observed llGRBs are always much closer to us. For Figures \ref{Fig:FluxLimitsLL} and \ref{Fig:FluxLimitsIronLL}, we have taken $z = 0.05$.

Inspection of Figures \ref{Fig:FluxLimitsLL} and \ref{Fig:FluxLimitsIronLL} shows that neither protons nor iron can be accelerated to $10^{20}$ eV or above in any part of the parameter space. \added{This result is valid for our fiducial parameters as given in Table \ref{tab:PhotosphereNumerics}; iron acceleration to $10^{20}$ eV would be possible given specific parameter values, see Section \ref{Sec:Discussion}.} All scenarios, except when $\Gamma = 10$, are severely limited by the requirement that the magnetic luminosity cannot be larger than the total luminosity; it is difficult to reconcile an emission region that is necessarily far out with a quite large magnetic field with a low total luminosity. Indeed, due to this constraint acceleration above $10^{20}$ eV is only permissible for iron in the burst where $\Gamma =10$ ($\Gamma = 50$), but this would overshoot the optical flux with an optical emission of $\gtrsim 3$ Jy ($\gtrsim 1$ Jy). \deleted{Indeed, }The predicted optical flux increases rapidly with decreasing $\Gamma$, as can be seen from the left-hand panels\deleted{ in Figures \ref{Fig:FluxLimitsLL} and \ref{Fig:FluxLimitsIronLL}}. This result, very high optical fluxes for a large part of the parameter space, is valid regardless of the emission mechanism of the prompt $\gamma$-rays, be it synchrotron, photospheric, or shock breakout. The latter is not discussed in this paper but suggested in the literature as a possible cause for the emission in llGRBs, see e.g. \citet{Suzuki2010}. We stress that the current small sample size might not give a good representation of the real distribution, and therefore this analysis should be repeated when there are more detections. 
However, the problem with the too large magnetic luminosity must be addressed in all llGRB - UHECR models.




%
%

\section{Discussion}\label{Sec:Discussion}

In the synchrotron model, the only strong dependence is on the acceleration efficiency $\eta$ and bulk outflow Lorentz factor $\Gamma$ as evident from Equations \eqref{varepsilonsync} - \eqref{varepsilonpgamma}. In high-luminosity bursts, proton acceleration to the highest observed energies can still not be obtained in the synchrotron model for $\eta = 1$ (observe that if $\eta > 1$, then the restriction that the Larmor radius needs to be smaller than the system size becomes relevant). Additionally, $\Gamma \geq 2000$ is required to reach $10^{20}$ eV. In low-luminosity bursts, increasing $\eta$ is not sufficient for proton acceleration above $10^{19}$ eV. Iron can reach energies of $10^{20}$ eV when $\eta = 1$, however it just falls short of reaching the highest observed energies ($10^{20.5}$ eV). 

\begin{figure*}
\begin{centering}
    \includegraphics[width=0.32\columnwidth]{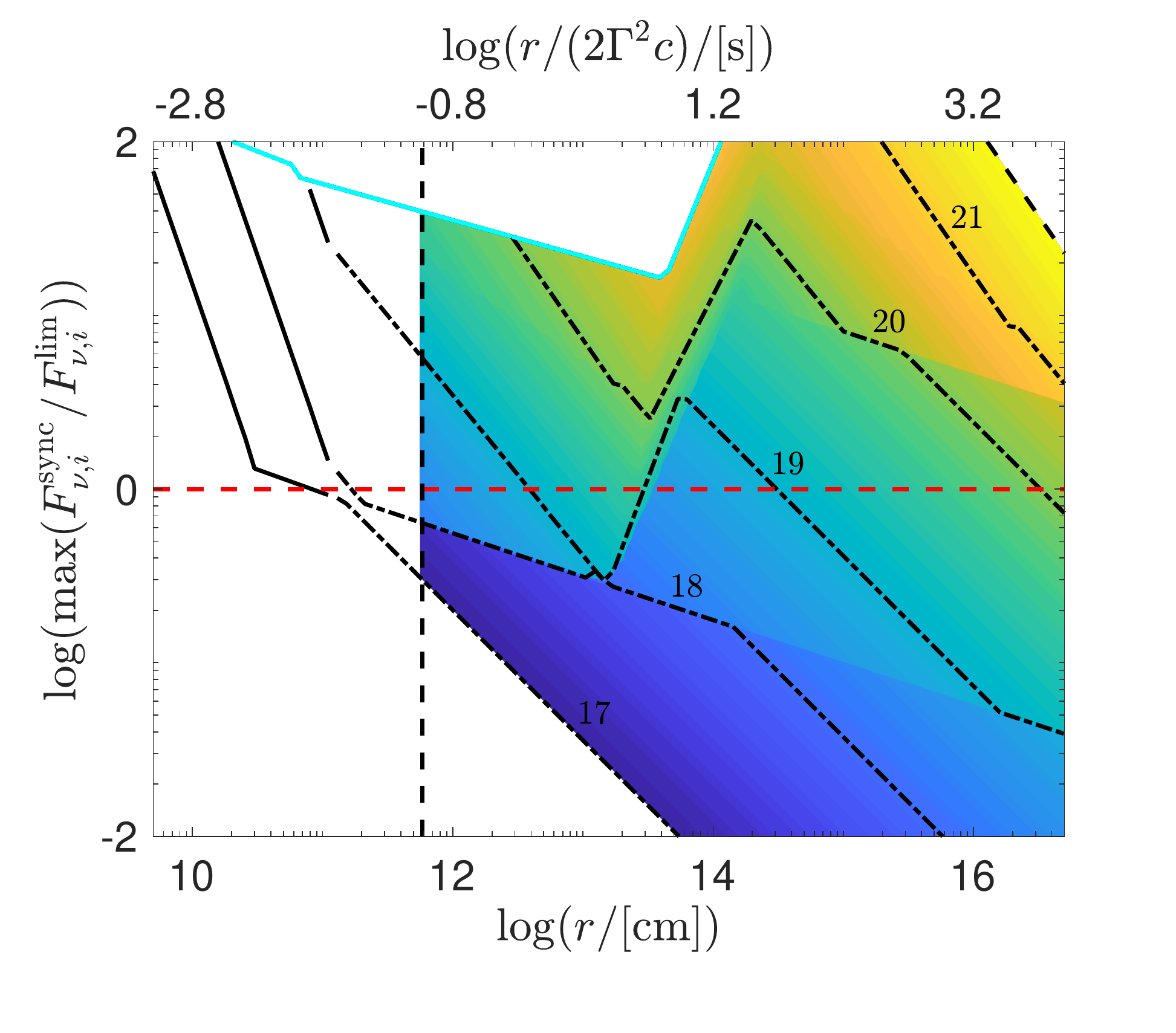}
    \includegraphics[width=0.32\columnwidth]{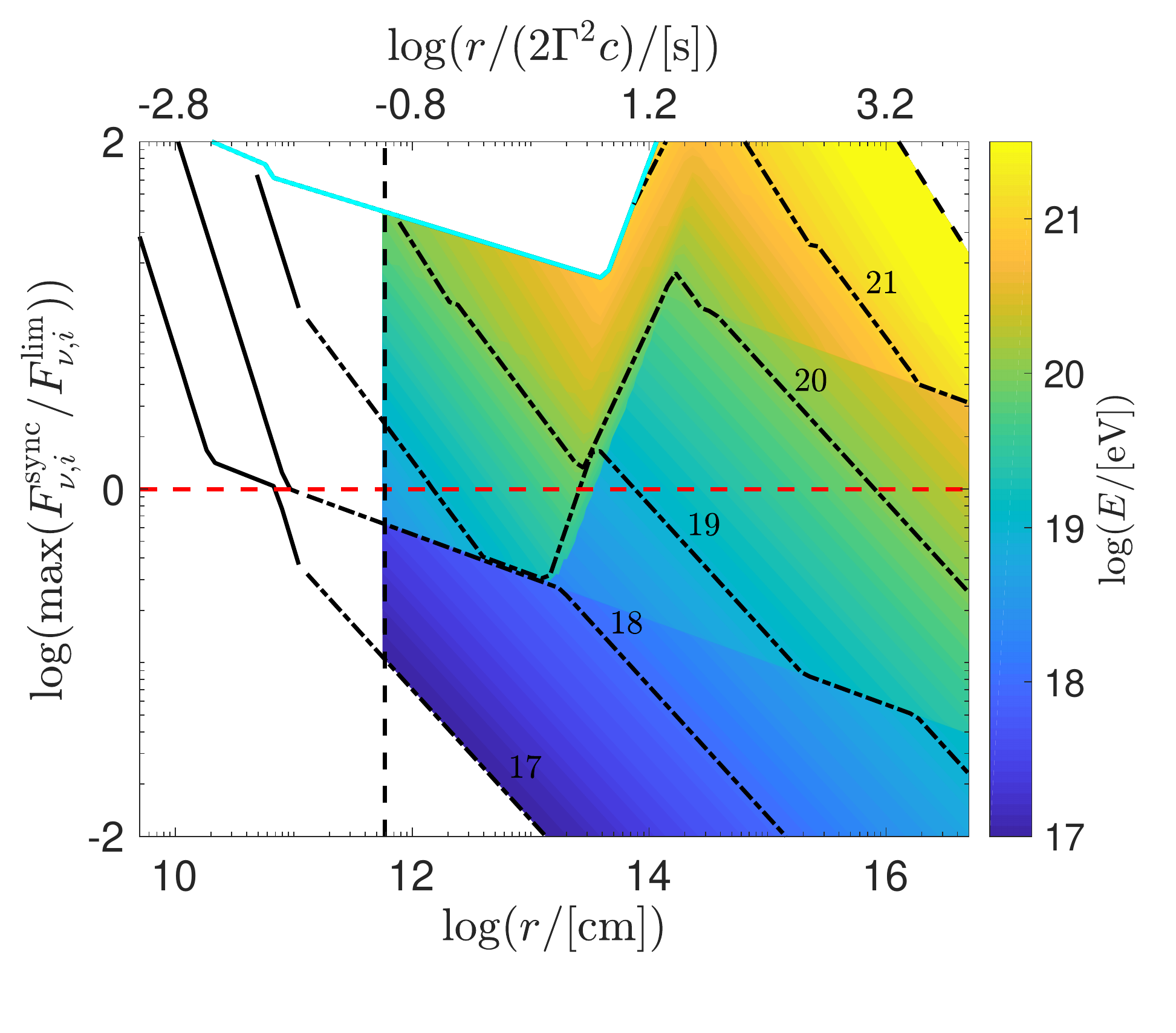}\\
    \includegraphics[width=0.32\columnwidth]{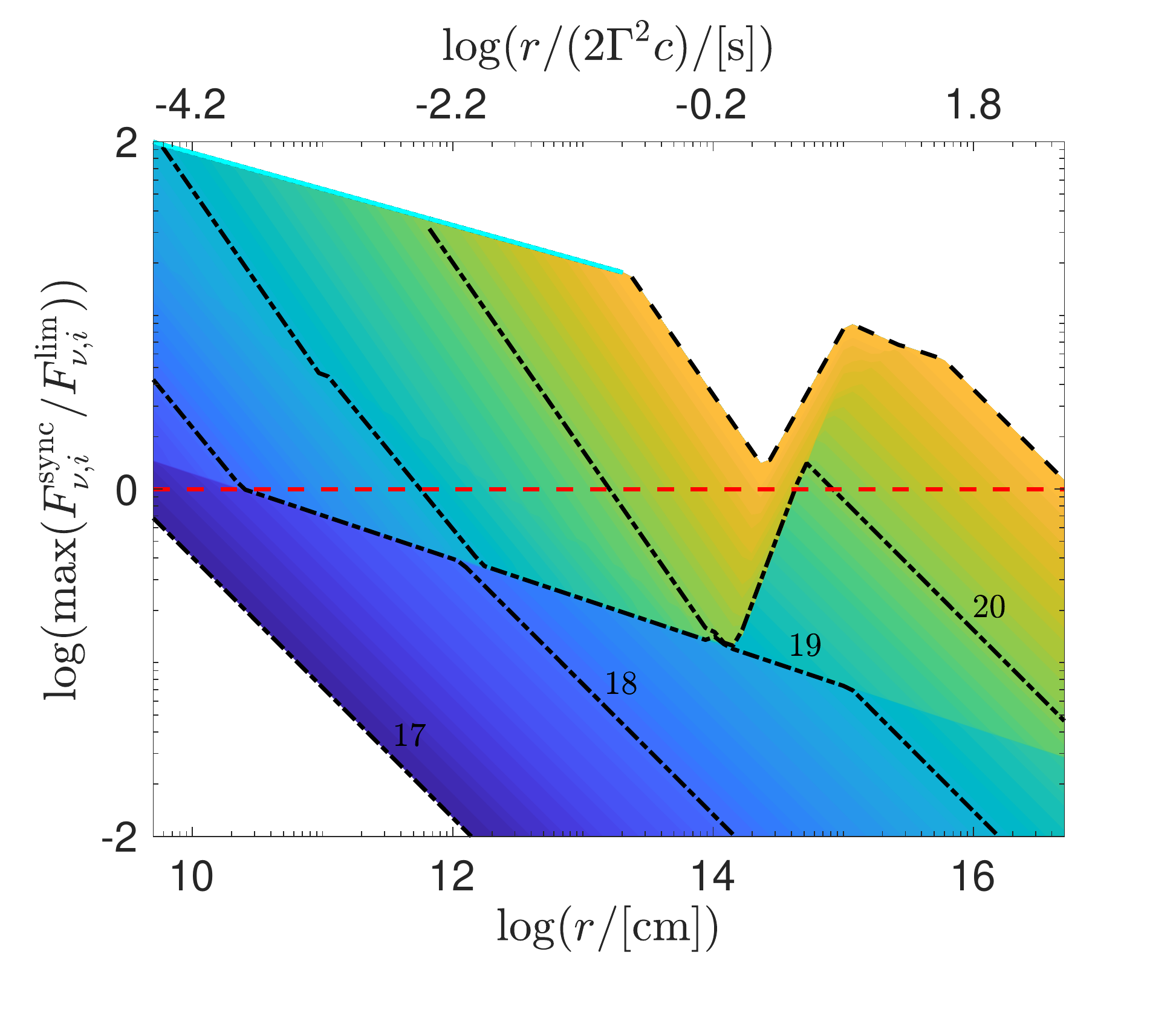}
    \includegraphics[width=0.32\columnwidth]{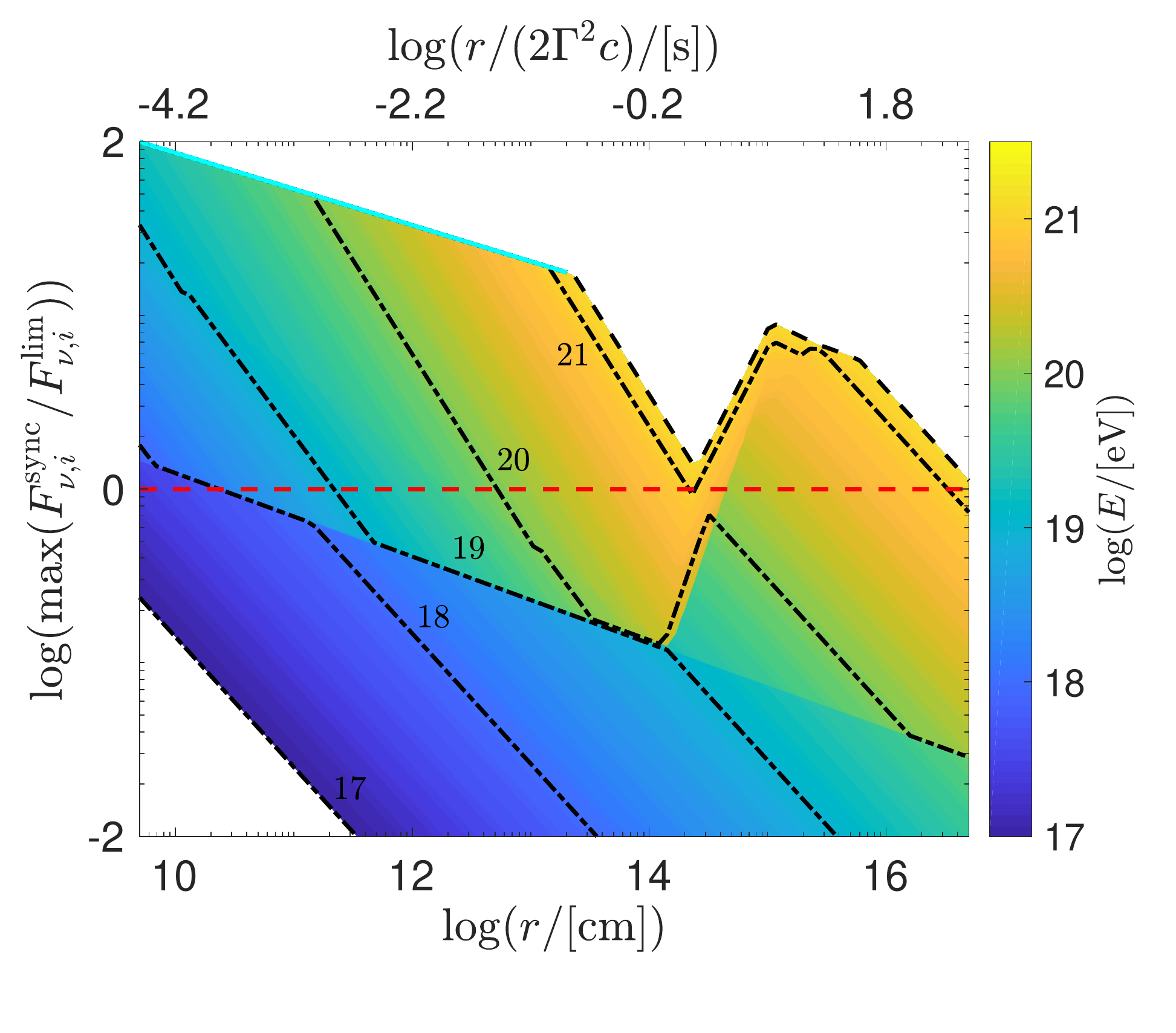}\\
    \includegraphics[width=0.32\columnwidth]{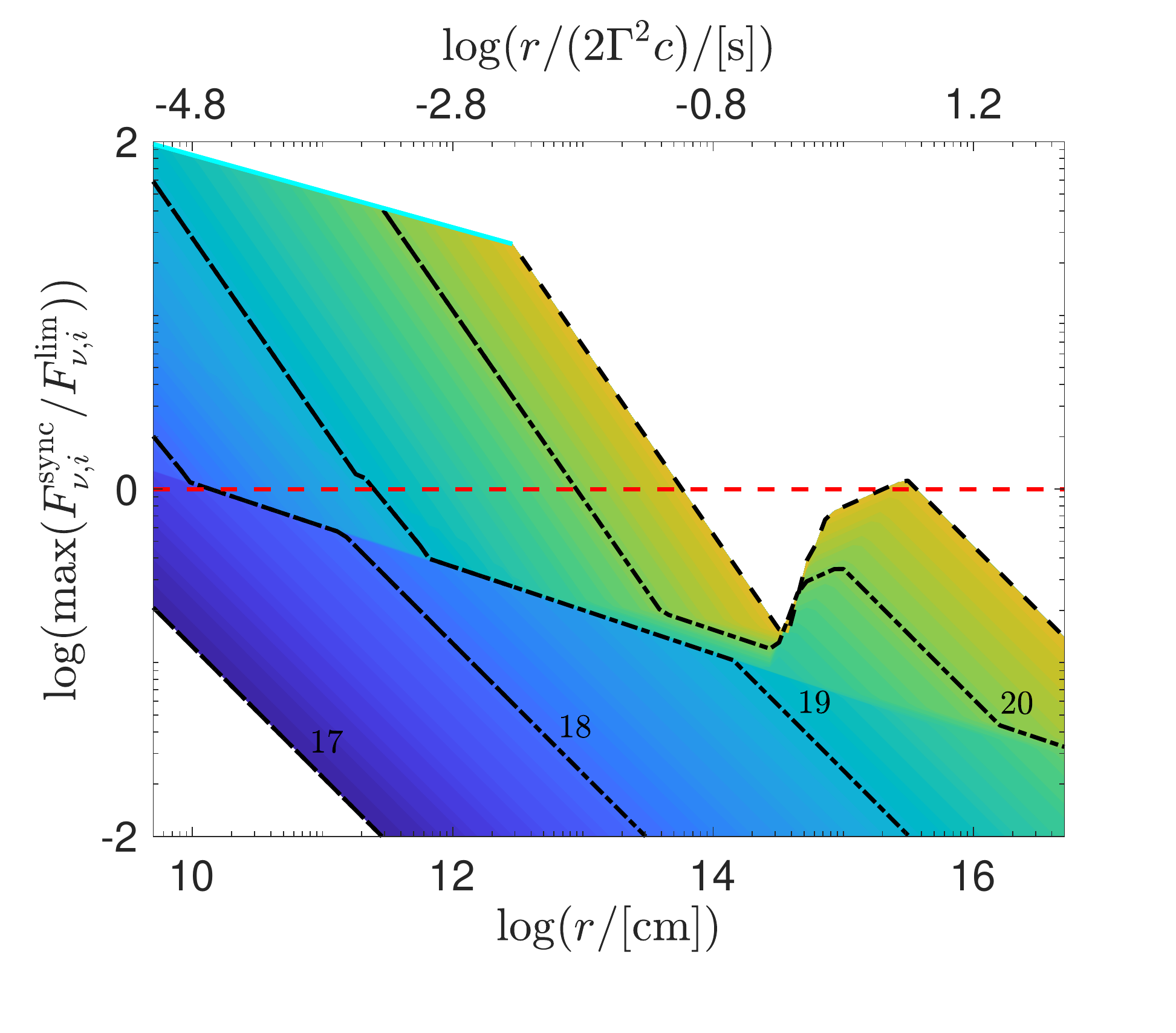}
    \includegraphics[width=0.32\columnwidth]{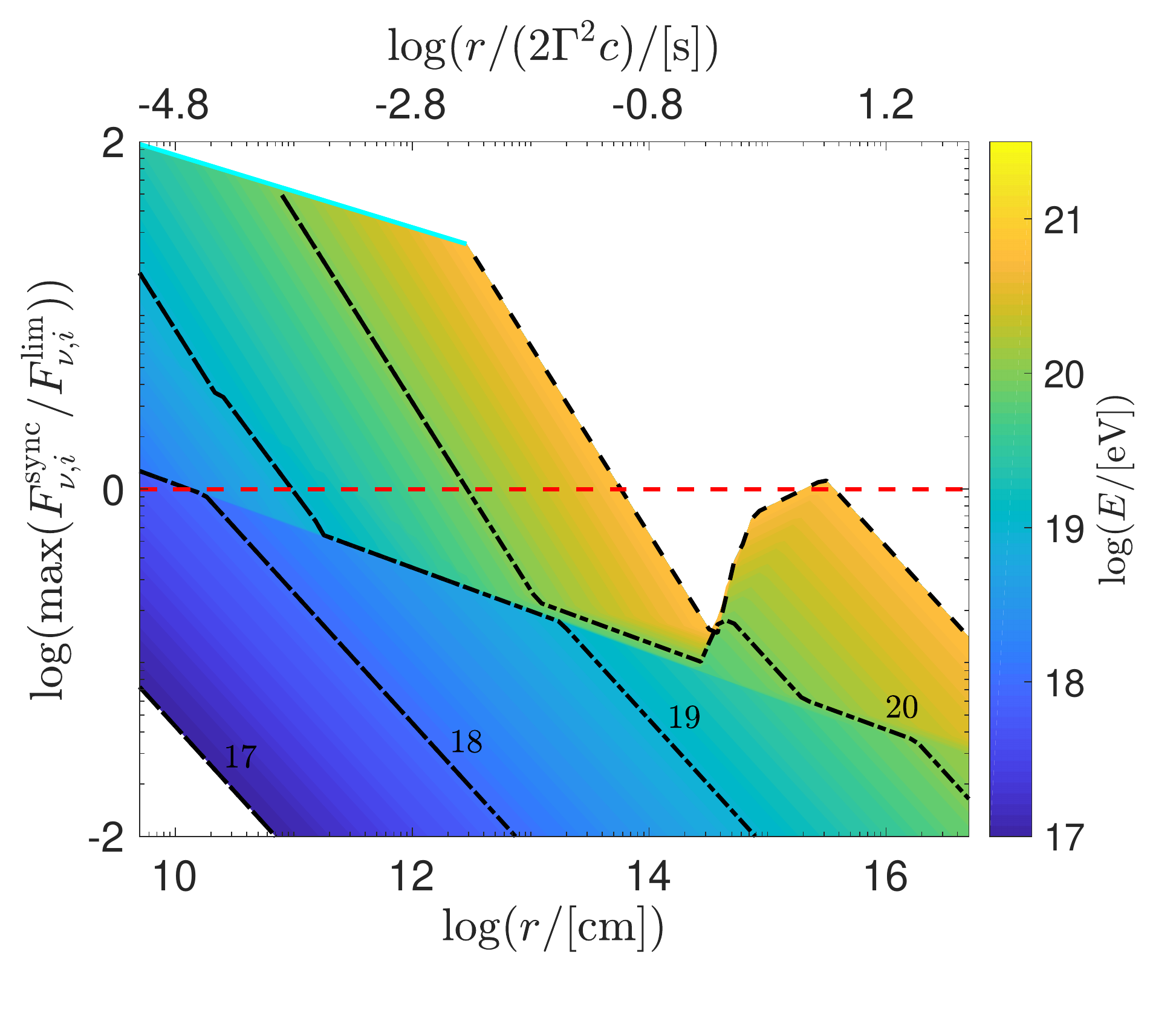}\\
    \includegraphics[width=0.32\columnwidth]{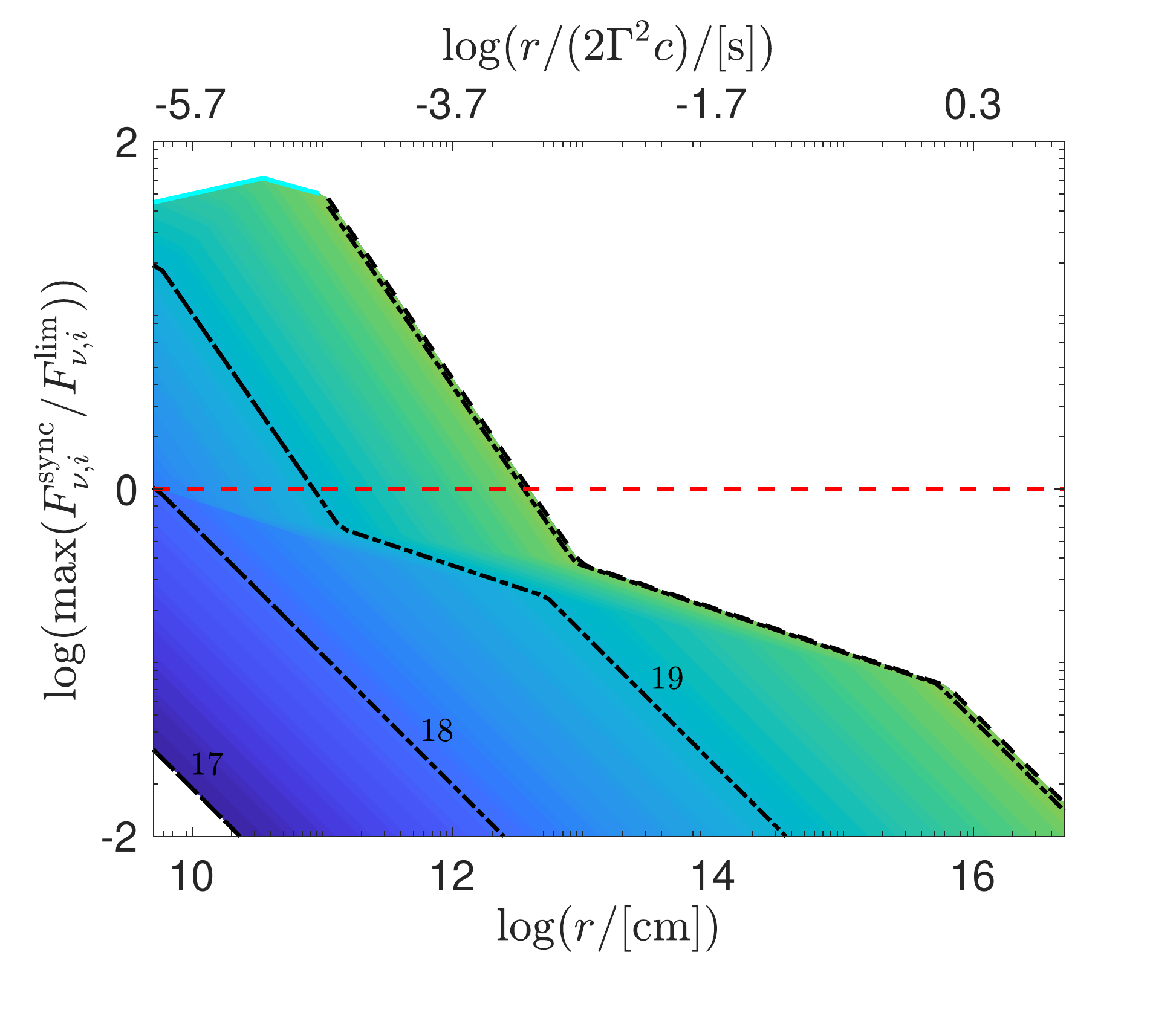}
    \includegraphics[width=0.32\columnwidth]{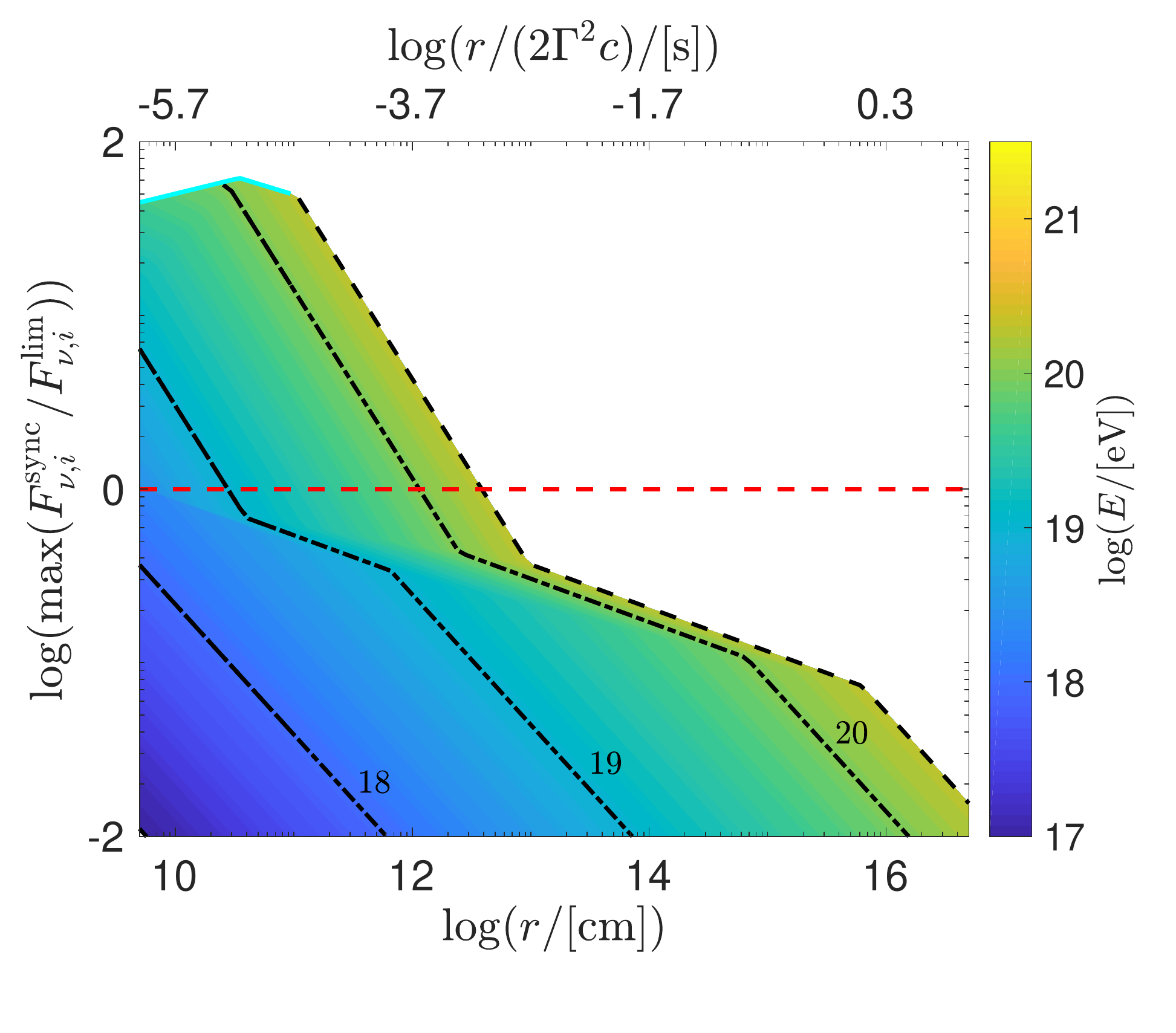}
\caption{Equivalent to right most column in Figure \ref{Fig:FluxLimitsIronLL}, but for acceleration efficiency $\eta = 0.5$ (left) and $\eta = 1$ (right). Plotted for $\Gamma = 10$, 50, 100, and 300 from top to bottom. Other numerical values used are given in Table \ref{tab:PhotosphereNumerics}.}
\label{Fig:VaryingEta}
\end{centering}
\end{figure*}

The dependencies in the photospheric model comes from the parameter dependence in the predicted synchrotron fluxes in optical and around the sub-MeV peak, and these are mostly dependent on the acceleration efficiency, the total luminosity, the fraction of electrons accelerated (see Equation \eqref{eq:N_e}), and the redshift of the burst. Increasing $\eta$ effectively decreases all lower limits shown in Figure \ref{Fig:FluxLimits} while the upper limits from the magnetic luminosity stays fixed, resulting in a larger allowed parameter space. Fixing $L_{\rm tot} = 10^{52}$ erg s$^{-1}$ and $z = 1$ as in Figure \ref{Fig:FluxLimits}, we find that when $\eta \geq 0.5$ there is a region where protons can be accelerated to the highest energies, requiring $r \sim 10^{15}$ cm, $\Gamma = 300$, and $B' \sim 1000$ G. When $\eta = 1$, 
bursts with $\Gamma = 100$ can not support acceleration up to $10^{20}$ eV but bursts with either $\Gamma = 300$ or $\Gamma = 1000$ can. 
In llGRBs putting $\eta = 1$ is not enough to reach proton acceleration to $10^{20}$ eV; the predicted fluxes in this case are similar to those shown in Figure \ref{Fig:FluxLimitsIronLL} but slightly more constraining. When considering iron, llGRBs can reach the highest observed energies for larger $\eta$ in the photospheric model. In Figure \ref{Fig:VaryingEta}, we plot the constraints from the predicted fluxes for iron in a llGRB similar to the panels on the right-hand side in Figure \ref{Fig:FluxLimitsIronLL}, but for $\eta = 0.5$ (left) and $\eta = 1$ (right). Especially in bursts with $\Gamma = 50$ or 100, there is quite a large region where the highest energies could be reached, given $\eta = 1$. 

\citet{Boncioli2019} argued for llGRBs as the primary source of UHECR. Similar to our Section \ref{MagneticFieldSection}, they investigate where acceleration would be possible by comparing the relevant time scales. They also calculate the neutrino production and consider the CR escape and propagation, and match the predicted neutrino and CR signal to observed spectra. Their analysis is based on intermediate mass nuclei (O and Si), which they accelerate very efficiently ($\eta = 1$) in a burst with outflow velocity $\Gamma \sim 10$. In the synchrotron model, the electrons at the acceleration region would need to be extremely fast cooling for such a GRB. In the photospheric model, one can look at Figure \ref{Fig:VaryingEta} to get an estimate of the fluxes in this case. For a burst to accelerate intermediate-mass particles to the highest energies given these conditions, the constraints from the fluxes would be most modest if the emission occurred at $r \gtrsim 10^{16}$ cm. Even so, the predicted optical spectral flux from such a burst would be $> 3$ Jy, so if llGRBs with a slow outflow are a real source of UHECR, then high prompt optical spectral fluxes are a necessary consequence.

\begin{figure*}
\begin{centering}
    \includegraphics[width=0.29\columnwidth]{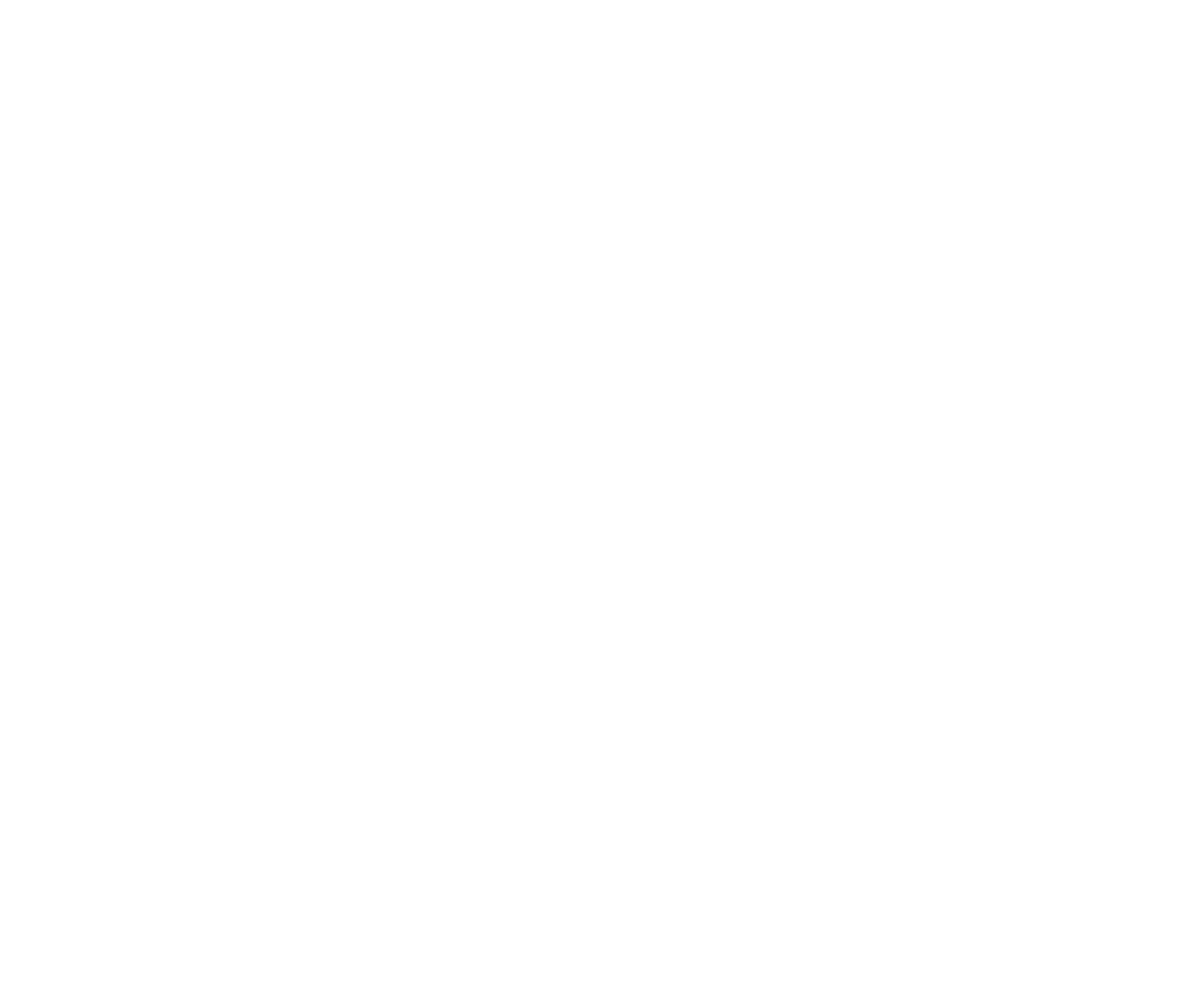}
    \includegraphics[width=0.29\columnwidth]{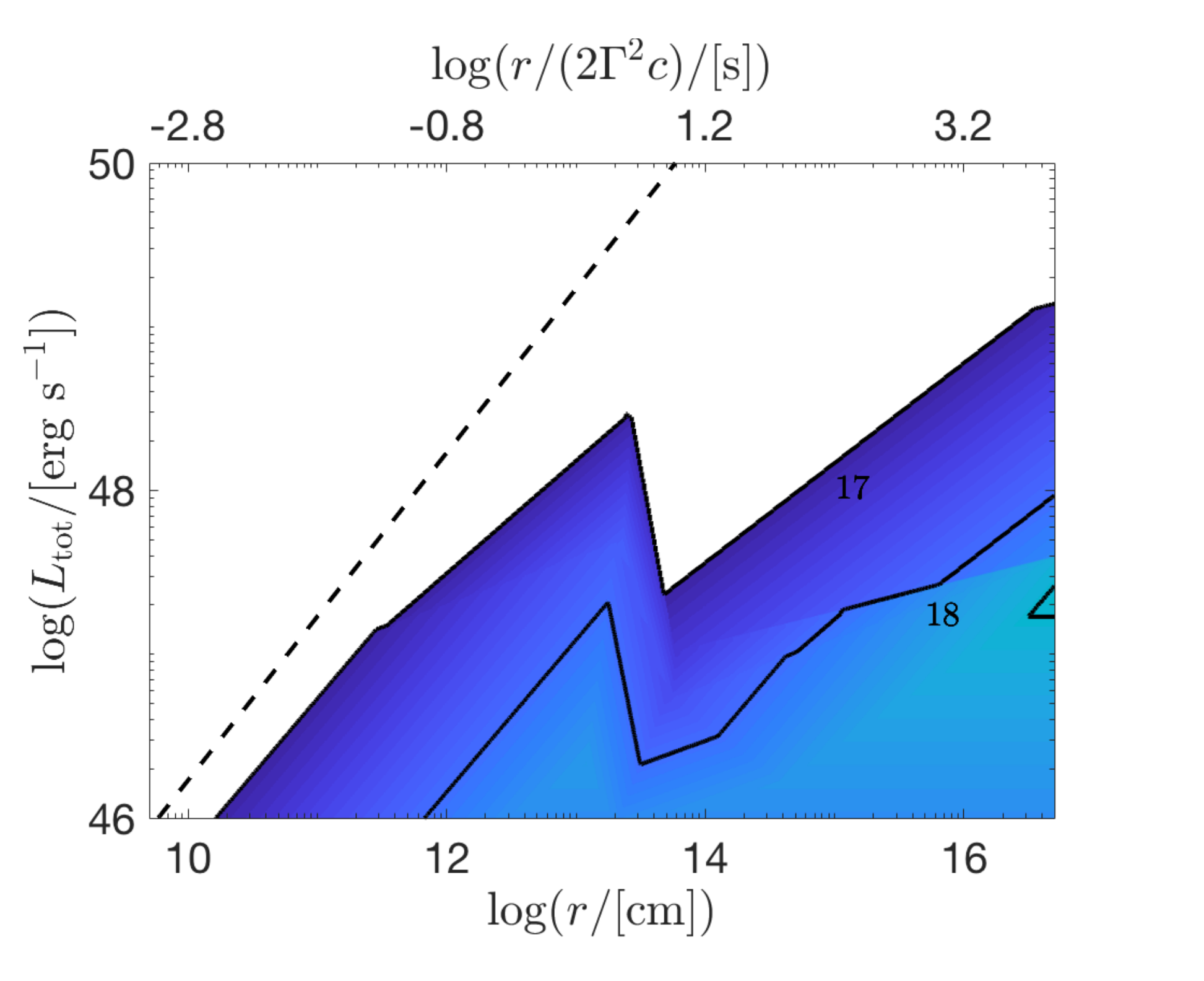}
    \includegraphics[width=0.29\columnwidth]{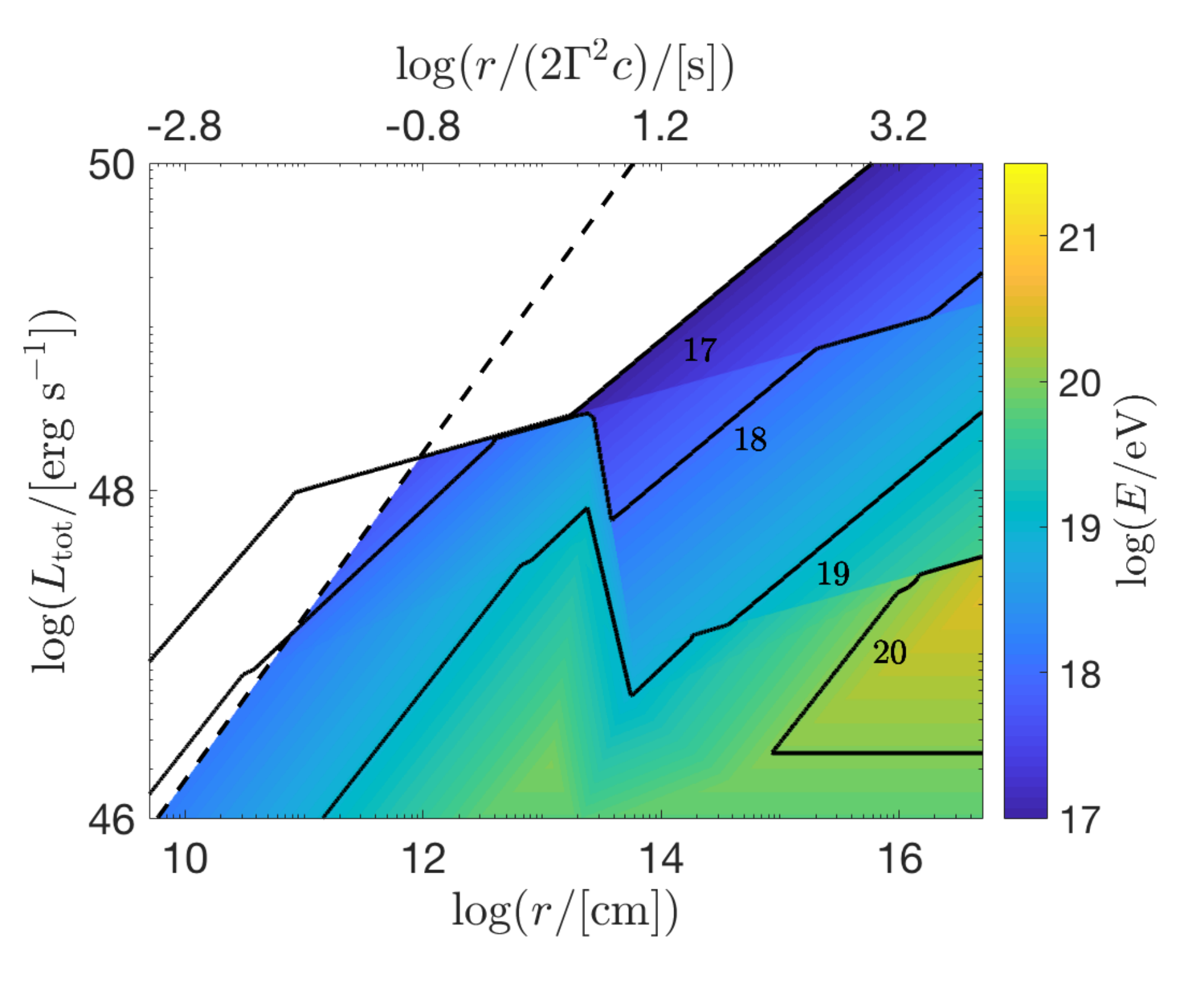}
    \includegraphics[width=0.29\columnwidth]{blank}
    \includegraphics[width=0.29\columnwidth]{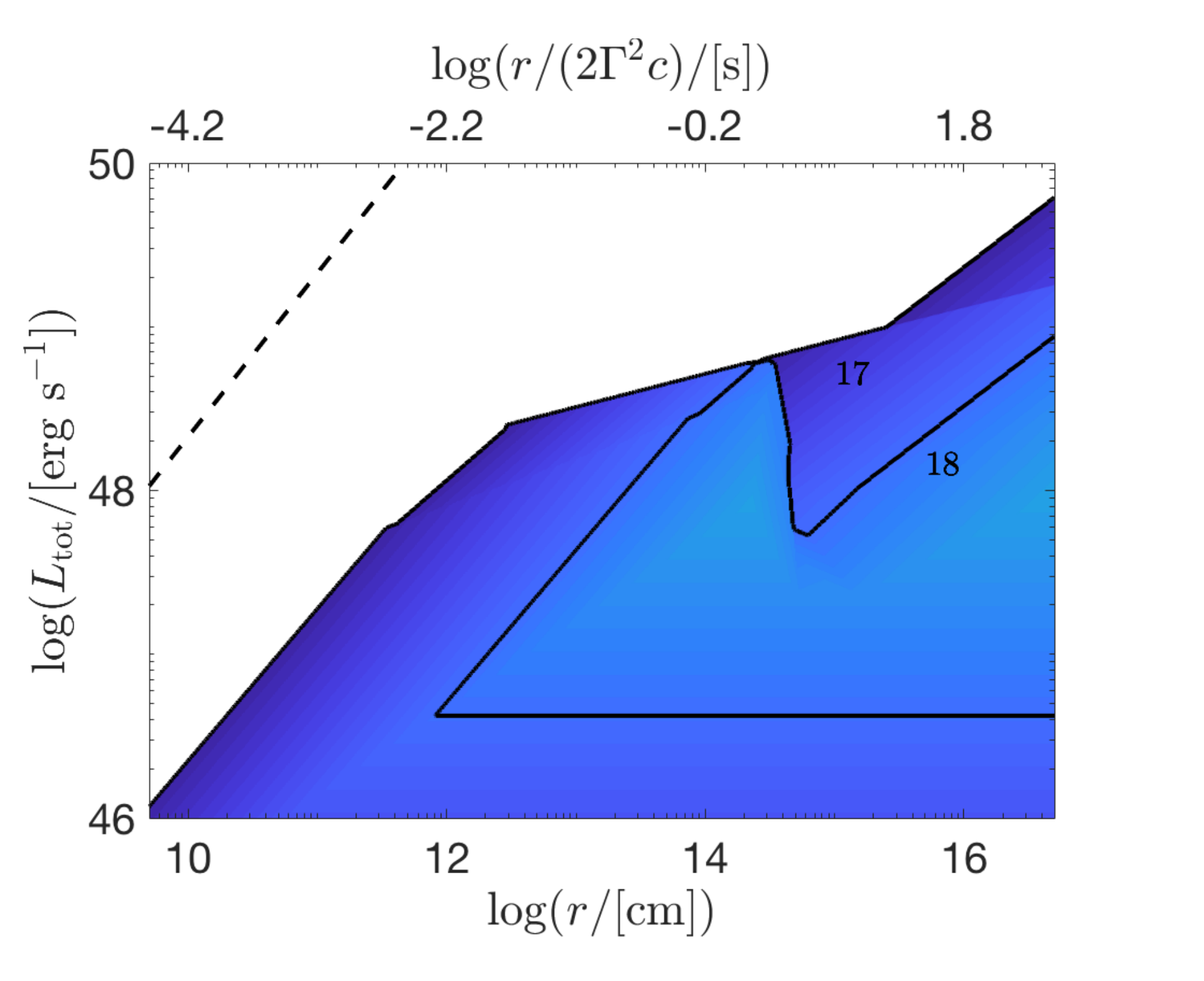}
    \includegraphics[width=0.29\columnwidth]{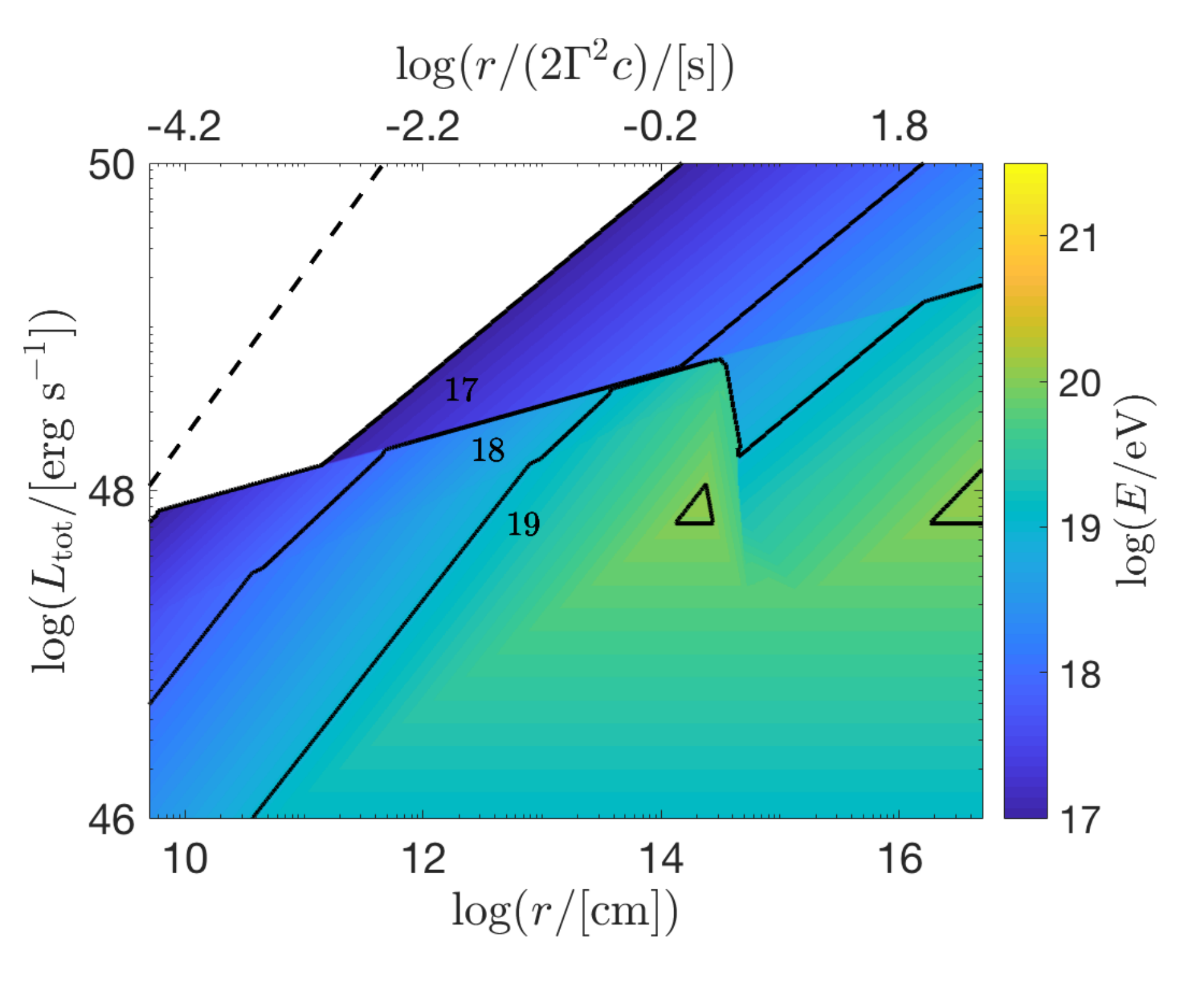}
    \includegraphics[width=0.29\columnwidth]{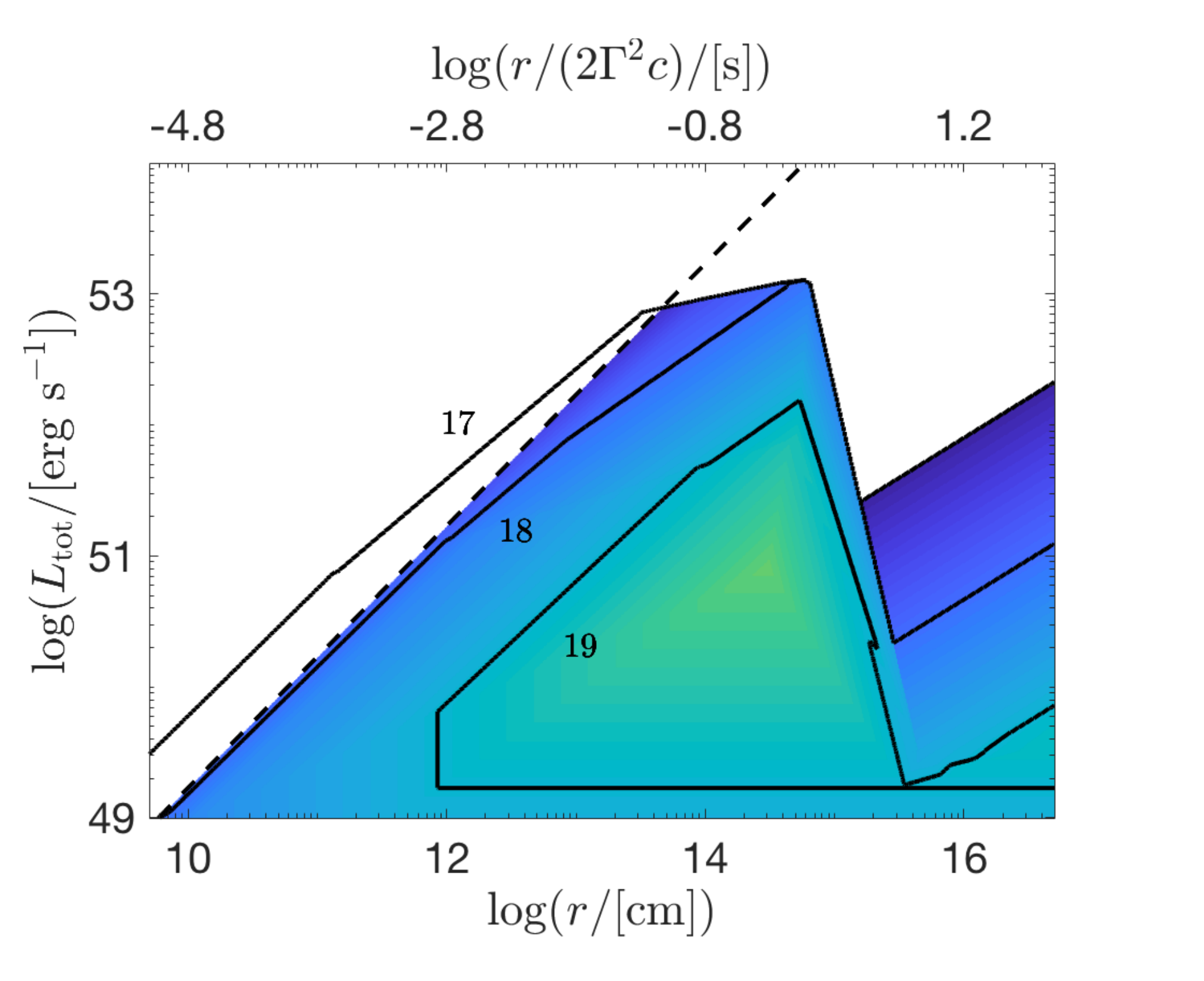}
    \includegraphics[width=0.29\columnwidth]{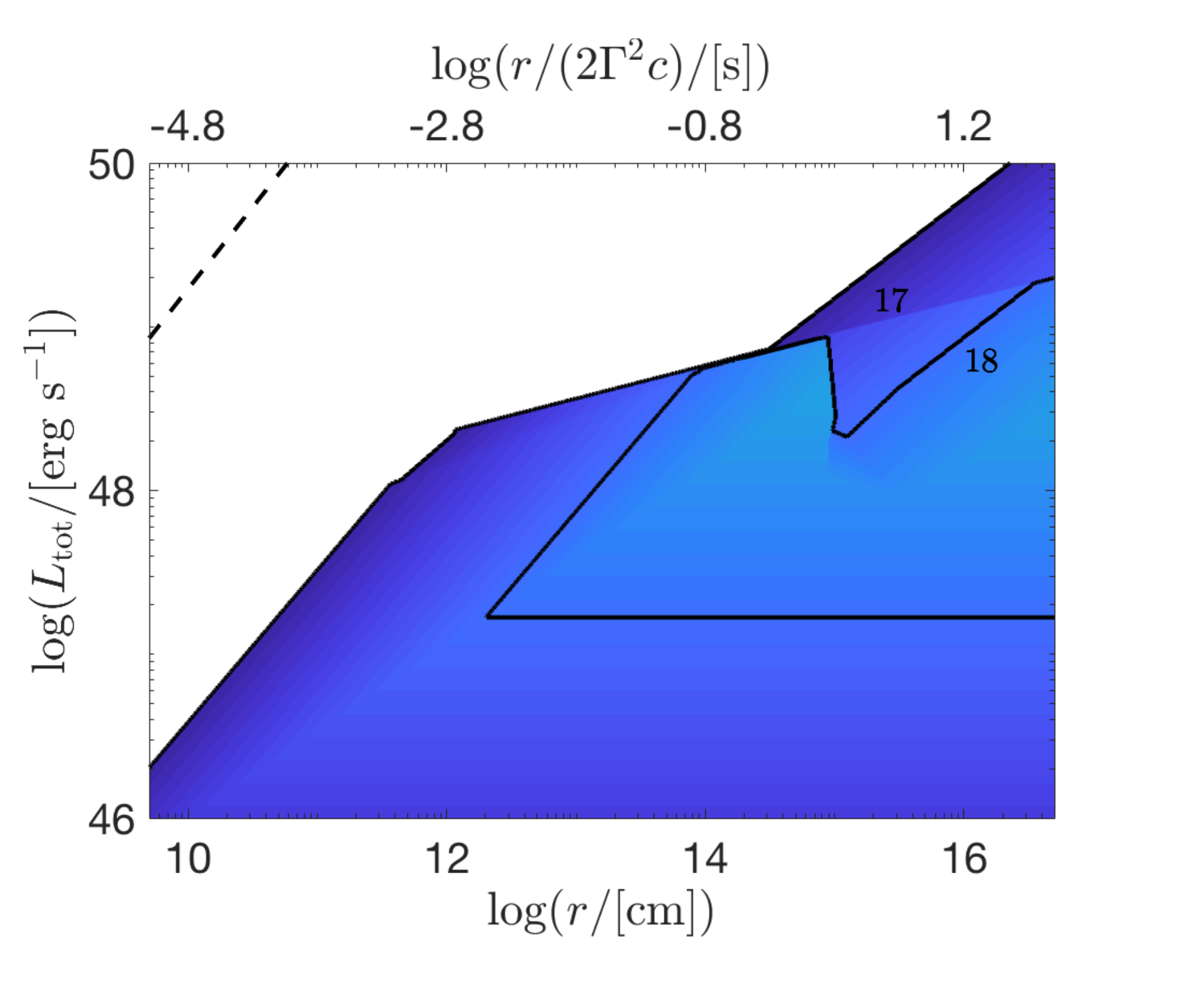}
    \includegraphics[width=0.29\columnwidth]{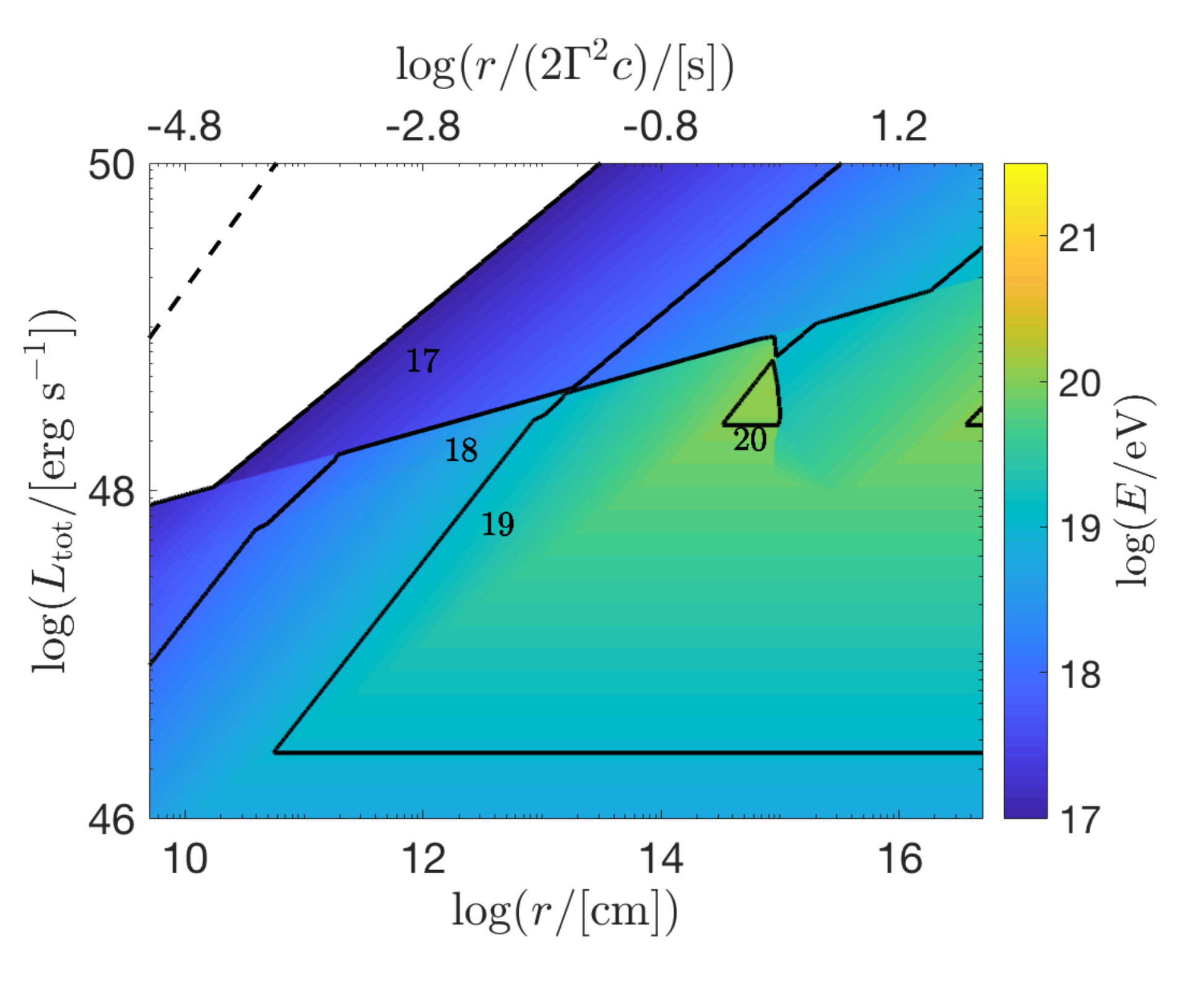}
    \includegraphics[width=0.29\columnwidth]{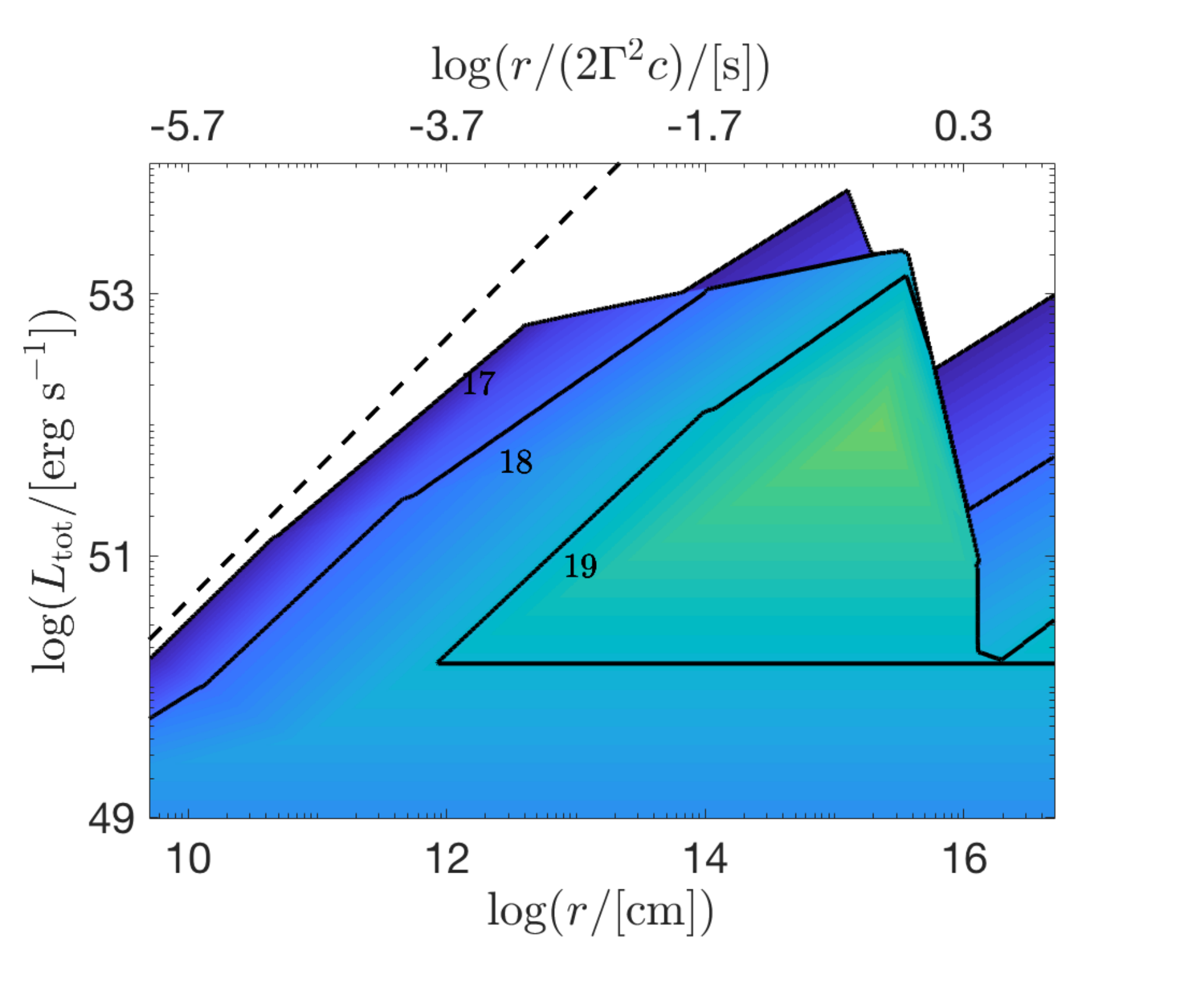}
    \includegraphics[width=0.29\columnwidth]{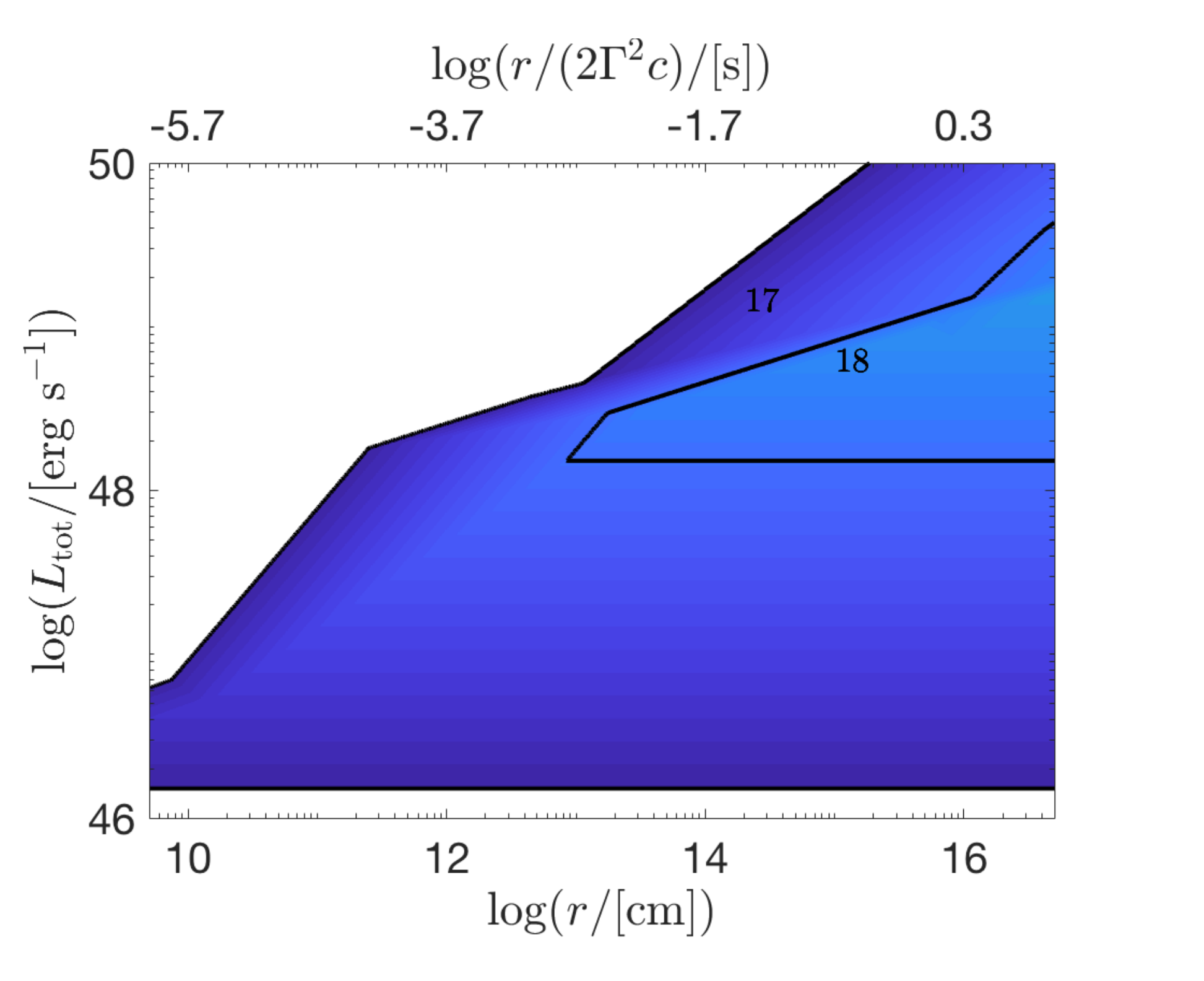}
    \includegraphics[width=0.29\columnwidth]{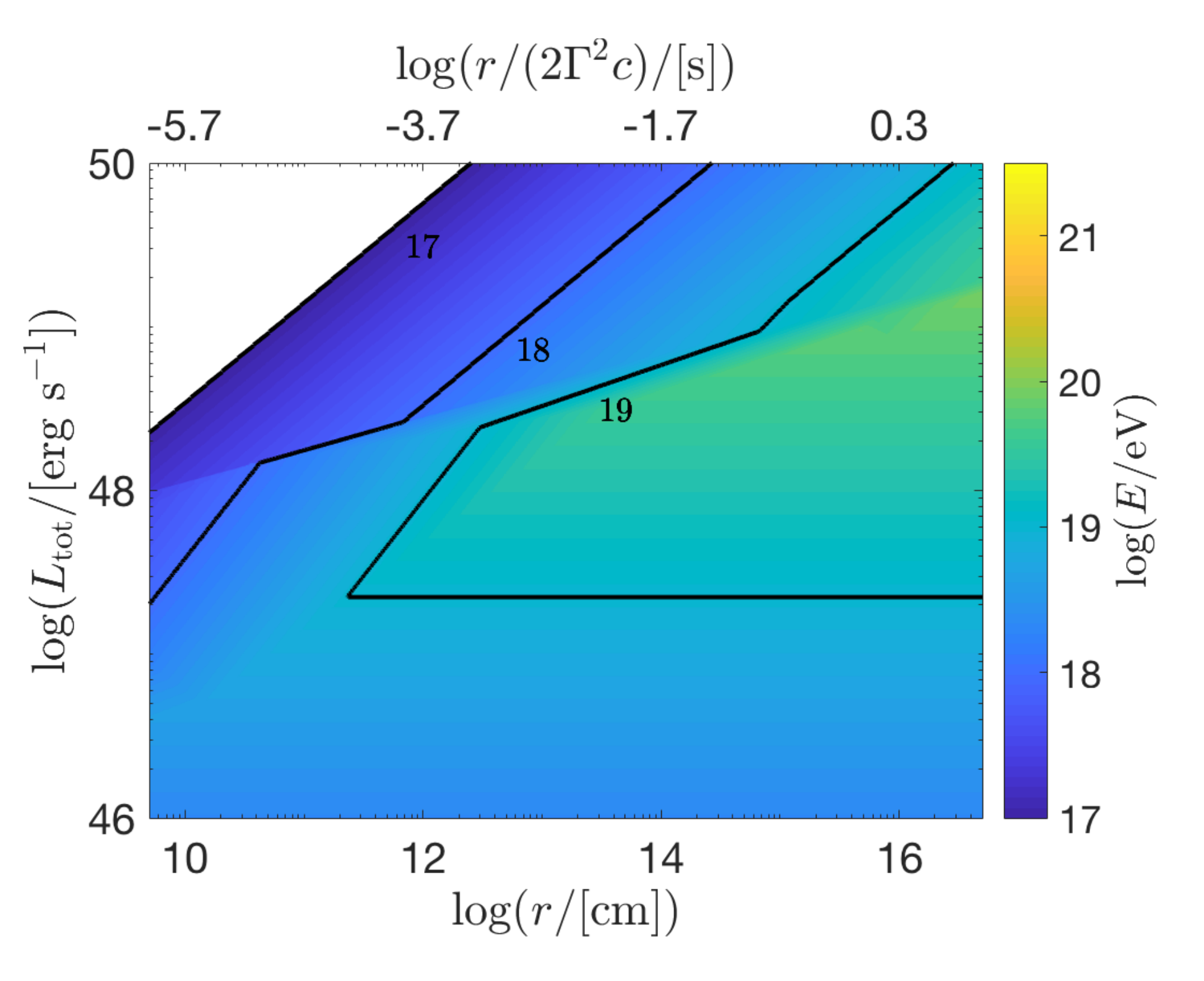}
    \includegraphics[width=0.29\columnwidth]{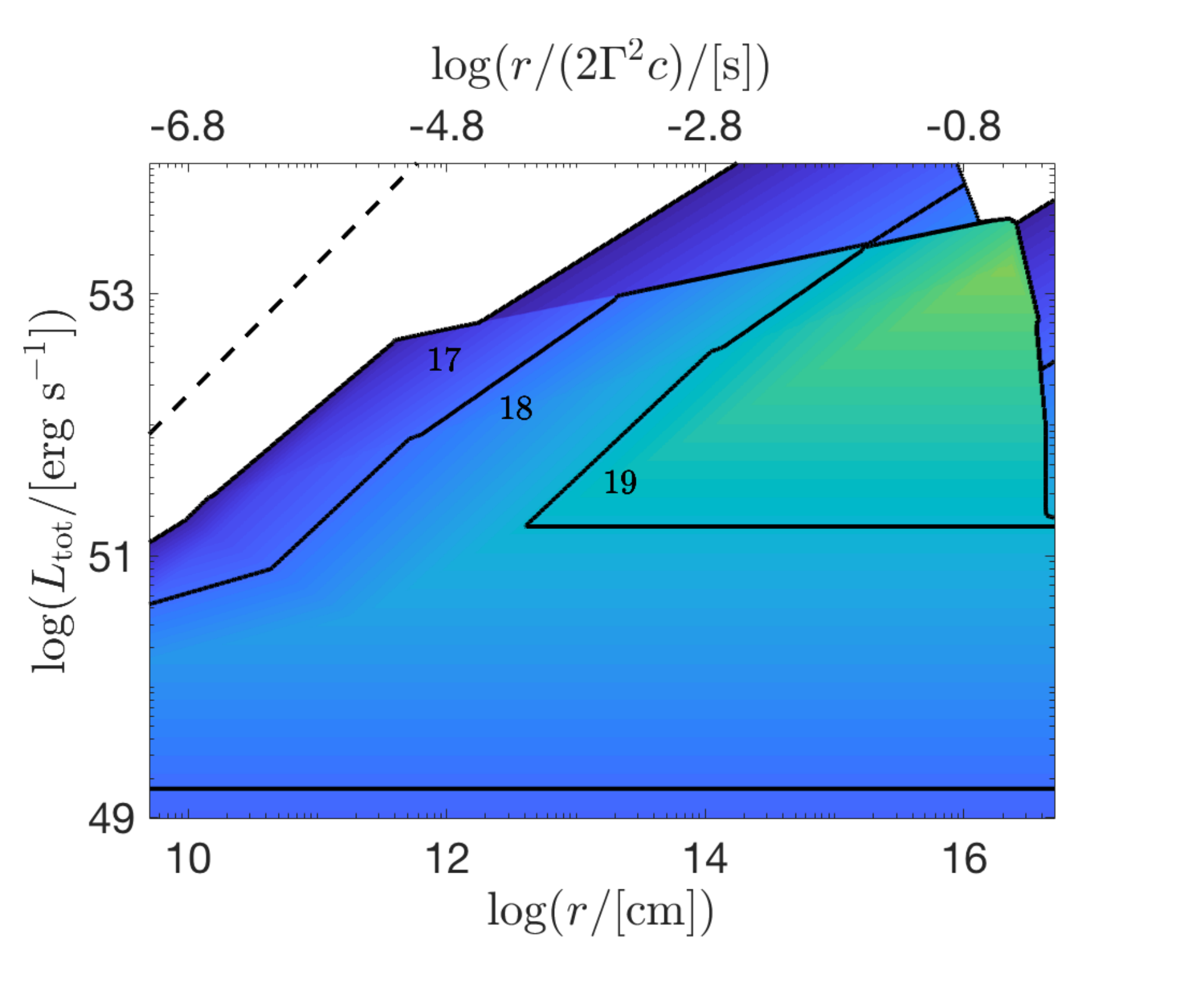}
    \includegraphics[width=0.29\columnwidth]{blank}
    \includegraphics[width=0.29\columnwidth]{blank}
\caption{Maximum attainable proton energy in a canonical GRB (left), in a llGRB (middle), and iron energy in a llGRB (right) as a function of $L_{\rm tot}$ and $r$. The Figures show $\Gamma = 10$, 50, 100, 300, and 1000 from top to bottom. The slanted dashed line indicates the photosphere. Other numerical values used are given in Table \ref{tab:PhotosphereNumerics}.}
\label{Fig:LuminosityVsRadius}
\end{centering}
\end{figure*}

The dependency of the result on the total luminosity of the burst is not trivial. Increasing $L_{\rm tot}$ makes the constraints on the fluxes in optical and around the observed peak harsher, but it makes the constraint on the magnetic luminosity less severe. We illustrate how the maximum UHECR energy varies with total luminosity in Figure \ref{Fig:LuminosityVsRadius}. 
Protons in a canonical GRB is shown on the left, protons in a llGRB in the middle, and iron in a llGRB on the right for $\Gamma = 10$, 50, 100, 300, and 1000 from top to bottom. The figure shows what maximum UHECR energy could be achieved for each ($L_{\rm tot}, r$)-pair for a given $\Gamma$. In other words, for each value in the range of luminosities, we check what maximum permitted (below the red dashed line) proton or iron energy can be achieved at each radius.\footnote{To generate Figure \ref{Fig:LuminosityVsRadius}, $L_\gamma$ is set to 10\% of the total luminosity.} 
From the Figure, we see that proton acceleration to the highest energies is not possible in canonical or llGRBs. This result is robust even if the redshift is changed in both cases. It is also robust to changes in the acceleration efficiency in the low-luminosity case. Furthermore, iron acceleration to $10^{20}$ eV and above is not possible in llGRBs except for a parameter space centered around $L_{\rm tot}= 10^{47}$ erg s$^{-1}$ and $r=10^{16}$ cm for $\Gamma = 10$, and some tiny regions for $\Gamma = 50$ and 100. However, this also requires ideal conditions within the burst in terms of magnetic field.

Varying the fraction of accelerated electrons $\xi_a$ has two effects. Firstly, the synchrotron spectral flux is directly proportional to the number of radiators, which leads to a larger permitted parameter space with decreasing $\xi_a$. Secondly, $\gamma'_{\rm m}$ is inversely proportional to $\xi_a$; if fewer electrons are accelerated they each get a larger portion of the available energy. An increase in $\gamma'_{\rm m}$ results in harsher constraints from the flux limits around the sub-MeV peak, shrinking the allowed parameter space. However, $\gamma'_{\rm m}$ is also proportional to $\epsilon_e$, so the observational constraints might be met by allowing $\epsilon_e$ to decrease. For relativistic outflows, $\epsilon_e \sim 0.1$ is well determined \citep{Wijers1999, Panaitescu2000, Santana2014, Beniamini2017}. \added{This value of $\epsilon_e$ or larger is also the most commonly used in studies of llGRBs, see e.g. \citep{Murase2006, Soderberg2006Nature, He2009, Liu2011, Senno2016, Xiao2017, ZhangMurase2018}. However, due to their slower outflows it is possible that the shocks in llGRBs are only mildly relativistic. \citet{Crumley2019} recently found using 2D PIC simulations that in this case, $\epsilon_e$ is much smaller $\sim 5\times 10^{-4}$, in agreement with 1D PIC simulation of non-relativistic shock made by \citet{Park2015}. If the prompt emission was dominated by these electrons, the total luminosity would be too high ($L_\textrm{tot} \sim L_\gamma / \epsilon_e$), which is problematic in terms of photodisintegration at the jet base, radiation efficiency, and observed flux. However, in our photospheric model the prompt emission is emanating from another part of the jet and such a small value of $\epsilon_e$ cannot be ruled out. }

\begin{figure*}
\begin{centering}
    \includegraphics[width=0.32\columnwidth]{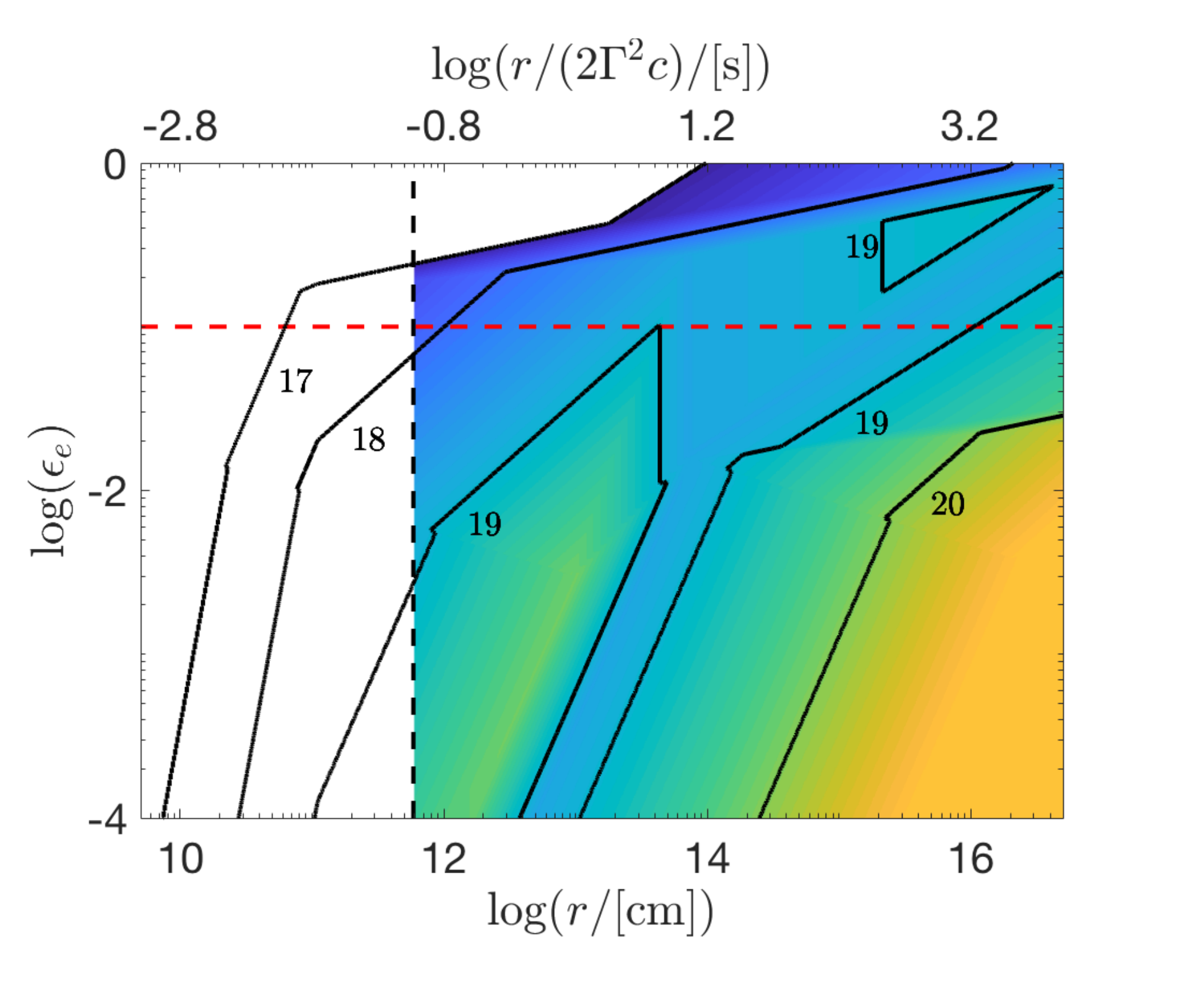}
    \includegraphics[width=0.32\columnwidth]{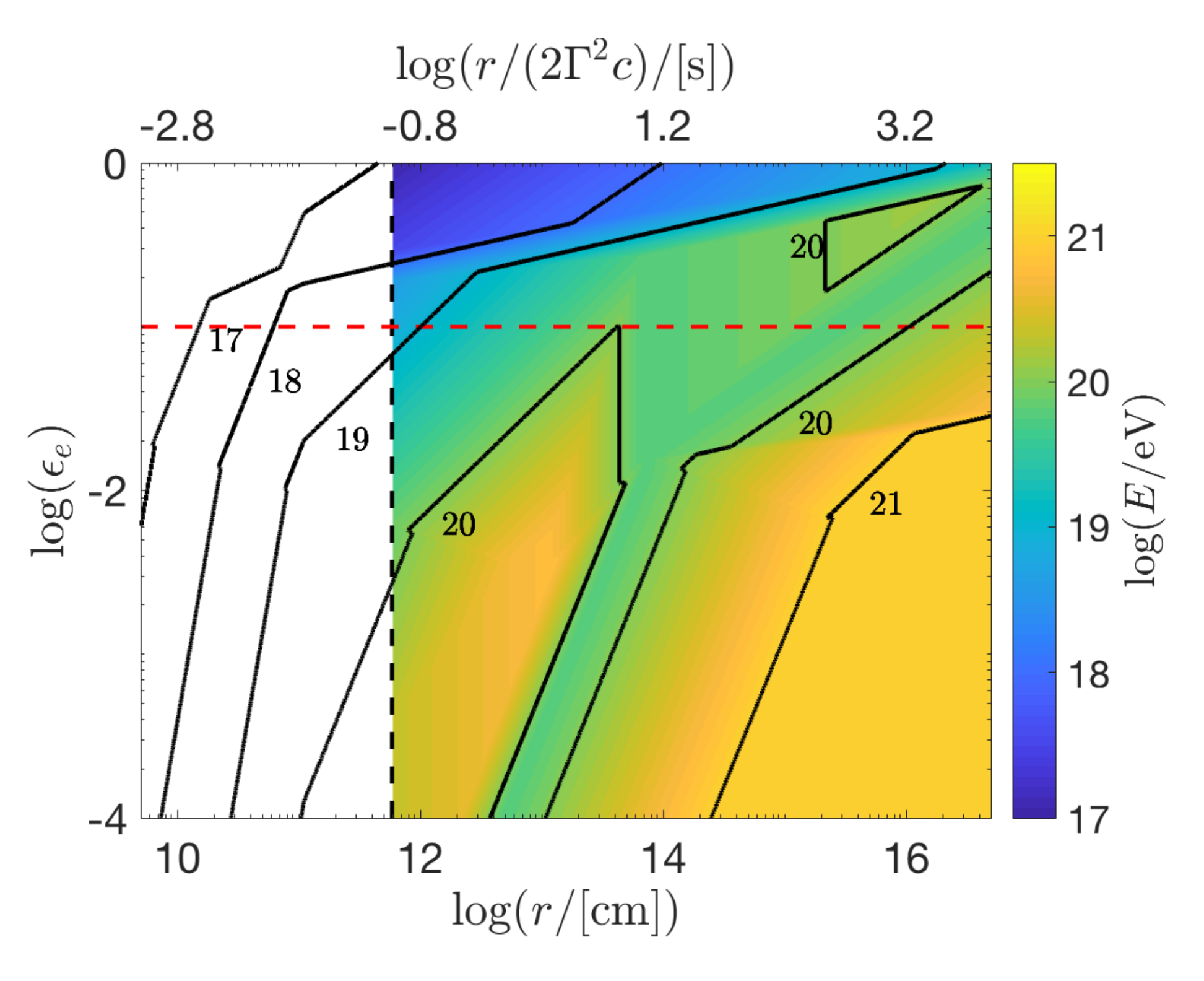}\\
    \includegraphics[width=0.32\columnwidth]{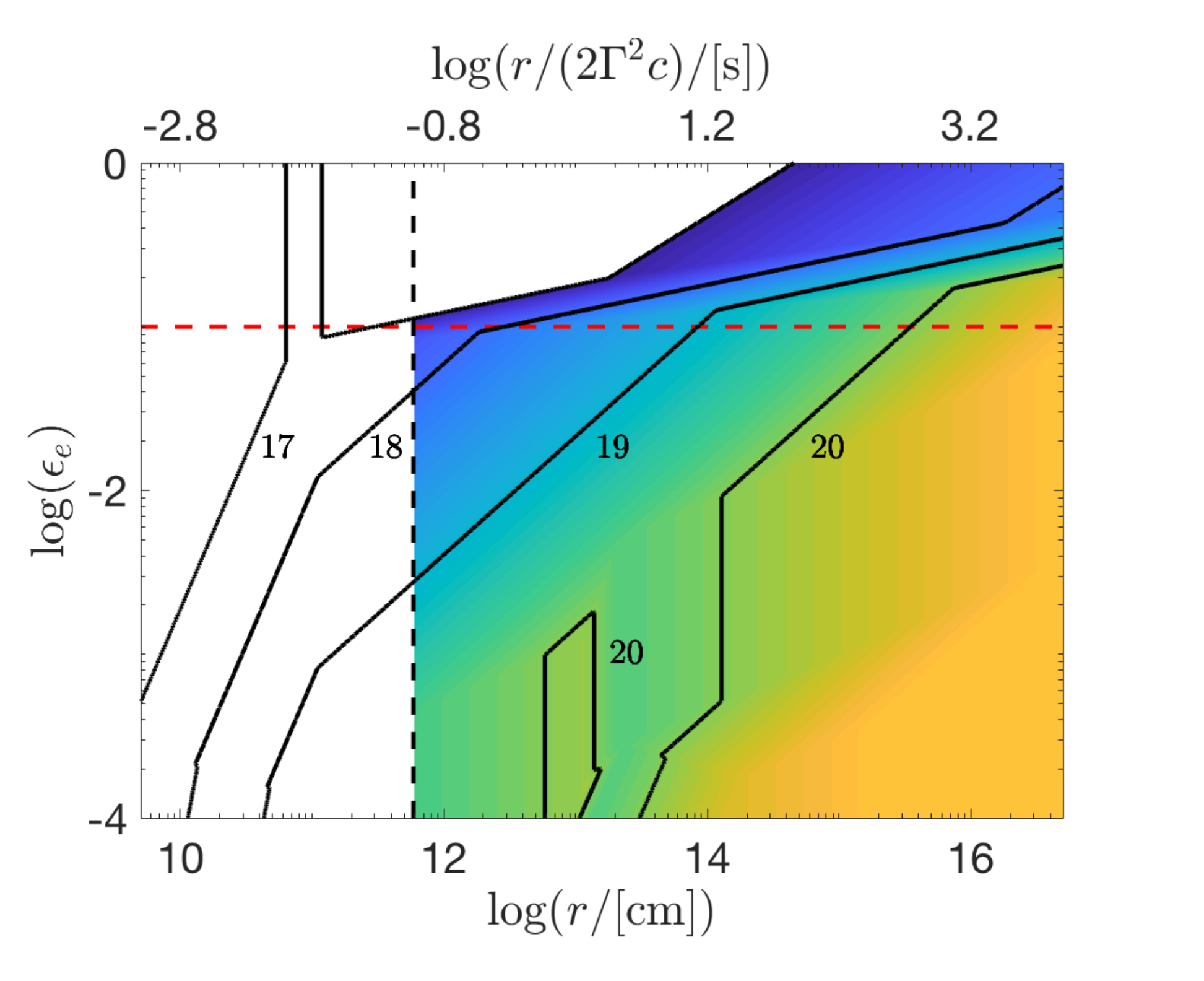}
    \includegraphics[width=0.32\columnwidth]{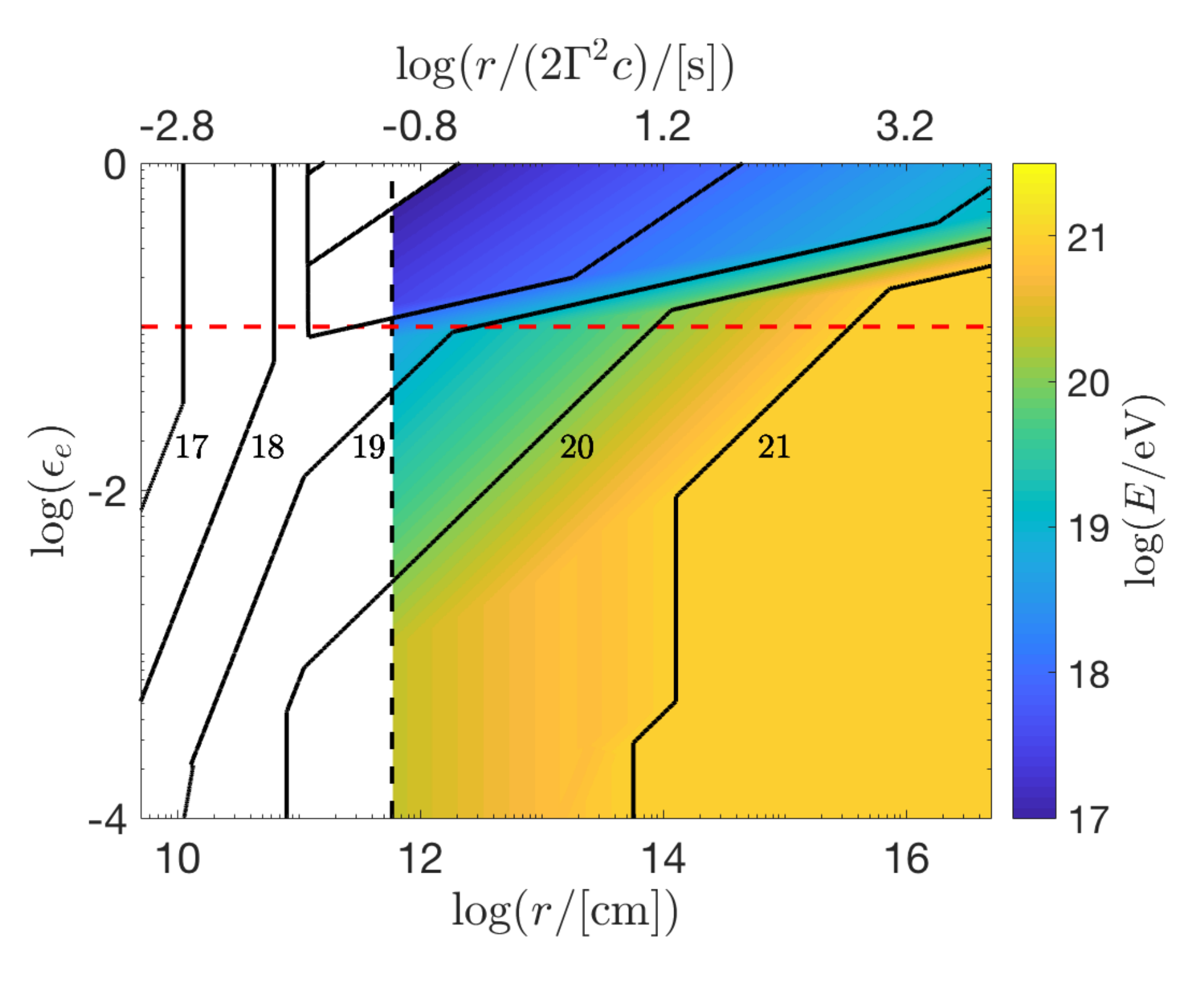}
\caption{Maximum iron energy in a llGRB with $\Gamma = 10$ as a function of $\epsilon_e$ and $r$. The plots on the left (right) have $\eta = 0.1$ ($\eta = 1$) and the plots on the top (bottom) have $\xi_a = 0.1$ ($\xi_a = 0.01$). The red dashed line shows $\epsilon_e = 0.1$ appropriate for relativistic outflows. Other numerical values used are given in Table \ref{tab:PhotosphereNumerics}.}
\label{Fig:epsilon_e_G10}
\end{centering}
\end{figure*}

\begin{figure*}
\begin{centering}
    \includegraphics[width=0.32\columnwidth]{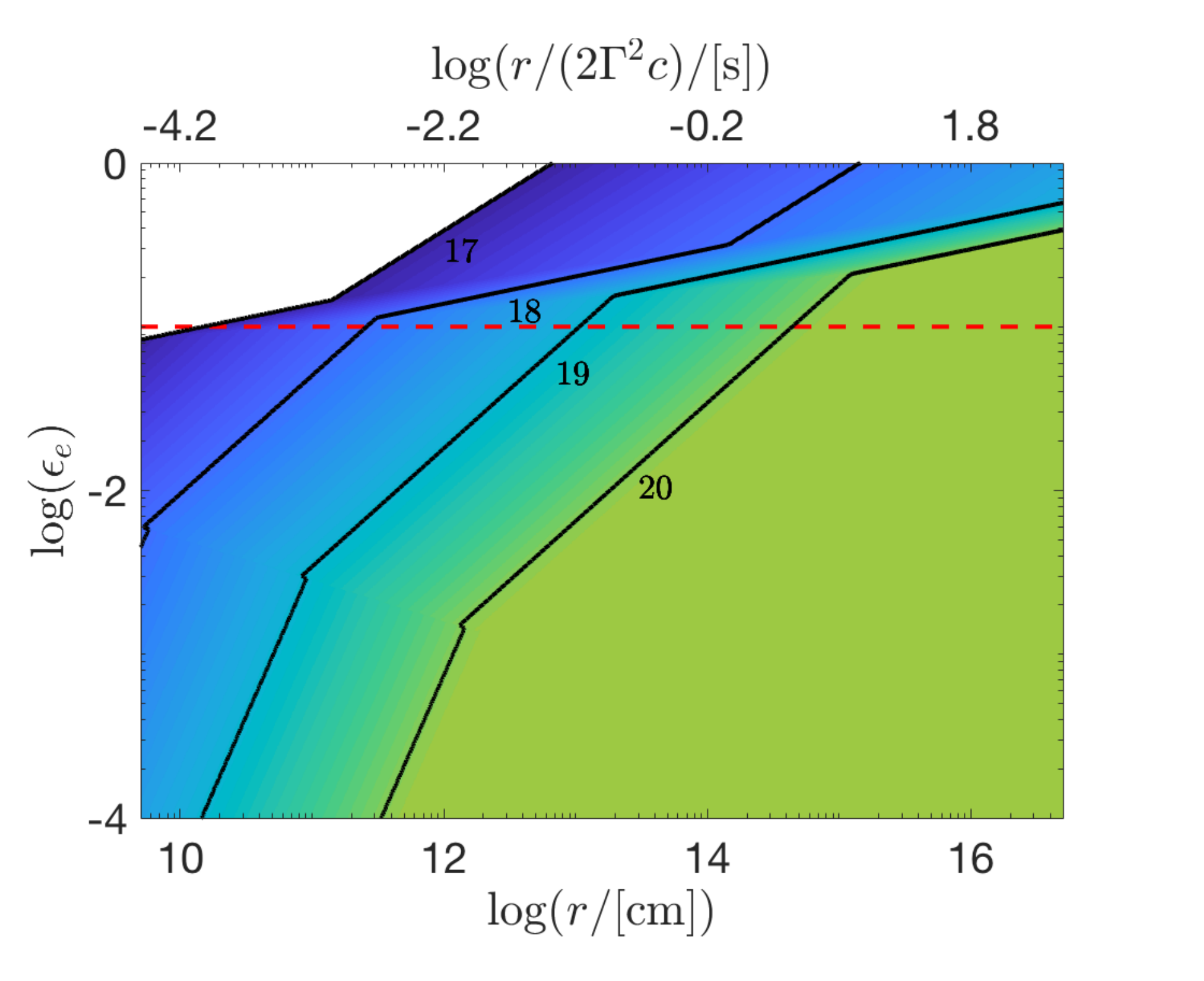}
    \includegraphics[width=0.32\columnwidth]{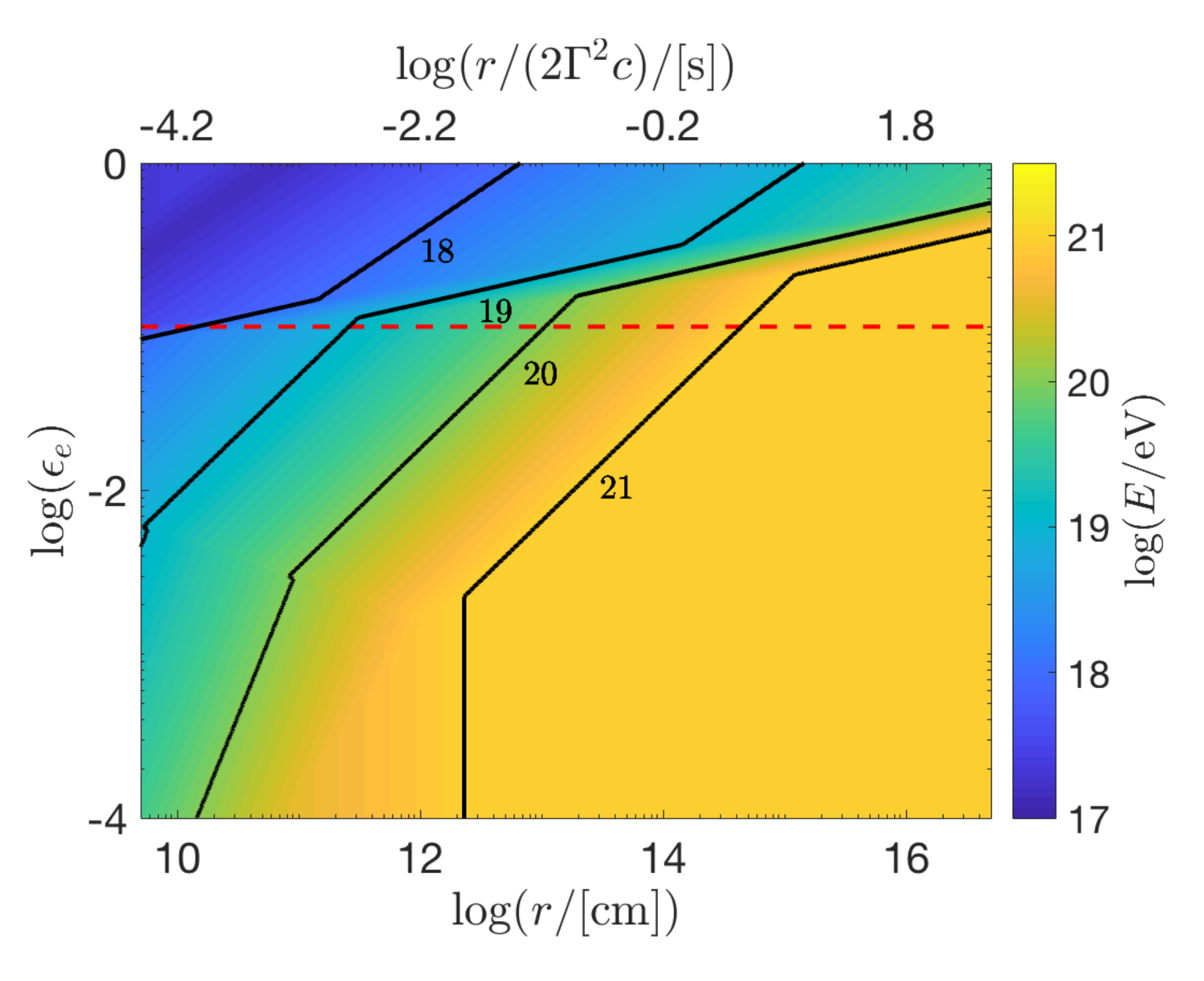}\\
    \includegraphics[width=0.32\columnwidth]{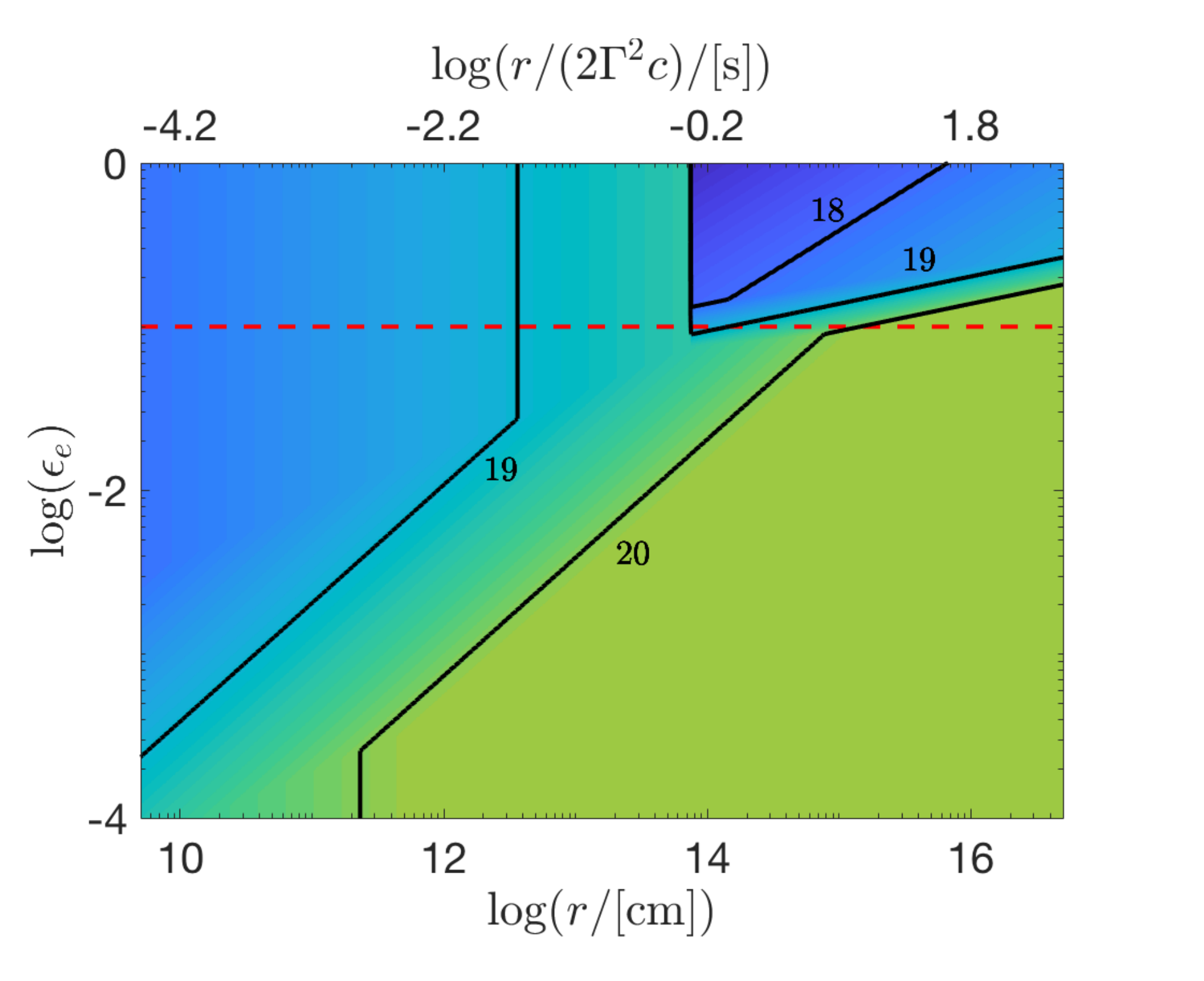}
    \includegraphics[width=0.32\columnwidth]{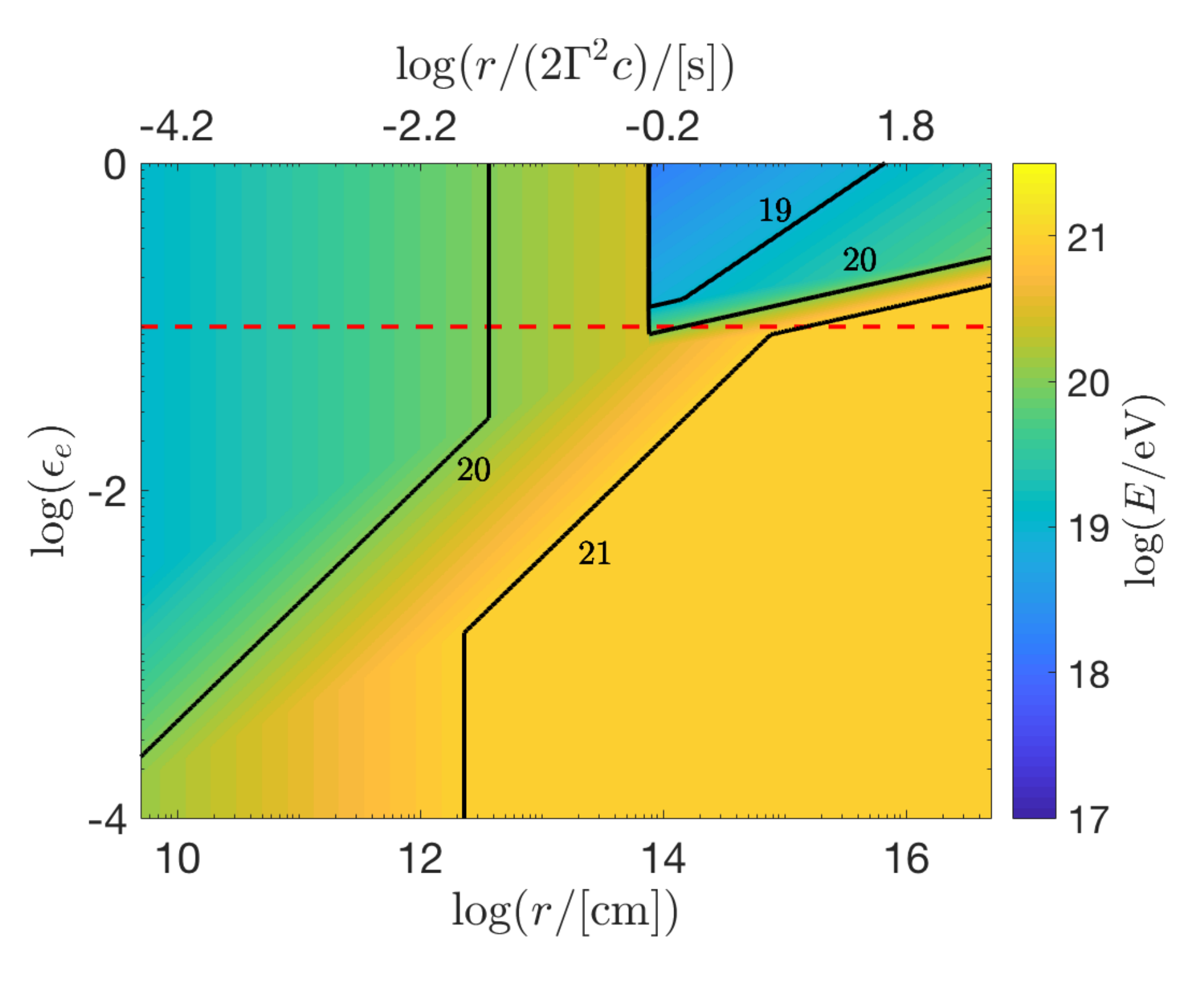}
\caption{Maximum iron energy in a llGRB with $\Gamma = 50$ as a function of $\epsilon_e$ and $r$. The plots on the left (right) have $\eta = 0.1$ ($\eta = 1$) and the plots on the top (bottom) have $\xi_a = 0.1$ ($\xi_a = 0.01$). The red dashed line shows $\epsilon_e = 0.1$ appropriate for relativistic outflows. Other numerical values used are given in Table \ref{tab:PhotosphereNumerics}.}
\label{Fig:epsilon_e_G50}
\end{centering}
\end{figure*}

Using an identical method to that used to generate the limits on the luminosity in Figure \ref{Fig:LuminosityVsRadius}, we obtain limits on $\epsilon_e$.
In Figures \ref{Fig:epsilon_e_G10} and \ref{Fig:epsilon_e_G50}, we show these limits as a function of radius and iron energy in a llGRB with $\Gamma = 10$ and 50 respectively. Both figures show the results for $\eta = 0.1$ and 1, and $\xi_a = 0.1$ and 0.01. We only show these plots for llGRBs for which the shocks could be non-relativistic. Furthermore, we only display iron as this is the least constraining case. The plots for protons look similar, but the maximum proton energy is \replaced{$\sim$ 300}{a factor $Z = 26$} times smaller than for iron for the same $\epsilon_e$ and $r$. 
The behavior in Figures \ref{Fig:epsilon_e_G10} and \ref{Fig:epsilon_e_G50} is quite complicated. 
The visible jumps are created when a case previously permitted, 
overshoots one of the flux limits leading to another solution being displayed, or when a less constraining solution 
becomes possible \added{(see Table \ref{tab:FluxCases})}. For some parts of the parameter space, no constraint can be put on $\epsilon_e$. This happens when either $F_{\nu, \rm{max}}^{\rm{sync}}$ is less then both flux limits, or when both the flux in optical and around the sub-MeV \added{peak} are independent of $\gamma'_{\rm m}$. It is apparent that when combining high acceleration efficiency and few accelerated electrons with a relatively small fractional energy, then the maximum iron energy is essentially unconstrained. This is because what is most constraining for outflows with $\Gamma \leq 50$ is the optical flux as evident from Figures \ref{Fig:FluxLimitsLL} and \ref{Fig:FluxLimitsIronLL}, and this is largely unaffected by the value of $\gamma'_{\rm m}$. So a decrease in $\xi_a$ gives a corresponding decrease of the optical flux. We note that we have been very conservative in our optical flux limit of 1 Jy, 
so once again we stress that high prompt optical spectral fluxes would be a characteristic signature of UHECR acceleration in llGRBs. Specifically in the case of GRB 060218, there are actually available optical observations that put the prompt peak V magnitude at 18.4 ($\sim 0.2$ mJy) \citep{GCN-4779}. This corresponds to a de-absorbed flux of $\sim 1$ mJy in which case the constraints on the maximum attainable iron energy at the source are much more severe ($E_{\rm max}^{\rm iron} < 10^{20}$ eV for $\Gamma = 10$, $\eta = 1$, $\xi_a = 0.01$, \added{and $\epsilon_e = 10^{-3}$}) (Samuelsson et al., in preparation).

Varying the redshift mainly effects the result as one would expect; higher redshift means longer distance traveled, hence lower observed fluxes. The effects due to changes of the relevant photon energies $\varepsilon_{\rm SSA}$, $\varepsilon_{\rm c}$, and $\varepsilon_{\rm m}$ with redshift, are small in comparison. Another parameter not discussed here that influence the results in the photospehric model is the electron injection slope $p$. Varying $p$ mostly influence the predicted flux around the sub-MeV peak. Smaller values of $p$, as expected in \textit{e.g.} magnetic reconnection models in the low-sigma regime \citep{Sironi2014}, lead to harsher constraints. 



\section{Conclusion}\label{Sec:Conslusion}
In this paper, we have studied if GRBs can be a source of UHECRs. First, we examined what parameter space would be allowed for the comoving magnetic field under the constraints that the acceleration time is shorter than typical loss times for UHECR. This part of the argumentation followed closely that of \citet{Waxman1995}, and many others after him, e.g. \citet{Murase2008, Guepin&Kotera2017}. We then investigated what such a magnetic field would imply for observations in the framework of optically thin models based on synchrotron radiation and of photospheric emission. We showed that for synchrotron models in high-luminosity GRBs, the electrons would be extremely fast cooling, in tension with observations because of the line-of-death problem. We also discussed the possibility of overcoming this problem with synchrotron self-Compton, and argued that this can not resolve the issue in this case. In models for high-luminosity GRBs where the observed sub-MeV peak is photospheric, we showed that the synchrotron flux from co-accelerated electrons in the UHECR acceleration region would overshoot the optical flux observed or outshine the photospheric component for most of the allowed parameter space. The acceleration is also limited by the fact that the magnetic luminosity cannot be larger than the total luminosity of the burst. In neither of the models could protons reach $10^{20}$ eV. This part of the paper only considered protons, as the radiation field at the base of the jet is so large that any heavier element is disintegrated \citep{Horiuchi2012,Zhang2018}. We then discussed the dependency of these results on the parameters. They were most sensitive to changes in the acceleration efficiency $\eta$, the total luminosity of the burst $L_{\rm tot}$, and the fraction of accelerated electrons $\xi_a$. Proton energies of $10^{20}$ eV could be reached in the photospheric model for canonical high-luminosity GRBs for $\eta = 1$ if $\Gamma \geq 300$.

Then we considered proton and iron acceleration in low-luminosity GRBs with typical luminosity $L_{\rm tot} = 10^{48}$ erg s$^{-1}$. These were severely constrained by the requirement that the magnetic luminosity cannot be larger than the total luminosity of the burst. To get high energy acceleration requires a rather high magnetic field far out in the jet implying a large magnetic luminosity, which is difficult to reconcile with a low total luminosity. \deleted{We found that}With our fiducial parameters, neither protons nor iron could reach $10^{20}$ eV in low-luminosity GRBs. 
\replaced{Additionally, we studied how the result varied with the parameters and}{However, we} found that iron could be accelerated to 10$^{20}$ eV and more, 
provided large values of $\eta$ ($\sim$ 1) \replaced{and}{and/or} small values of $\xi_a$ ($\sim$ 0.01-0.1) and $\epsilon_e$ ($\sim$ 0.01-0.1). \added{The latter values could be typical in low-luminosity GRBs if the shocks are not relativistic \citep{Park2015, Crumley2019}.} \replaced{We argued that this is due to our conservative choice for the optical flux limit, as in almost all cases and especially for slower outflows ($\Gamma \leq 50$), we find that the association of GRBs with UHECR would result in very high prompt optical fluxes.}{This acceleration would result in very high prompt optical fluxes, especially for slower outflows ($\Gamma \leq 50$). Detecting high optical fluxes early from a low-luminosity GRB could therefore indicate a successful UHECR source.} \added{However, we note that we have been very conservative in our choice of the optical flux limit of 1 Jy; the low-luminosity burst GRB 060218 where optical measurements are available, had a de-absorbed optical flux of 1 mJy \citep{Campana2006, Ghisellini2007}. In this specific case the constraints are much more severe, as we show in a forthcoming paper.}

\deleted{Our results calls for caution when evaluating the contribution of GRBs to UHECR with energy around 10$^{20}$ eV. Indeed, the photon spectrum and flux strongly constrain the maximum energy of such CRs. We find that for our fiducial parameters (given in Table \ref{tab:PhotosphereNumerics}), 
acceleration to $10^{20}$ eV is not achievable; their production requires fine tuned plasma parameters and emission model scenario under the constraints considered in this paper. In a forthcoming paper, we show that a direct application to GRB 060218 for which optical measurements exists, rule out the production of iron to 10$^{20}$ eV, unless an exotic model is invocated with $\eta = 1$, $\xi_a < 10^{-3}$, and $\epsilon_e < 10^{-3}$.}

\section*{Acknowledgements}
We thank Francesca Capel and Kohta Murase for interesting and fruitful discussions. 
We acknowledge support from the Swedish National Space Agency and the Swedish Research Council (Vetenskapsr{\aa}det). 
DB is supported by the Deutsche Forschungsgemeinschaft (SFB 1258).
FR is supported by the  G\"oran Gustafsson Foundation for Research in Natural Sciences and Medicine.
AP is partially supported by the European Research Council via ERC consolidating grant \#773062 (acronym O.M.J.).
Finally, FS thanks Angelica Alamaa for her support and patience with him, even when he is writing his first paper. 



\newpage
\bibliographystyle{mnras}
\bibliography{References_UHECR_2017} 

\begin{thebibliography}{}
\makeatletter
\relax
\def\mn@urlcharsother{\let\do\@makeother \do\$\do\&\do\#\do\^\do\_\do\%\do\~}
\def\mn@doi{\begingroup\mn@urlcharsother \@ifnextchar [ {\mn@doi@}
  {\mn@doi@[]}}
\def\mn@doi@[#1]#2{\def\@tempa{#1}\ifx\@tempa\@empty \href
  {http://dx.doi.org/#2} {doi:#2}\else \href {http://dx.doi.org/#2} {#1}\fi
  \endgroup}
\def\mn@eprint#1#2{\mn@eprint@#1:#2::\@nil}
\def\mn@eprint@arXiv#1{\href {http://arxiv.org/abs/#1} {{\tt arXiv:#1}}}
\def\mn@eprint@dblp#1{\href {http://dblp.uni-trier.de/rec/bibtex/#1.xml}
  {dblp:#1}}
\def\mn@eprint@#1:#2:#3:#4\@nil{\def\@tempa {#1}\def\@tempb {#2}\def\@tempc
  {#3}\ifx \@tempc \@empty \let \@tempc \@tempb \let \@tempb \@tempa \fi \ifx
  \@tempb \@empty \def\@tempb {arXiv}\fi \@ifundefined
  {mn@eprint@\@tempb}{\@tempb:\@tempc}{\expandafter \expandafter \csname
  mn@eprint@\@tempb\endcsname \expandafter{\@tempc}}}

\bibitem[\protect\citeauthoryear{{Aab} et~al.,}{{Aab}
  et~al.}{2018}]{PierreAugerCollaboration2018}
{Aab} A.,  et~al., 2018, \mn@doi [\apjl] {10.3847/2041-8213/aaa66d}, \href
  {http://adsabs.harvard.edu/abs/2018ApJ...853L..29A} {853, L29}

\bibitem[\protect\citeauthoryear{{Aartsen} et~al.,}{{Aartsen}
  et~al.}{2015}]{IceCubeCollaboration2015GRBLimit}
{Aartsen} M.~G.,  et~al., 2015, \mn@doi [\apjl] {10.1088/2041-8205/805/1/L5},
  \href {http://adsabs.harvard.edu/abs/2015ApJ...805L...5A} {805, L5}

\bibitem[\protect\citeauthoryear{Alfaro et~al.,}{Alfaro
  et~al.}{2017}]{HAWC2018}
Alfaro R.,  et~al., 2017, The Astrophysical Journal, 843, 88

\bibitem[\protect\citeauthoryear{{Baerwald}, {Bustamante}  \&
  {Winter}}{{Baerwald} et~al.}{2015}]{Baerwald2015}
{Baerwald} P.,  {Bustamante} M.,   {Winter} W.,  2015, \mn@doi [Astroparticle
  Physics] {10.1016/j.astropartphys.2014.07.007}, \href
  {http://adsabs.harvard.edu/abs/2015APh....62...66B} {62, 66}

\bibitem[\protect\citeauthoryear{{B{\'e}gu{\'e}}, {Pe'er}  \&
  {Lyubarsky}}{{B{\'e}gu{\'e}} et~al.}{2017}]{Begue2017}
{B{\'e}gu{\'e}} D.,  {Pe'er} A.,   {Lyubarsky} Y.,  2017, \mn@doi [\mnras]
  {10.1093/mnras/stx237}, \href
  {http://adsabs.harvard.edu/abs/2017MNRAS.467.2594B} {467, 2594}

\bibitem[\protect\citeauthoryear{{Beloborodov}}{{Beloborodov}}{2010}]{Beloborodov2010}
{Beloborodov} A.~M.,  2010, \mn@doi [\mnras]
  {10.1111/j.1365-2966.2010.16770.x}, \href
  {http://adsabs.harvard.edu/abs/2010MNRAS.407.1033B} {407, 1033}

\bibitem[\protect\citeauthoryear{{Beloborodov}}{{Beloborodov}}{2013}]{Beloborodov2013}
{Beloborodov} A.~M.,  2013, \mn@doi [\apj] {10.1088/0004-637X/764/2/157}, \href
  {http://adsabs.harvard.edu/abs/2013ApJ...764..157B} {764, 157}

\bibitem[\protect\citeauthoryear{{Beloborodov}}{{Beloborodov}}{2017}]{Beloborodov2017}
{Beloborodov} A.~M.,  2017, \mn@doi [\apj] {10.3847/1538-4357/aa5c8c}, \href
  {http://adsabs.harvard.edu/abs/2017ApJ...838..125B} {838, 125}

\bibitem[\protect\citeauthoryear{{Beniamini} \& {Giannios}}{{Beniamini} \&
  {Giannios}}{2017}]{Beniamini&Giannios2017}
{Beniamini} P.,  {Giannios} D.,  2017, \mn@doi [\mnras] {10.1093/mnras/stx717},
  \href {http://adsabs.harvard.edu/abs/2017MNRAS.468.3202B} {468, 3202}

\bibitem[\protect\citeauthoryear{{Beniamini} \& {Piran}}{{Beniamini} \&
  {Piran}}{2013}]{Beniamini2013}
{Beniamini} P.,  {Piran} T.,  2013, \mn@doi [\apj]
  {10.1088/0004-637X/769/1/69}, \href
  {http://adsabs.harvard.edu/abs/2013ApJ...769...69B} {769, 69}

\bibitem[\protect\citeauthoryear{{Beniamini} \& {van der Horst}}{{Beniamini} \&
  {van der Horst}}{2017}]{Beniamini2017}
{Beniamini} P.,  {van der Horst} A.~J.,  2017, \mn@doi [\mnras]
  {10.1093/mnras/stx2203}, \href
  {http://adsabs.harvard.edu/abs/2017MNRAS.472.3161B} {472, 3161}

\bibitem[\protect\citeauthoryear{{Beniamini}, {Barniol Duran}  \&
  {Giannios}}{{Beniamini} et~al.}{2018}]{Beniamini2018}
{Beniamini} P.,  {Barniol Duran} R.,   {Giannios} D.,  2018, preprint, \href
  {http://adsabs.harvard.edu/abs/2018arXiv180104944B} {} (\mn@eprint {arXiv}
  {1801.04944})

\bibitem[\protect\citeauthoryear{{Biehl}, {Boncioli}, {Lunardini}  \&
  {Winter}}{{Biehl} et~al.}{2017}]{Biehl2017}
{Biehl} D.,  {Boncioli} D.,  {Lunardini} C.,   {Winter} W.,  2017, preprint,
  \href {http://adsabs.harvard.edu/abs/2017arXiv171103555B} {} (\mn@eprint
  {arXiv} {1711.03555})

\bibitem[\protect\citeauthoryear{{Biehl}, {Boncioli}, {Fedynitch}  \&
  {Winter}}{{Biehl} et~al.}{2018}]{Biehl2018}
{Biehl} D.,  {Boncioli} D.,  {Fedynitch} A.,   {Winter} W.,  2018, \mn@doi
  [\aap] {10.1051/0004-6361/201731337}, \href
  {http://adsabs.harvard.edu/abs/2018A%26A...611A.101B} {611, A101}

\bibitem[\protect\citeauthoryear{{Blandford} \& {Znajek}}{{Blandford} \&
  {Znajek}}{1977}]{BlandfordZnajek1977}
{Blandford} R.~D.,  {Znajek} R.~L.,  1977, \mn@doi [\mnras]
  {10.1093/mnras/179.3.433}, \href
  {http://adsabs.harvard.edu/abs/1977MNRAS.179..433B} {179, 433}

\bibitem[\protect\citeauthoryear{{Blasi}, {Epstein}  \& {Olinto}}{{Blasi}
  et~al.}{2000}]{Blasi2000}
{Blasi} P.,  {Epstein} R.~I.,   {Olinto} A.~V.,  2000, \mn@doi [\apjl]
  {10.1086/312626}, \href {http://adsabs.harvard.edu/abs/2000ApJ...533L.123B}
  {533, L123}

\bibitem[\protect\citeauthoryear{{Boncioli}, {Biehl}  \& {Winter}}{{Boncioli}
  et~al.}{2019}]{Boncioli2019}
{Boncioli} D.,  {Biehl} D.,   {Winter} W.,  2019, \mn@doi [\apj]
  {10.3847/1538-4357/aafda7}, \href
  {http://adsabs.harvard.edu/abs/2019ApJ...872..110B} {872, 110}

\bibitem[\protect\citeauthoryear{{Bromberg}, {Nakar}  \& {Piran}}{{Bromberg}
  et~al.}{2011}]{Bromberg2011}
{Bromberg} O.,  {Nakar} E.,   {Piran} T.,  2011, \mn@doi [\apjl]
  {10.1088/2041-8205/739/2/L55}, \href
  {http://adsabs.harvard.edu/abs/2011ApJ...739L..55B} {739, L55}

\bibitem[\protect\citeauthoryear{{Budnik}, {Katz}, {MacFadyen}  \&
  {Waxman}}{{Budnik} et~al.}{2008}]{Budnik2008}
{Budnik} R.,  {Katz} B.,  {MacFadyen} A.,   {Waxman} E.,  2008, \mn@doi [\apj]
  {10.1086/524923}, \href {http://adsabs.harvard.edu/abs/2008ApJ...673..928B}
  {673, 928}

\bibitem[\protect\citeauthoryear{{Budnik}, {Katz}, {Sagiv}  \&
  {Waxman}}{{Budnik} et~al.}{2010}]{Budnik2010}
{Budnik} R.,  {Katz} B.,  {Sagiv} A.,   {Waxman} E.,  2010, \mn@doi [\apj]
  {10.1088/0004-637X/725/1/63}, \href
  {http://adsabs.harvard.edu/abs/2010ApJ...725...63B} {725, 63}

\bibitem[\protect\citeauthoryear{{Burgess}, {B{\'e}gu{\'e}}, {Bacelj},
  {Giannios}, {Berlato}  \& {Greiner}}{{Burgess} et~al.}{2018}]{Burgess2018}
{Burgess} J.~M.,  {B{\'e}gu{\'e}} D.,  {Bacelj} A.,  {Giannios} D.,  {Berlato}
  F.,   {Greiner} J.,  2018, arXiv e-prints, \href
  {http://adsabs.harvard.edu/abs/2018arXiv181006965B} {}

\bibitem[\protect\citeauthoryear{{Campana} et~al.,}{{Campana}
  et~al.}{2006}]{Campana2006}
{Campana} S.,  et~al., 2006, \mn@doi [\nat] {10.1038/nature04892}, \href
  {http://adsabs.harvard.edu/abs/2006Natur.442.1008C} {442, 1008}

\bibitem[\protect\citeauthoryear{{Caprioli} \& {Spitkovsky}}{{Caprioli} \&
  {Spitkovsky}}{2014}]{Caprioli2014}
{Caprioli} D.,  {Spitkovsky} A.,  2014, \mn@doi [\apj]
  {10.1088/0004-637X/794/1/47}, \href
  {http://adsabs.harvard.edu/abs/2014ApJ...794...47C} {794, 47}

\bibitem[\protect\citeauthoryear{{Crider} et~al.,}{{Crider}
  et~al.}{1997}]{Crider1997}
{Crider} A.,  et~al., 1997, \mn@doi [\apjl] {10.1086/310574}, \href
  {http://adsabs.harvard.edu/abs/1997ApJ...479L..39C} {479, L39}

\bibitem[\protect\citeauthoryear{{Crumley}, {Caprioli}, {Markoff}  \&
  {Spitkovsky}}{{Crumley} et~al.}{2019}]{Crumley2019}
{Crumley} P.,  {Caprioli} D.,  {Markoff} S.,   {Spitkovsky} A.,  2019, \mn@doi
  [\mnras] {10.1093/mnras/stz232}, \href
  {http://adsabs.harvard.edu/abs/2019MNRAS.tmp..277C} {}

\bibitem[\protect\citeauthoryear{{Daigne} \& {Mochkovitch}}{{Daigne} \&
  {Mochkovitch}}{1998}]{Daigne1998}
{Daigne} F.,  {Mochkovitch} R.,  1998, \mn@doi [\mnras]
  {10.1046/j.1365-8711.1998.01305.x}, \href
  {http://adsabs.harvard.edu/abs/1998MNRAS.296..275D} {296, 275}

\bibitem[\protect\citeauthoryear{{Daigne}, {Bo{\v s}njak}  \& {Dubus}}{{Daigne}
  et~al.}{2011}]{Daigne2011}
{Daigne} F.,  {Bo{\v s}njak} {\v Z}.,   {Dubus} G.,  2011, \mn@doi [\aap]
  {10.1051/0004-6361/201015457}, \href
  {http://adsabs.harvard.edu/abs/2011A%26A...526A.110D} {526, A110}

\bibitem[\protect\citeauthoryear{{Denton} \& {Tamborra}}{{Denton} \&
  {Tamborra}}{2018}]{Denton2018}
{Denton} P.~B.,  {Tamborra} I.,  2018, \mn@doi [\apj]
  {10.3847/1538-4357/aaab4a}, \href
  {http://adsabs.harvard.edu/abs/2018ApJ...855...37D} {855, 37}

\bibitem[\protect\citeauthoryear{{Dereli}, {Bo{\"e}r}, {Gendre}, {Amati},
  {Dichiara}  \& {Orange}}{{Dereli} et~al.}{2017}]{Dereli2017}
{Dereli} H.,  {Bo{\"e}r} M.,  {Gendre} B.,  {Amati} L.,  {Dichiara} S.,
  {Orange} N.~B.,  2017, \mn@doi [\apj] {10.3847/1538-4357/aa947d}, \href
  {http://adsabs.harvard.edu/abs/2017ApJ...850..117D} {850, 117}

\bibitem[\protect\citeauthoryear{{Drenkhahn} \& {Spruit}}{{Drenkhahn} \&
  {Spruit}}{2002}]{Drenkhahn2002}
{Drenkhahn} G.,  {Spruit} H.~C.,  2002, \mn@doi [\aap]
  {10.1051/0004-6361:20020839}, \href
  {http://adsabs.harvard.edu/abs/2002A%26A...391.1141D} {391, 1141}

\bibitem[\protect\citeauthoryear{{Eichler} \& {Waxman}}{{Eichler} \&
  {Waxman}}{2005}]{Eichler2005}
{Eichler} D.,  {Waxman} E.,  2005, \mn@doi [\apj] {10.1086/430596}, \href
  {http://adsabs.harvard.edu/abs/2005ApJ...627..861E} {627, 861}

\bibitem[\protect\citeauthoryear{{Fan}, {Zhang}, {Xu}, {Liang}  \&
  {Zhang}}{{Fan} et~al.}{2011}]{Fan2011}
{Fan} Y.-Z.,  {Zhang} B.-B.,  {Xu} D.,  {Liang} E.-W.,   {Zhang} B.,  2011,
  \mn@doi [\apj] {10.1088/0004-637X/726/1/32}, \href
  {http://adsabs.harvard.edu/abs/2011ApJ...726...32F} {726, 32}

\bibitem[\protect\citeauthoryear{{Fishman} \& {Meegan}}{{Fishman} \&
  {Meegan}}{1995}]{Fishman&Meegan1995}
{Fishman} G.~J.,  {Meegan} C.~A.,  1995, \mn@doi [\araa]
  {10.1146/annurev.aa.33.090195.002215}, \href
  {http://adsabs.harvard.edu/abs/1995ARA%26A..33..415F} {33, 415}

\bibitem[\protect\citeauthoryear{{Ghirlanda}, {Nava}, {Ghisellini}, {Celotti}
  \& {Firmani}}{{Ghirlanda} et~al.}{2009}]{Ghirlanda2009}
{Ghirlanda} G.,  {Nava} L.,  {Ghisellini} G.,  {Celotti} A.,   {Firmani} C.,
  2009, \mn@doi [\aap] {10.1051/0004-6361/200811209}, \href
  {http://adsabs.harvard.edu/abs/2009A%26A...496..585G} {496, 585}

\bibitem[\protect\citeauthoryear{{Ghisellini}, {Ghirlanda}  \&
  {Tavecchio}}{{Ghisellini} et~al.}{2007}]{Ghisellini2007}
{Ghisellini} G.,  {Ghirlanda} G.,   {Tavecchio} F.,  2007, \mn@doi [\mnras]
  {10.1111/j.1745-3933.2006.00270.x}, \href
  {http://adsabs.harvard.edu/abs/2007MNRAS.375L..36G} {375, L36}

\bibitem[\protect\citeauthoryear{{Giannios}}{{Giannios}}{2006}]{Giannos2006}
{Giannios} D.,  2006, \mn@doi [\aap] {10.1051/0004-6361:20065000}, \href
  {http://adsabs.harvard.edu/abs/2006A%26A...457..763G} {457, 763}

\bibitem[\protect\citeauthoryear{{Globus}, {Allard}, {Mochkovitch}  \&
  {Parizot}}{{Globus} et~al.}{2015}]{Globus2015}
{Globus} N.,  {Allard} D.,  {Mochkovitch} R.,   {Parizot} E.,  2015, \mn@doi
  [\mnras] {10.1093/mnras/stv893}, \href
  {http://adsabs.harvard.edu/abs/2015MNRAS.451..751G} {451, 751}

\bibitem[\protect\citeauthoryear{{Goldstein} et~al.,}{{Goldstein}
  et~al.}{2012}]{Goldstein2012}
{Goldstein} A.,  et~al., 2012, \mn@doi [\apjs] {10.1088/0067-0049/199/1/19},
  \href {http://adsabs.harvard.edu/abs/2012ApJS..199...19G} {199, 19}

\bibitem[\protect\citeauthoryear{{Goodman}}{{Goodman}}{1986}]{Goodman1986}
{Goodman} J.,  1986, \mn@doi [\apjl] {10.1086/184741}, \href
  {http://adsabs.harvard.edu/abs/1986ApJ...308L..47G} {308, L47}

\bibitem[\protect\citeauthoryear{{Greiner} et~al.,}{{Greiner}
  et~al.}{1996}]{Greiner1996}
{Greiner} J.,  et~al., 1996, in {Kouveliotou} C.,  {Briggs} M.~F.,   {Fishman}
  G.~J.,  eds,  American Institute of Physics Conference Series Vol. 384,
  American Institute of Physics Conference Series. pp 622--626,
  \mn@doi{10.1063/1.51581}

\bibitem[\protect\citeauthoryear{{Gu{\'e}pin} \& {Kotera}}{{Gu{\'e}pin} \&
  {Kotera}}{2017}]{Guepin&Kotera2017}
{Gu{\'e}pin} C.,  {Kotera} K.,  2017, \mn@doi [\aap]
  {10.1051/0004-6361/201630326}, \href
  {http://adsabs.harvard.edu/abs/2017A%26A...603A..76G} {603, A76}

\bibitem[\protect\citeauthoryear{{Guetta} \& {Della Valle}}{{Guetta} \& {Della
  Valle}}{2007}]{Guetta2007}
{Guetta} D.,  {Della Valle} M.,  2007, \mn@doi [\apjl] {10.1086/511417}, \href
  {http://adsabs.harvard.edu/abs/2007ApJ...657L..73G} {657, L73}

\bibitem[\protect\citeauthoryear{{He}, {Wang}, {Yu}  \&
  {M{\'e}sz{\'a}ros}}{{He} et~al.}{2009}]{He2009}
{He} H.-N.,  {Wang} X.-Y.,  {Yu} Y.-W.,   {M{\'e}sz{\'a}ros} P.,  2009, \mn@doi
  [\apj] {10.1088/0004-637X/706/2/1152}, \href
  {http://adsabs.harvard.edu/abs/2009ApJ...706.1152H} {706, 1152}

\bibitem[\protect\citeauthoryear{{Horiuchi}, {Murase}, {Ioka}  \&
  {M{\'e}sz{\'a}ros}}{{Horiuchi} et~al.}{2012}]{Horiuchi2012}
{Horiuchi} S.,  {Murase} K.,  {Ioka} K.,   {M{\'e}sz{\'a}ros} P.,  2012,
  \mn@doi [\apj] {10.1088/0004-637X/753/1/69}, \href
  {http://adsabs.harvard.edu/abs/2012ApJ...753...69H} {753, 69}

\bibitem[\protect\citeauthoryear{{H{\"u}mmer}, {R{\"u}ger}, {Spanier}  \&
  {Winter}}{{H{\"u}mmer} et~al.}{2010}]{Hummer2010}
{H{\"u}mmer} S.,  {R{\"u}ger} M.,  {Spanier} F.,   {Winter} W.,  2010, \mn@doi
  [\apj] {10.1088/0004-637X/721/1/630}, \href
  {http://adsabs.harvard.edu/abs/2010ApJ...721..630H} {721, 630}

\bibitem[\protect\citeauthoryear{{IceCube Collaboration} et~al.,}{{IceCube
  Collaboration} et~al.}{2018}]{IceCubeCollaboration2018}
{IceCube Collaboration} et~al., 2018, \mn@doi [Science]
  {10.1126/science.aat1378}, \href
  {http://adsabs.harvard.edu/abs/2018Sci...361.1378I} {361, eaat1378}

\bibitem[\protect\citeauthoryear{{Kehoe} et~al.,}{{Kehoe}
  et~al.}{2002}]{Kehoe2002}
{Kehoe} R.,  et~al., 2002, \mn@doi [\apj] {10.1086/342231}, \href
  {http://adsabs.harvard.edu/abs/2002ApJ...577..845K} {577, 845}

\bibitem[\protect\citeauthoryear{{Keivani} et~al.,}{{Keivani}
  et~al.}{2018}]{Keivani2018}
{Keivani} A.,  et~al., 2018, preprint, \href
  {http://adsabs.harvard.edu/abs/2018arXiv180704537K} {} (\mn@eprint {arXiv}
  {1807.04537})

\bibitem[\protect\citeauthoryear{{Klotz}, {Bo{\"e}r}, {Atteia}  \&
  {Gendre}}{{Klotz} et~al.}{2009}]{Klotz2009}
{Klotz} A.,  {Bo{\"e}r} M.,  {Atteia} J.~L.,   {Gendre} B.,  2009, \mn@doi
  [\aj] {10.1088/0004-6256/137/5/4100}, \href
  {http://adsabs.harvard.edu/abs/2009AJ....137.4100K} {137, 4100}

\bibitem[\protect\citeauthoryear{{Kobayashi}, {Piran}  \& {Sari}}{{Kobayashi}
  et~al.}{1999}]{Kobayashi1999}
{Kobayashi} S.,  {Piran} T.,   {Sari} R.,  1999, \mn@doi [\apj]
  {10.1086/306868}, \href {http://adsabs.harvard.edu/abs/1999ApJ...513..669K}
  {513, 669}

\bibitem[\protect\citeauthoryear{{Koyama}, {Petre}, {Gotthelf}, {Hwang},
  {Matsuura}, {Ozaki}  \& {Holt}}{{Koyama} et~al.}{1995}]{Koyama1995}
{Koyama} K.,  {Petre} R.,  {Gotthelf} E.~V.,  {Hwang} U.,  {Matsuura} M.,
  {Ozaki} M.,   {Holt} S.~S.,  1995, \mn@doi [\nat] {10.1038/378255a0}, \href
  {http://adsabs.harvard.edu/abs/1995Natur.378..255K} {378, 255}

\bibitem[\protect\citeauthoryear{{Lagage} \& {Cesarsky}}{{Lagage} \&
  {Cesarsky}}{1983}]{Lagage1983}
{Lagage} P.~O.,  {Cesarsky} C.~J.,  1983, \aap, \href
  {http://adsabs.harvard.edu/abs/1983A%26A...125..249L} {125, 249}

\bibitem[\protect\citeauthoryear{{Liu}, {Wang}  \& {Dai}}{{Liu}
  et~al.}{2011}]{Liu2011}
{Liu} R.-Y.,  {Wang} X.-Y.,   {Dai} Z.-G.,  2011, \mn@doi [\mnras]
  {10.1111/j.1365-2966.2011.19590.x}, \href
  {http://adsabs.harvard.edu/abs/2011MNRAS.418.1382L} {418, 1382}

\bibitem[\protect\citeauthoryear{{Lyutikov} \& {Blandford}}{{Lyutikov} \&
  {Blandford}}{2003}]{Lyutikov2003}
{Lyutikov} M.,  {Blandford} R.,  2003, ArXiv Astrophysics e-prints, \href
  {http://adsabs.harvard.edu/abs/2003astro.ph.12347L} {}

\bibitem[\protect\citeauthoryear{{Margutti}, {Guidorzi}  \&
  {Chincarini}}{{Margutti} et~al.}{2011}]{Margutti2011}
{Margutti} R.,  {Guidorzi} C.,   {Chincarini} G.,  2011, \mn@doi [International
  Journal of Modern Physics D] {10.1142/S0218271811020020}, \href
  {http://adsabs.harvard.edu/abs/2011IJMPD..20.1969M} {20, 1969}

\bibitem[\protect\citeauthoryear{{Marshall}, {Immler}  \&
  {Cusumano}}{{Marshall} et~al.}{2006}]{GCN-4779}
{Marshall} F.,  {Immler} S.,   {Cusumano} G.,  2006, GRB Coordinates Network,
  \href {http://adsabs.harvard.edu/abs/2006GCN..4779....1M} {4779}

\bibitem[\protect\citeauthoryear{{M{\'e}sz{\'a}ros}}{{M{\'e}sz{\'a}ros}}{2006}]{Meszaros2006}
{M{\'e}sz{\'a}ros} P.,  2006, \mn@doi [Reports on Progress in Physics]
  {10.1088/0034-4885/69/8/R01}, \href
  {http://adsabs.harvard.edu/abs/2006RPPh...69.2259M} {69, 2259}

\bibitem[\protect\citeauthoryear{{M{\"u}cke}, {Rachen}, {Engel}, {Protheroe}
  \& {Stanev}}{{M{\"u}cke} et~al.}{1999}]{Mucke1999}
{M{\"u}cke} A.,  {Rachen} J.~P.,  {Engel} R.,  {Protheroe} R.~J.,   {Stanev}
  T.,  1999, \mn@doi [\pasa] {10.1071/AS99160}, \href
  {http://adsabs.harvard.edu/abs/1999PASA...16..160M} {16, 160}

\bibitem[\protect\citeauthoryear{{Murase} \& {Ioka}}{{Murase} \&
  {Ioka}}{2013}]{Murase2013}
{Murase} K.,  {Ioka} K.,  2013, \mn@doi [Physical Review Letters]
  {10.1103/PhysRevLett.111.121102}, \href
  {http://adsabs.harvard.edu/abs/2013PhRvL.111l1102M} {111, 121102}

\bibitem[\protect\citeauthoryear{{Murase}, {Ioka}, {Nagataki}  \&
  {Nakamura}}{{Murase} et~al.}{2006}]{Murase2006}
{Murase} K.,  {Ioka} K.,  {Nagataki} S.,   {Nakamura} T.,  2006, \mn@doi
  [\apjl] {10.1086/509323}, \href
  {http://adsabs.harvard.edu/abs/2006ApJ...651L...5M} {651, L5}

\bibitem[\protect\citeauthoryear{{Murase}, {Ioka}, {Nagataki}  \&
  {Nakamura}}{{Murase} et~al.}{2008}]{Murase2008}
{Murase} K.,  {Ioka} K.,  {Nagataki} S.,   {Nakamura} T.,  2008, \mn@doi [\prd]
  {10.1103/PhysRevD.78.023005}, \href
  {http://adsabs.harvard.edu/abs/2008PhRvD..78b3005M} {78, 023005}

\bibitem[\protect\citeauthoryear{{Murase}, {Dermer}, {Takami}  \&
  {Migliori}}{{Murase} et~al.}{2012}]{Murase2012}
{Murase} K.,  {Dermer} C.~D.,  {Takami} H.,   {Migliori} G.,  2012, \mn@doi
  [\apj] {10.1088/0004-637X/749/1/63}, \href
  {http://adsabs.harvard.edu/abs/2012ApJ...749...63M} {749, 63}

\bibitem[\protect\citeauthoryear{{Nakar}, {Ando}  \& {Sari}}{{Nakar}
  et~al.}{2009}]{Nakar2009}
{Nakar} E.,  {Ando} S.,   {Sari} R.,  2009, \mn@doi [\apj]
  {10.1088/0004-637X/703/1/675}, \href
  {http://adsabs.harvard.edu/abs/2009ApJ...703..675N} {703, 675}

\bibitem[\protect\citeauthoryear{{Neronov}}{{Neronov}}{2017}]{Neronov2017}
{Neronov} A.,  2017, \mn@doi [Physical Review Letters]
  {10.1103/PhysRevLett.119.191102}, \href
  {http://adsabs.harvard.edu/abs/2017PhRvL.119s1102N} {119, 191102}

\bibitem[\protect\citeauthoryear{{Paczy{\'n}ski}}{{Paczy{\'n}ski}}{1986}]{Paczynski1986}
{Paczy{\'n}ski} B.,  1986, \mn@doi [\apjl] {10.1086/184740}, \href
  {http://adsabs.harvard.edu/abs/1986ApJ...308L..43P} {308, L43}

\bibitem[\protect\citeauthoryear{{Panaitescu} \& {Kumar}}{{Panaitescu} \&
  {Kumar}}{2000}]{Panaitescu2000}
{Panaitescu} A.,  {Kumar} P.,  2000, \mn@doi [\apj] {10.1086/317090}, \href
  {http://adsabs.harvard.edu/abs/2000ApJ...543...66P} {543, 66}

\bibitem[\protect\citeauthoryear{{Park}, {Caprioli}  \& {Spitkovsky}}{{Park}
  et~al.}{2015}]{Park2015}
{Park} J.,  {Caprioli} D.,   {Spitkovsky} A.,  2015, \mn@doi [Physical Review
  Letters] {10.1103/PhysRevLett.114.085003}, \href
  {http://adsabs.harvard.edu/abs/2015PhRvL.114h5003P} {114, 085003}

\bibitem[\protect\citeauthoryear{{Patrignani}, et al.  \& {(Particle data
  group)}}{{Patrignani} et~al.}{2016}]{CrossSectionPGamma2016}
{Patrignani} C.,  et al.  {(Particle data group)} 2016, Chinese Physics C, 40

\bibitem[\protect\citeauthoryear{{Pe'er}}{{Pe'er}}{2015}]{Peer2015}
{Pe'er} A.,  2015, \mn@doi [Advances in Astronomy] {10.1155/2015/907321}, \href
  {http://adsabs.harvard.edu/abs/2015AdAst2015E..22P} {2015, 907321}

\bibitem[\protect\citeauthoryear{{Pe'er}, {M{\'e}sz{\'a}ros}  \&
  {Rees}}{{Pe'er} et~al.}{2005}]{Peer2005}
{Pe'er} A.,  {M{\'e}sz{\'a}ros} P.,   {Rees} M.~J.,  2005, \mn@doi [\apj]
  {10.1086/497360}, \href {http://adsabs.harvard.edu/abs/2005ApJ...635..476P}
  {635, 476}

\bibitem[\protect\citeauthoryear{{Piran}, {Shemi}  \& {Narayan}}{{Piran}
  et~al.}{1993}]{Piran1993}
{Piran} T.,  {Shemi} A.,   {Narayan} R.,  1993, \mn@doi [\mnras]
  {10.1093/mnras/263.4.861}, \href
  {http://adsabs.harvard.edu/abs/1993MNRAS.263..861P} {263, 861}

\bibitem[\protect\citeauthoryear{{Preece}, {Briggs}, {Mallozzi}, {Pendleton},
  {Paciesas}  \& {Band}}{{Preece} et~al.}{1998}]{Preece1998}
{Preece} R.~D.,  {Briggs} M.~S.,  {Mallozzi} R.~S.,  {Pendleton} G.~N.,
  {Paciesas} W.~S.,   {Band} D.~L.,  1998, \mn@doi [\apjl] {10.1086/311644},
  \href {http://adsabs.harvard.edu/abs/1998ApJ...506L..23P} {506, L23}

\bibitem[\protect\citeauthoryear{{Protheroe} \& {Clay}}{{Protheroe} \&
  {Clay}}{2004}]{Protheroe2004}
{Protheroe} R.~J.,  {Clay} R.~W.,  2004, \mn@doi [\pasa] {10.1071/AS03047},
  \href {http://adsabs.harvard.edu/abs/2004PASA...21....1P} {21, 1}

\bibitem[\protect\citeauthoryear{{Rau}, {Greiner}  \& {Schwarz}}{{Rau}
  et~al.}{2006}]{Rau2006}
{Rau} A.,  {Greiner} J.,   {Schwarz} R.,  2006, \mn@doi [\aap]
  {10.1051/0004-6361:20054317}, \href
  {http://adsabs.harvard.edu/abs/2006A%26A...449...79R} {449, 79}

\bibitem[\protect\citeauthoryear{{Rees} \& {M{\'e}sz{\'a}ros}}{{Rees} \&
  {M{\'e}sz{\'a}ros}}{1992}]{Rees&Meszaros1992}
{Rees} M.~J.,  {M{\'e}sz{\'a}ros} P.,  1992, \mn@doi [\mnras]
  {10.1093/mnras/258.1.41P}, \href
  {http://adsabs.harvard.edu/abs/1992MNRAS.258P..41R} {258, 41P}

\bibitem[\protect\citeauthoryear{{Rees} \& {M{\'e}sz{\'a}ros}}{{Rees} \&
  {M{\'e}sz{\'a}ros}}{1994}]{Rees&Meszaros1994}
{Rees} M.~J.,  {M{\'e}sz{\'a}ros} P.,  1994, \mn@doi [\apjl] {10.1086/187446},
  \href {http://adsabs.harvard.edu/abs/1994ApJ...430L..93R} {430, L93}

\bibitem[\protect\citeauthoryear{{Rees} \& {M{\'e}sz{\'a}ros}}{{Rees} \&
  {M{\'e}sz{\'a}ros}}{2005}]{Rees&Meszaros2005}
{Rees} M.~J.,  {M{\'e}sz{\'a}ros} P.,  2005, \mn@doi [\apj] {10.1086/430818},
  \href {http://adsabs.harvard.edu/abs/2005ApJ...628..847R} {628, 847}

\bibitem[\protect\citeauthoryear{{Rieger}, {Bosch-Ramon}  \& {Duffy}}{{Rieger}
  et~al.}{2007}]{Rieger2007}
{Rieger} F.~M.,  {Bosch-Ramon} V.,   {Duffy} P.,  2007, \mn@doi [\apss]
  {10.1007/s10509-007-9466-z}, \href
  {http://adsabs.harvard.edu/abs/2007Ap%26SS.309..119R} {309, 119}

\bibitem[\protect\citeauthoryear{{Ruffini}, {Siutsou}  \&
  {Vereshchagin}}{{Ruffini} et~al.}{2013}]{Ruffini2013}
{Ruffini} R.,  {Siutsou} I.~A.,   {Vereshchagin} G.~V.,  2013, \mn@doi [\apj]
  {10.1088/0004-637X/772/1/11}, \href
  {http://adsabs.harvard.edu/abs/2013ApJ...772...11R} {772, 11}

\bibitem[\protect\citeauthoryear{{Rybicki} \& {Lightman}}{{Rybicki} \&
  {Lightman}}{1979}]{Rybicki&Lightman}
{Rybicki} G.~B.,  {Lightman} A.~P.,  1979, {Radiative processes in
  astrophysics}

\bibitem[\protect\citeauthoryear{{Ryde}}{{Ryde}}{2004}]{Ryde2004}
{Ryde} F.,  2004, \mn@doi [\apj] {10.1086/423782}, \href
  {http://adsabs.harvard.edu/abs/2004ApJ...614..827R} {614, 827}

\bibitem[\protect\citeauthoryear{{Ryde} et~al.,}{{Ryde}
  et~al.}{2010}]{Ryde2010}
{Ryde} F.,  et~al., 2010, \mn@doi [\apjl] {10.1088/2041-8205/709/2/L172}, \href
  {http://adsabs.harvard.edu/abs/2010ApJ...709L.172R} {709, L172}

\bibitem[\protect\citeauthoryear{{Ryde}, {Lundman}  \& {Acuner}}{{Ryde}
  et~al.}{2017}]{Ryde2017}
{Ryde} F.,  {Lundman} C.,   {Acuner} Z.,  2017, \mn@doi [\mnras]
  {10.1093/mnras/stx2019}, \href
  {http://adsabs.harvard.edu/abs/2017MNRAS.472.1897R} {472, 1897}

\bibitem[\protect\citeauthoryear{{Rykoff} et~al.,}{{Rykoff}
  et~al.}{2005}]{Rykoff2005}
{Rykoff} E.~S.,  et~al., 2005, \mn@doi [\apj] {10.1086/432832}, \href
  {http://adsabs.harvard.edu/abs/2005ApJ...631.1032R} {631, 1032}

\bibitem[\protect\citeauthoryear{{Santana}, {Barniol Duran}  \&
  {Kumar}}{{Santana} et~al.}{2014}]{Santana2014}
{Santana} R.,  {Barniol Duran} R.,   {Kumar} P.,  2014, \mn@doi [\apj]
  {10.1088/0004-637X/785/1/29}, \href
  {http://adsabs.harvard.edu/abs/2014ApJ...785...29S} {785, 29}

\bibitem[\protect\citeauthoryear{{Sari}, {Piran}  \& {Narayan}}{{Sari}
  et~al.}{1998}]{Sari1998}
{Sari} R.,  {Piran} T.,   {Narayan} R.,  1998, \mn@doi [\apjl]
  {10.1086/311269}, \href {http://adsabs.harvard.edu/abs/1998ApJ...497L..17S}
  {497, L17}

\bibitem[\protect\citeauthoryear{{Senno}, {Murase}  \&
  {M{\'e}sz{\'a}ros}}{{Senno} et~al.}{2016}]{Senno2016}
{Senno} N.,  {Murase} K.,   {M{\'e}sz{\'a}ros} P.,  2016, \mn@doi [\prd]
  {10.1103/PhysRevD.93.083003}, \href
  {http://adsabs.harvard.edu/abs/2016PhRvD..93h3003S} {93, 083003}

\bibitem[\protect\citeauthoryear{{Sironi} \& {Spitkovsky}}{{Sironi} \&
  {Spitkovsky}}{2014}]{Sironi2014}
{Sironi} L.,  {Spitkovsky} A.,  2014, \mn@doi [\apjl]
  {10.1088/2041-8205/783/1/L21}, \href
  {http://adsabs.harvard.edu/abs/2014ApJ...783L..21S} {783, L21}

\bibitem[\protect\citeauthoryear{{Soderberg} et~al.,}{{Soderberg}
  et~al.}{2006}]{Soderberg2006Nature}
{Soderberg} A.~M.,  et~al., 2006, \mn@doi [\nat] {10.1038/nature05087}, \href
  {http://adsabs.harvard.edu/abs/2006Natur.442.1014S} {442, 1014}

\bibitem[\protect\citeauthoryear{{Spruit}, {Daigne}  \& {Drenkhahn}}{{Spruit}
  et~al.}{2001}]{Spruit&Daigne&Drenkhahn2001}
{Spruit} H.~C.,  {Daigne} F.,   {Drenkhahn} G.,  2001, \mn@doi [\aap]
  {10.1051/0004-6361:20010131}, \href
  {http://adsabs.harvard.edu/abs/2001A%26A...369..694S} {369, 694}

\bibitem[\protect\citeauthoryear{{Starling} et~al.,}{{Starling}
  et~al.}{2011}]{Starling2011}
{Starling} R.~L.~C.,  et~al., 2011, \mn@doi [\mnras]
  {10.1111/j.1365-2966.2010.17879.x}, \href
  {http://adsabs.harvard.edu/abs/2011MNRAS.411.2792S} {411, 2792}

\bibitem[\protect\citeauthoryear{{Stecker}}{{Stecker}}{2000}]{Stecker2000}
{Stecker} F.~W.,  2000, \mn@doi [Astroparticle Physics]
  {10.1016/S0927-6505(00)00122-5}, \href
  {http://adsabs.harvard.edu/abs/2000APh....14..207S} {14, 207}

\bibitem[\protect\citeauthoryear{{Sudoh}, {Totani}  \& {Kawanaka}}{{Sudoh}
  et~al.}{2018}]{Sudoh2018}
{Sudoh} T.,  {Totani} T.,   {Kawanaka} N.,  2018, \mn@doi [\pasj]
  {10.1093/pasj/psy039}, \href
  {http://adsabs.harvard.edu/abs/2018PASJ..tmp...50S} {}

\bibitem[\protect\citeauthoryear{{Sun}, {Zhang}  \& {Li}}{{Sun}
  et~al.}{2015}]{Sun2015}
{Sun} H.,  {Zhang} B.,   {Li} Z.,  2015, \mn@doi [\apj]
  {10.1088/0004-637X/812/1/33}, \href
  {http://adsabs.harvard.edu/abs/2015ApJ...812...33S} {812, 33}

\bibitem[\protect\citeauthoryear{{Suzuki} \& {Shigeyama}}{{Suzuki} \&
  {Shigeyama}}{2010}]{Suzuki2010}
{Suzuki} A.,  {Shigeyama} T.,  2010, \mn@doi [\apj]
  {10.1088/0004-637X/719/1/881}, \href
  {http://adsabs.harvard.edu/abs/2010ApJ...719..881S} {719, 881}

\bibitem[\protect\citeauthoryear{{Telescope Array Collaboration}
  et~al.,}{{Telescope Array Collaboration} et~al.}{2018}]{TelescopeArray2018}
{Telescope Array Collaboration} et~al., 2018, preprint, \href
  {http://adsabs.harvard.edu/abs/2018arXiv180803680T} {} (\mn@eprint {arXiv}
  {1808.03680})

\bibitem[\protect\citeauthoryear{{The Fermi LAT Collaboration}}{{The Fermi LAT
  Collaboration}}{2018}]{FermiSwift2018}
{The Fermi LAT Collaboration} 2018, preprint, \href
  {http://adsabs.harvard.edu/abs/2018arXiv180801683T} {} (\mn@eprint {arXiv}
  {1808.01683})

\bibitem[\protect\citeauthoryear{{The Pierre Auger Collaboration} et~al.,}{{The
  Pierre Auger Collaboration} et~al.}{2017}]{PierreAugerICRC2017}
{The Pierre Auger Collaboration} et~al., 2017, preprint, \href
  {http://adsabs.harvard.edu/abs/2017arXiv170806592T} {} (\mn@eprint {arXiv}
  {1708.06592})

\bibitem[\protect\citeauthoryear{{Thoudam}, {Rachen}, {van Vliet},
  {Achterberg}, {Buitink}, {Falcke}  \& {H{\"o}randel}}{{Thoudam}
  et~al.}{2016}]{Thoudam2016}
{Thoudam} S.,  {Rachen} J.~P.,  {van Vliet} A.,  {Achterberg} A.,  {Buitink}
  S.,  {Falcke} H.,   {H{\"o}randel} J.~R.,  2016, \mn@doi [\aap]
  {10.1051/0004-6361/201628894}, \href
  {http://adsabs.harvard.edu/abs/2016A%26A...595A..33T} {595, A33}

\bibitem[\protect\citeauthoryear{{Usov}}{{Usov}}{1992}]{Usov1992}
{Usov} V.~V.,  1992, \mn@doi [\nat] {10.1038/357472a0}, \href
  {http://adsabs.harvard.edu/abs/1992Natur.357..472U} {357, 472}

\bibitem[\protect\citeauthoryear{{Vereshchagin}}{{Vereshchagin}}{2014}]{Vereshchagin2014}
{Vereshchagin} G.~V.,  2014, \mn@doi [International Journal of Modern Physics
  D] {10.1142/S0218271814300031}, \href
  {http://adsabs.harvard.edu/abs/2014IJMPD..2330003V} {23, 1430003}

\bibitem[\protect\citeauthoryear{{Vurm}, {Lyubarsky}  \& {Piran}}{{Vurm}
  et~al.}{2013}]{Vurm2013}
{Vurm} I.,  {Lyubarsky} Y.,   {Piran} T.,  2013, \mn@doi [\apj]
  {10.1088/0004-637X/764/2/143}, \href
  {http://adsabs.harvard.edu/abs/2013ApJ...764..143V} {764, 143}

\bibitem[\protect\citeauthoryear{{Waxman}}{{Waxman}}{1995}]{Waxman1995}
{Waxman} E.,  1995, \mn@doi [Physical Review Letters]
  {10.1103/PhysRevLett.75.386}, \href
  {http://adsabs.harvard.edu/abs/1995PhRvL..75..386W} {75, 386}

\bibitem[\protect\citeauthoryear{{Wijers} \& {Galama}}{{Wijers} \&
  {Galama}}{1999}]{Wijers1999}
{Wijers} R.~A.~M.~J.,  {Galama} T.~J.,  1999, \mn@doi [\apj] {10.1086/307705},
  \href {http://adsabs.harvard.edu/abs/1999ApJ...523..177W} {523, 177}

\bibitem[\protect\citeauthoryear{{Xiao}, {Liu}, {Dai}  \& {Wu}}{{Xiao}
  et~al.}{2017}]{Xiao2017}
{Xiao} D.,  {Liu} L.-D.,  {Dai} Z.-G.,   {Wu} X.-F.,  2017, \mn@doi [\apjl]
  {10.3847/2041-8213/aa9b2b}, \href
  {http://adsabs.harvard.edu/abs/2017ApJ...850L..41X} {850, L41}

\bibitem[\protect\citeauthoryear{{Yonetoku}, {Murakami}, {Tsutsui}, {Nakamura},
  {Morihara}  \& {Takahashi}}{{Yonetoku} et~al.}{2010}]{Yonetoku2010}
{Yonetoku} D.,  {Murakami} T.,  {Tsutsui} R.,  {Nakamura} T.,  {Morihara} Y.,
  {Takahashi} K.,  2010, \mn@doi [\pasj] {10.1093/pasj/62.6.1495}, \href
  {http://adsabs.harvard.edu/abs/2010PASJ...62.1495Y} {62, 1495}

\bibitem[\protect\citeauthoryear{{Yost} et~al.,}{{Yost}
  et~al.}{2007}]{Yost2007}
{Yost} S.~A.,  et~al., 2007, \mn@doi [\apj] {10.1086/521668}, \href
  {http://adsabs.harvard.edu/abs/2007ApJ...669.1107Y} {669, 1107}

\bibitem[\protect\citeauthoryear{{Yu} et~al.,}{{Yu} et~al.}{2016}]{Yu2016}
{Yu} H.-F.,  et~al., 2016, \mn@doi [\aap] {10.1051/0004-6361/201527509}, \href
  {http://adsabs.harvard.edu/abs/2016A%26A...588A.135Y} {588, A135}

\bibitem[\protect\citeauthoryear{{Zhang} \& {Murase}}{{Zhang} \&
  {Murase}}{2018}]{ZhangMurase2018}
{Zhang} B.~T.,  {Murase} K.,  2018, arXiv e-prints, \href
  {http://adsabs.harvard.edu/abs/2018arXiv181210289Z} {}

\bibitem[\protect\citeauthoryear{{Zhang} \& {Yan}}{{Zhang} \&
  {Yan}}{2011}]{Zhang&Yan2011}
{Zhang} B.,  {Yan} H.,  2011, \mn@doi [\apj] {10.1088/0004-637X/726/2/90},
  \href {http://adsabs.harvard.edu/abs/2011ApJ...726...90Z} {726, 90}

\bibitem[\protect\citeauthoryear{{Zhang}, {Murase}, {Kimura}, {Horiuchi}  \&
  {M{\'e}sz{\'a}ros}}{{Zhang} et~al.}{2018}]{Zhang2018}
{Zhang} B.~T.,  {Murase} K.,  {Kimura} S.~S.,  {Horiuchi} S.,
  {M{\'e}sz{\'a}ros} P.,  2018, \mn@doi [\prd] {10.1103/PhysRevD.97.083010},
  \href {http://adsabs.harvard.edu/abs/2018PhRvD..97h3010Z} {97, 083010}

\makeatother
\end{thebibliography}



\appendix
\section{Obtaining $F_{\nu, \rm opt}^{\rm sync}$ and $F_{\nu, \rm peak}^{\rm sync}$}\label{App:SSA}
The maximum spectral flux is given by
\begin{equation}\label{eq:F_nuc}
	F_{\nu, \rm max}^{\rm sync} = \frac{P_{\nu, \rm max} N_e}{4\pi d_l(z)^2}.
\end{equation}
where $N_e$ is the number of emitting electrons, $d_l(z)$ is the luminosity distance as a function of redshift $z$, and $P_{\nu, \rm max}$ is the maximum power per frequency given by
\begin{equation}
	P_{\nu, \rm max} =  \frac{\sigma_T m_e c^2}{3e} \Gamma B',
\end{equation}
\citep{Sari1998}. The number of emitting electrons at the emission region can be written as \citep{Peer2015}
\begin{equation}\label{eq:N_e}
	N_e = \xi_a n_e'V'  = \xi_a n_p'V' = \frac{\xi_a L_{\rm tot} }{4\pi r^2m_pc^3\Gamma^2} \times 4\pi r^2 \frac{r}{\Gamma} = \frac{\xi_a L_{\rm tot} r}{m_p c^3 \Gamma^3}.
\end{equation}
Because it is unknown what fraction of electrons injected from the central engine that are actually accelerated, this term comes with some acceleration fraction $\xi_a$. This $\xi_a$ can also incorporate the electrons produced from pair-production, although this number should be small far out in the ejecta \citep{Nakar2009}. The luminosity in the equation above is really the kinetic luminosity and not the total luminosity. However, above the saturation radius most of the available energy will be kinetic, and the kinetic luminosity will therefore be of the same order as the total luminosity at the UHECR acceleration region. This is also the case for magnetically dominated outflows, as can be seen in Figure 1 in \citet{Drenkhahn2002}. The fraction of order unity difference between the two luminosities can be included in $\xi_a$. As discussed in Section \ref{Sec:Discussion}, decreasing $\xi_a$ has the effect of lowering the observed fluxes as less electrons are accelerated, without affecting the constraint given on the magnetic luminosity in Equation \eqref{Blum}. Therefore, a smaller $\xi_a$ results in a larger parameter space. However, the fraction of accelerated electrons cannot be arbitrarily small, as this should cause the number of UHECR accelerated to also be small. There would then be a problem with not supplying enough power to match the observed total CR flux. If the number of UHECR accelerated is still high even though $\xi_a$ is small, then this deviation from symmetry requires an explanation. 

The cooling break frequency $\varepsilon_{\rm c}$ is calculated as in Equation \eqref{eq:varepsilon_c} and corrected for redshift, and $\varepsilon_{\rm m}$ is obtained by inserting $\gamma'_{\rm m} = a \frac{\epsilon_e}{\xi_a} \frac{m_p}{m_e}$ \citep{Meszaros2006} into Equation \eqref{varepsilon}, also corrected for redshift. Here, $\epsilon_e$ is the fraction of internal energy given to electrons, and $\xi_a$ is once again the fraction of electrons accelerated. The dependency on $\epsilon_e$ and $\xi_a$ is expected; if $\epsilon_e$ is small, less energy is available for the electrons and $\gamma_{\rm m}'$ is reduced, while if $\xi_a$ is small, fewer electrons share the energy and $\gamma_{\rm m }'$ will increase. The parameter $a$ is a constant of order a few for internal collisions and $\sim \Gamma$ for external collision. We use $a = 1$, which is certainly lower than the real value. The effect of increasing $a$ is to reduce the allowed parameter space, as this lead to larger values of $\varepsilon_{\rm m}$ resulting in higher predicted $F^{\rm sync}(\varepsilon_{\rm peak}^{\rm obs})$. 
The effect of varying $\epsilon_e$ is similar, in the sense that in our analysis, it only affects $\gamma_{\rm m}'$. Reducing $\epsilon_e$ reduces the peak flux by lowering the value of $\varepsilon_{\rm m}$. We use $\epsilon_e = 0.1$. 

The absorption coefficient for a power law distribution of electrons is given by \citep{Rybicki&Lightman} 
\begin{equation}
	\alpha_{\varepsilon'} = \frac{(s+2)c^2}{8\pi}\frac{h^2}{(\varepsilon')^2} \int d\gamma' \, P(\nu, \, \gamma') \frac{n'(\gamma')}{\gamma'm_ec^2},
\end{equation}
%
where $P(\nu, \, \gamma')$ is the power radiated in photons with observed frequency $\nu$ from an electron with comoving Lorentz factor $\gamma'$, and $n'$ is the comoving number density. The factor $s$ is the electron distribution power law index (corresponding to $p$ in \citet{Rybicki&Lightman}) and is given by $s = 2$ if $ \varepsilon'_{\rm c} < \varepsilon' < \varepsilon'_{\rm m}$, it is equal to the injection index $s = p$ if $\varepsilon'_{\rm m} < \varepsilon' < \varepsilon'_{\rm c}$, and $s = p+1$ if both $ \varepsilon'_{\rm c}, \varepsilon'_{\rm m} < \varepsilon'$. The integral above is the same as they solve in \citet{Sari1998} to get their parameterization, except the electron power-law slope is shifted down by 1, through the division of $\gamma'$. One can therefore use their parameterization in this case. Three things to note: 1) There is an additional factor $(\varepsilon')^{-2}$ in the expression for $\alpha_{\varepsilon'}$ outside of the integral, 2) lowering the index of the electron power law by 1, only lowers the index of the absorption spectrum by one half and 3) the lower part of the spectrum with slope 1/3 is from the low energy tail of the synchrotron emission, which is unaffected by the new power law slope. Thus,
\begin{equation}\label{eq:absorptionCoefficient}
\begin{split}
	\alpha_{\varepsilon'} = \frac{(s+2)c^2}{8\pi } \frac{h^2}{(\varepsilon'_{\rm c})^2}\xi_a n'_e \tilde P_{\nu, \rm max} \times
	\begin{cases}
		 	\left(\frac{\varepsilon'}{\varepsilon'_{\rm c}}\right)^{-5/3}  \quad & \varepsilon' < \varepsilon'_{\rm c}, \\
		\left(\frac{\varepsilon'}{\varepsilon'_{\rm c}}\right)^{-3} \quad &  \varepsilon'_{\rm c} <  \varepsilon' <  \varepsilon'_{\rm m}, \\
		\left(\frac{\varepsilon'_{\rm m}}{\varepsilon'_{\rm c}}\right)^{-3} \left(\frac{\varepsilon'}{\varepsilon'_{\rm m}}\right)^{-(p+5)/2} \quad & \varepsilon'_{\rm m} <  \varepsilon',
	\end{cases}
\end{split}
\end{equation}
for fast cooling, where $\tilde P_{\nu, \rm max} = P_{\nu, \rm max}/(\gamma'_{\rm c} m_ec^2)$. 
For slow cooling, it becomes 
\begin{equation}\label{eq:absorptionCoefficient_SC}
\begin{split}
	\alpha_{\varepsilon'} = \frac{(s+2)c^2}{8\pi } \frac{h^2}{(\varepsilon'_{\rm m})^2} \xi_a n'_e \tilde P_{\nu, \rm max} \times
	\begin{cases}
		 \left(\frac{\varepsilon'}{\varepsilon'_{\rm m}}\right)^{-5/3}  \quad & \varepsilon' <  \varepsilon'_{\rm m}, \\
		\left(\frac{\varepsilon'}{\varepsilon'_{\rm m}}\right)^{-(p+4)/2} \quad &  \varepsilon'_{\rm m} <  \varepsilon' < \varepsilon'_{\rm c}, \\
		\left(\frac{\varepsilon'_{\rm c}}{\varepsilon'_{\rm m}}\right)^{-(p+4)/2} \left(\frac{\varepsilon'}{\varepsilon'_{\rm c}}\right)^{-(p+5)/2} \quad & \varepsilon'_{\rm c} <  \varepsilon',
	\end{cases}
\end{split}
\end{equation}
where in this scenario $\tilde P_{\nu, \rm max} = P_{\nu, \rm max}/(\gamma'_{\rm m} m_ec^2)$. The material becomes optically thick to synchrotron self absorption (SSA) once the optical depth becomes unity, $\tau_{\rm SSA} = 1$. Assuming constant magnetic field over the width of the acceleration region, this gives $\alpha_{\varepsilon'_{\rm SSA}} = \tau_{SSA}\Gamma/r$. 

If SSA was not important, then the maximum synchrotron flux $F_{\nu, \rm max}^{\rm sync}$ would occur at $\min [\varepsilon_{\rm c}, \varepsilon_{\rm m}]$, because most electrons have Lorentz factor $\min [\gamma'_{\rm c}, \gamma'_{\rm m}]$. However, if $\varepsilon_{\rm SSA}$ is larger than this value, then the electrons pile up at $\gamma'_{\rm SSA}$ instead, as electrons with $\gamma'_e < \gamma'_{\rm SSA}$ are heated. This justifies the radiative index of 2 below $\varepsilon_{\rm SSA}$ \citep{Rybicki&Lightman}. The maximum flux from synchrotron emission therefore occurs at $\max [\varepsilon_{\rm SSA}, \ \min [\varepsilon_{\rm c}, \varepsilon_{\rm m}]]$. Because of the steep decline in the number of electrons with increasing energy, almost all electrons emit at this energy and using a delta approximation for these electrons' synchrotron emission, one can use Equation \eqref{eq:F_nuc}. Knowing $F_{\nu, \rm max}^{\rm sync}$, $\varepsilon_{\rm SSA}$, $\varepsilon_{\rm c}$, and $\varepsilon_{\rm m}$, the flux at the optical band and the $\sim$ 100 keV band can be calculated. In Table \ref{tab:FluxCases}, we have summarized how we calculate the spectral flux for all different permutations of $\varepsilon_{\rm SSA}$, $\varepsilon_{\rm c}$, and $\varepsilon_{\rm m}$ \citep{Sari1998}. There are 24 possible permutations including the energy of interest. However, many of these cases are similar. 

%
%
%
%

\begin{table}
\begin{center}
\caption{Numerical values used}
  \begin{tabular}{ l | l | r }\label{tab:PhotosphereNumerics}
  	Quantity & Symbol & Value used \\ \hline
  	\hline
    Acceleration efficiency & $\eta$ & $0.1$ \\ [2.3mm]
    Typical observed photon energy & $\left< \varepsilon \right>$ & $300$ keV\\ [2.3mm]
  	Optical energy & $\varepsilon_{\rm opt}$ & 2 eV \\[2.3mm]
    Observed peak energy & $\varepsilon_{\rm peak}^{\rm obs}$ & 300 keV\\[2.3mm]
  	Observed optical flux & $F_{\nu, \rm opt}^{\rm lim}$ & 100 mJy\\[2.3mm]
    Observed peak flux & $F_{\nu, \rm peak}^{\rm lim}$ & 2 mJy \\[2.3mm]
  	Redshift & $z$ & $1$ \\ [2.3mm]
    Observed low-luminosity peak energy & $\varepsilon_{\rm ll, peak}^{\rm obs}$ & 100 keV \\[2.3mm]
    Observed low-luminosity optical flux & $F_{\rm ll, \nu, opt}^{\rm lim}$ & 1 Jy\\[2.3mm]
    Observed low-luminosity peak flux & $F_{\rm ll,\nu, peak}^{\rm lim}$ & 0.1 mJy \\[2.3mm]
  	Redshift low-luminosity & $z_{\rm ll}$ & $0.05$ \\ [2.3mm]
  	Electron energy fraction & $\epsilon_e$ & 0.1 \\[2.3mm]
  	Electron slope & $p$ & 2.5 \\[2.3mm]
    Electron acceleration fraction & $\xi_a$ & 1 \\[2.3mm]  
    Constant in $\gamma'_{\rm m}$ & $a$ & 1
  \end{tabular}
\end{center}
\end{table}
\begin{table}
\begin{center}
\caption{How the spectral flux is calculated in each case.}
\label{tab:FluxCases}
\begin{tabular}{ l | l | l }
	Case number \\ top to bottom & Relative positions & Calculated spectral flux\\ 
    \hline
	\hline
	& & \\
	$1-4$ & $\varepsilon_{\rm SSA} < \varepsilon_{\rm c} < \varepsilon_{\rm m} $ & 
	$
	F_\nu^{\rm sync} ( \varepsilon) = F_{\nu, \rm{max}}^{\rm{sync}}\times
	\begin{cases}
		\left(\frac{\varepsilon_{\rm SSA}}{\varepsilon_{\rm c}}\right)^{1/3} \left(\frac{\varepsilon}{\varepsilon_{\rm SSA}}		 		\right)^2, \quad &\varepsilon < \varepsilon_{\rm SSA} \\
		\left(\frac{\varepsilon}{\varepsilon_{\rm c}}\right)^{1/3}, \quad &\varepsilon_{\rm SSA} < \varepsilon < \varepsilon_{\rm 		c} \\
		\left(\frac{\varepsilon}{\varepsilon_{\rm c}}\right)^{-1/2}, \quad &\varepsilon_{\rm c} < \varepsilon < \varepsilon_{\rm m} 		\\
		\left(\frac{\varepsilon_{\rm m}}{\varepsilon_{\rm c}}\right)^{-1/2} \left(\frac{\varepsilon}{\varepsilon_{\rm m}}\right)^{-p/2}, 		\quad &\varepsilon_{\rm m}  < \varepsilon
	\end{cases}
	$	
	\\ 
	& & \\
	\hline
	& & \\
	$5-8$ & $\varepsilon_{\rm SSA} < \varepsilon_{\rm m} < \varepsilon_{\rm c} $ & 
	$
	F_\nu^{\rm sync}  ( \varepsilon) = F_{\nu, \rm{max}}^{\rm{sync}}\times
	\begin{cases}
		\left(\frac{\varepsilon_{\rm SSA}}{\varepsilon_{\rm m}}\right)^{1/3} \left(\frac{\varepsilon}{\varepsilon_{\rm SSA}}		 		\right)^2, \quad &\varepsilon < \varepsilon_{\rm SSA} \\
		\left(\frac{\varepsilon}{\varepsilon_{\rm m}}\right)^{1/3}, \quad &\varepsilon_{\rm SSA} < \varepsilon < \varepsilon_{\rm 		m} \\ 
		\left(\frac{\varepsilon}{\varepsilon_{\rm m}}\right)^{-(p-1)/2}, \quad &\varepsilon_{\rm m} < \varepsilon < 					\varepsilon_{\rm c} \\
		\left(\frac{\varepsilon_{\rm c}}{\varepsilon_{\rm m}}\right)^{-(p-1)/2} \left(\frac{\varepsilon}{\varepsilon_{\rm c}}\right)^{-p/		2}, \quad &\varepsilon_{\rm c}  < \varepsilon
	\end{cases}
	$	
	\\ 
	& & \\
	\hline
	& &  \\
	$9-11$ & $\varepsilon_{\rm c} < \varepsilon_{\rm SSA} < \varepsilon_{\rm m} $ & 
	$
	F_\nu^{\rm sync}  ( \varepsilon) = F_{\nu, \rm{max}}^{\rm{sync}}\times
	\begin{cases}
		\left(\frac{\varepsilon}{\varepsilon_{\rm SSA}} \right)^2, \quad &\varepsilon < \varepsilon_{\rm SSA} \\
		\left(\frac{\varepsilon}{\varepsilon_{\rm SSA}}\right)^{-1/2}, \quad &\varepsilon_{\rm SSA} < \varepsilon < 					\varepsilon_{\rm m} \\
		\left(\frac{\varepsilon_{\rm m}}{\varepsilon_{\rm SSA}}\right)^{-1/2} \left(\frac{\varepsilon}{\varepsilon_{\rm m}}\right)^{-		p/2}, \quad &\varepsilon_{\rm m}  < \varepsilon
	\end{cases}
	$	
	\\ 
	& & \\
	\hline
	& & \\
	$12-14$ & $\varepsilon_{\rm m} < \varepsilon_{\rm SSA} < \varepsilon_{\rm c} $ & 
	$
	F_\nu^{\rm sync}  ( \varepsilon) = F_{\nu, \rm{max}}^{\rm{sync}}\times
	\begin{cases}
		\left(\frac{\varepsilon}{\varepsilon_{\rm SSA}} \right)^2, \quad &\varepsilon < \varepsilon_{\rm SSA} \\
		\left(\frac{\varepsilon}{\varepsilon_{\rm SSA}}\right)^{-(p-1)/2}, \quad &\varepsilon_{\rm SSA} < \varepsilon < 					\varepsilon_{\rm c} \\
		\left(\frac{\varepsilon_{\rm c}}{\varepsilon_{\rm SSA}}\right)^{-(p-1)/2} \left(\frac{\varepsilon}{\varepsilon_{\rm c}}				\right)^{-p/2}, \quad &\varepsilon_{\rm c}  < \varepsilon
	\end{cases}
	$	
	\\ 
	& & \\
	\hline
	& & \\
	$15-16$ & $\varepsilon_{\rm m},\, \varepsilon_{\rm c} < \varepsilon_{\rm SSA} $ & 
	$
	F_\nu^{\rm sync}  ( \varepsilon) = F_{\nu, \rm{max}}^{\rm{sync}}\times
	\begin{cases}
		\left(\frac{\varepsilon}{\varepsilon_{\rm SSA}} \right)^2, \quad &\varepsilon < \varepsilon_{\rm SSA} \\
		\left(\frac{\varepsilon}{\varepsilon_{\rm SSA}}\right)^{-p/2}, \quad &\varepsilon_{\rm SSA}  < \varepsilon
	\end{cases}
	$
	\\
	& & \\
\end{tabular}
\end{center}
\end{table}
\label{lastpage}
\end{document}